\pdfoutput=1
\documentclass[journal]{IEEEtai}

\usepackage{balance} 
\usepackage{microtype}
\usepackage{graphicx}
\usepackage{subfigure}
\usepackage{booktabs} 
\usepackage{algorithm}
\usepackage{algpseudocode}
\usepackage{algorithmicx}
\usepackage{float}
\usepackage{amsfonts}
\usepackage{enumerate}
\usepackage{amsmath,bm}
\usepackage[marginal]{footmisc}
\usepackage{caption}
\usepackage{multirow}
\usepackage{makecell}
\usepackage[table]{xcolor}
\usepackage[colorlinks,urlcolor=blue,linkcolor=blue,citecolor=blue]{hyperref}
\usepackage{color,array}
\usepackage{graphicx,cite}

\usepackage[english]{babel}
\usepackage{amsthm}

\makeatletter
\let\remark\@undefined
\newtheorem{remark}{Remark}
\makeatother
\newtheorem{theorem}{Theorem}

\begin{document}

\title{Learning Multi-agent Skills for Tabular Reinforcement Learning using Factor Graphs}

\author{Jiayu Chen, Jingdi Chen, Tian Lan, and Vaneet Aggarwal 
\thanks{Jiayu Chen and Vaneet Aggarwal are with the School of Industrial Engineering, Purdue University, West Lafayette IN 47907, USA, email: \{chen3686, vaneet\}@purdue.edu. V. Aggarwal is also with Computer Science, King Abdullah University Of Science And Technology,  Thuwal 23955, Saudi Arabia.} 
\thanks{Jingdi Chen and Tian Lan are with the Department of Electrical and Computer Engineering, George Washington University, Washinton DC 20052, USA, email:\{jingdic, tlan\}@gwu.edu.}
\thanks{This work was supported in part by Meta Platforms, Inc., Cisco Systems, Inc., and the U.S. National Science Foundation under grant 2114415.}}

\maketitle 

\begin{abstract}

Covering skill (a.k.a., option) discovery has been developed to improve the exploration of reinforcement learning in single-agent scenarios, where only sparse reward signals are available. It aims to connect the most distant states identified through the Fiedler vector of the state transition graph. However, the approach cannot be directly extended to multi-agent scenarios, since the joint state space grows exponentially with the number of agents thus prohibiting efficient option computation. Existing research adopting options in multi-agent scenarios still relies on single-agent algorithms and fails to directly discover joint options that can improve the connectivity of the joint state space. 

In this paper, we propose a new algorithm to directly compute multi-agent options with collaborative exploratory behaviors while still enjoying the ease of decomposition. Our key idea is to approximate the joint state space as the Kronecker product of individual agents' state spaces, based on which we can directly estimate the Fiedler vector of the joint state space using the Laplacian spectrum of individual agents' transition graphs. This decomposition enables us to efficiently construct multi-agent joint options by encouraging agents to connect the sub-goal joint states which are corresponding to the minimum or maximum of the estimated joint Fiedler vector. Evaluation on multi-agent collaborative tasks shows that 
our algorithm can successfully identify multi-agent options, and significantly outperforms prior works using single-agent options or no options, in terms of both faster exploration and higher cumulative rewards.

\end{abstract}

\begin{IEEEImpStatement}
Multi-agent reinforcement learning (MARL) has become increasingly important due to growing complexity of real-world decision making problems. A key performance bottleneck for MARL is the lack of efficient coordinated exploration among multiple agents. The proposed multi-agent option discovery approach addresses this problem by alleviating the exponential complexity involved in multi-agent explorations. The approach achieves significantly improved exploration and higher cumulative rewards in challenging multi-agent decision making scenarios. 
\end{IEEEImpStatement}


\section{Introduction}
\IEEEPARstart{O}{ption} discovery \cite{DBLP:journals/ai/SuttonPS99} enables temporally-abstract actions to be constructed and utilized in the reinforcement learning (RL) process, which can significantly improve the performance of RL agents. Among recent developments on the topic, \textit{Covering Option Discovery} \cite{DBLP:conf/icml/JinnaiAHLK19, DBLP:conf/icml/JinnaiPAK19} has been shown to be an effective approach to improve the exploration in environments with sparse reward signals. In particular, it first computes the second smallest eigenvalue and the corresponding eigenvector (i.e., Fiedler vector \cite{fiedler1973algebraic}) of the Laplacian matrix extracted from the state transition process in RL. Then, options are built to connect the states corresponding to the minimum or maximum values in the Fiedler vector, which has been proved to greedily improve the algebraic connectivity of the state space \cite{fast_graphs}. With these options, the accessibility from each state to the other states (especially the infrequently visited states) will be enhanced, due to which the exploration in the state space can be accelerated a lot.

In this paper, we consider the problem of constructing and utilizing covering options in multi-agent reinforcement learning (MARL). Due to the exponentially-large state space in multi-agent scenarios, a commonly-adopted way to solve this problem \cite{DBLP:conf/atal/AmatoKK14, amato2019modeling, shen2006multi, DBLP:conf/atal/ChakravortyWRCB20, DBLP:conf/iclr/LeeYL20} is to construct the single-agent options as if in a single-agent environment first, and then learn to collectively leverage these individual options to tackle multi-agent tasks. This method fails to consider the coordination among agents in the option discovery process, and thus can suffer from very poor behavior in multi-agent collaborative tasks. To this end, in our work, we propose a framework that makes novel use of Kronecker product of factor graphs to directly construct the multi-agent options in the joint state space, and adopt them to accelerate the joint exploration of agents in MARL. We show through experiments that agents leveraging our multi-agent options significantly outperform agents with single-agent options or no options in MARL tasks. For some challenging tasks, the adoption of multi-agent options can improve the convergence speed by two orders of magnitude and the episodic cumulative reward by $\sim 100\%$. Also, instead of directly adopting the \textit{Covering Option Discovery} to the joint state space since its size grows exponentially with the number of agents, we build multi-agent options based on the individual state transition graphs, making our method much more scalable.

Specifically, the main contributions are as follows: (1) We propose \textit{Multi-agent Covering Option Discovery} -- it approximates the joint state transition graph as a Kronecker product of the individual ones, so that we can estimate the Fiedler vector of the joint state space based on the Laplacian spectrum of the individual state spaces to enjoy the ease of decomposition. Then, the joint options composed of multiple agents’ temporal action sequences can be directly constructed to connect the joint states corresponding to the minimum or maximum in the Fiedler vector, resulting in a greedy improvement of the joint state space's algebraic connectivity. (2) We propose that the multi-agent options can be adopted to MARL in either a decentralized or centralized manner. For the centralized manner, different agents jointly decide on their options. In contrast, for the decentralized manner, agents can choose their options independently and select different options to execute simultaneously. Further, we compare the decentralized or centralized approaches through experiments.
(3) We propose group-based multi-agent option discovery that first groups the agents based on their sub-tasks, and then discover the multi-agent options within each sub-group. This sub-group division not only makes our approach more scalable but also makes the option discovery more accurate.

The rest of this paper is organized as follows. Section \ref{related} introduces some related works and highlights the innovation of this paper. Section \ref{background} presents the required basic conceptions and background knowledge on option discovery. Section \ref{alg} explains the proposed approach in detail, including how to discover and utilize multi-agent options in MARL and the theoretical foundations of our approach. Section \ref{evaluation} describes the simulation setup, and thoroughly compares our algorithm with two baselines from prior works: MARL without option discovery, and MARL with single-agent option discovery. Section \ref{conc} concludes this paper with a discussion of the future directions.

\section{Related Work} \label{related}
\noindent \textbf{Option Discovery: }The option framework was proposed in \cite{DBLP:journals/ai/SuttonPS99}, which extends the usual notion of actions to include options — the closed-loop policies for taking actions over a period of time. In the literature, lots of option discovery algorithms have been proposed. Some of them are based on task-related reward signals, such as \cite{DBLP:conf/icml/McGovernB01, Menache02q-cut-, DBLP:conf/nips/MankowitzMM16, Harb2018WhenWI}. Specifically, they directly define or learn through gradient descent the options that can lead the agent to the rewarding states in the environments, and then utilize these trajectory segments (options) to compose the completed trajectory toward the goal state. These methods rely on dense reward signals, which are usually hard to acquire in real-life tasks. Other works define the sub-goal states (termination states of the options) based on the visitation frequency of the states. For example, in \cite{stolle2002learning, DBLP:conf/icml/SimsekWB05, DBLP:conf/nips/SimsekB08}, they discover the options by recognizing the bottleneck states in the environment, through which the agent can transfer between the sub-areas that are loosely connected in the state space, and they define these options as betweenness options. Recently, there are some state-of-the-art option generation methods based on the Laplacian spectrum of the state-transition graph, such as \cite{DBLP:journals/corr/MachadoBB17, DBLP:conf/iclr/MachadoRGLTC18, DBLP:conf/icml/JinnaiAHLK19, DBLP:conf/icml/JinnaiPAK19}, since the eigenvectors of the Laplacian of the state space can provide embeddings in lower-dimensional space, based on which we can obtain good measurements of the accessibility from one state to another. Especially, in \cite{DBLP:conf/icml/JinnaiPAK19}, they propose covering options and prove that their option generation method has higher exploration speed and better performance compared with other Laplacian-based approaches and the betweenness options mentioned above.

Note that all the approaches mentioned above are for single-agent scenarios, and in this paper we will extend the construction and adoption of covering options to MARL.

\noindent\textbf{Adopting options in multi-agent scenarios: }Most of the researches about adopting options in MARL, such as \cite{DBLP:conf/atal/AmatoKK14, amato2019modeling, shen2006multi, DBLP:conf/atal/ChakravortyWRCB20, DBLP:conf/iclr/LeeYL20, DBLP:conf/atal/YangBZ20}, try to first learn the options for each individual agent with the option discovery methods we mentioned above, and then learn to collaboratively utilize these individual options. Therefore, the options they use are still single-agent options, and the coordination in the multi-agent system can only be shown/utilized in the option-choosing process while not the option discovery process. We can classify these works by the option discovery methods they use: the algorithms in \cite{DBLP:conf/atal/AmatoKK14,amato2019modeling} directly define the options based on their task without the learning process; the algorithms in \cite{shen2006multi, DBLP:conf/atal/ChakravortyWRCB20, DBLP:conf/iclr/LeeYL20} learn the options based on the task-related reward signals generated by the environment; the algorithm in \cite{DBLP:conf/atal/YangBZ20} trains the options based on a reward function that is a weighted summation of the environment reward and the information theoretic reward term proposed in \cite{DBLP:conf/iclr/EysenbachGIL19}.

In this paper, we propose constructing multi-agent covering options from joint state transition graphs based on the aforementioned Laplacian-based framework to encourage efficient exploration in the joint state space, and explore how to utilize the multi-agent options in MARL tasks effectively. Further, we propose to model the joint state transition graph as a Kronecker graph \cite{leskovec_kronecker_2010} -- the Kronecker product of the agents' individual state transition graphs, so that we can discover the joint options based on the individual state spaces, which makes our approach much more scalable.

\begin{figure*}[t]
	\centering
	\subfigure[The joint state transition graph of agent 1 and agent 2]{
	\label{fig:-1(a)}
	\includegraphics[height=1.0in, width=3.4in]{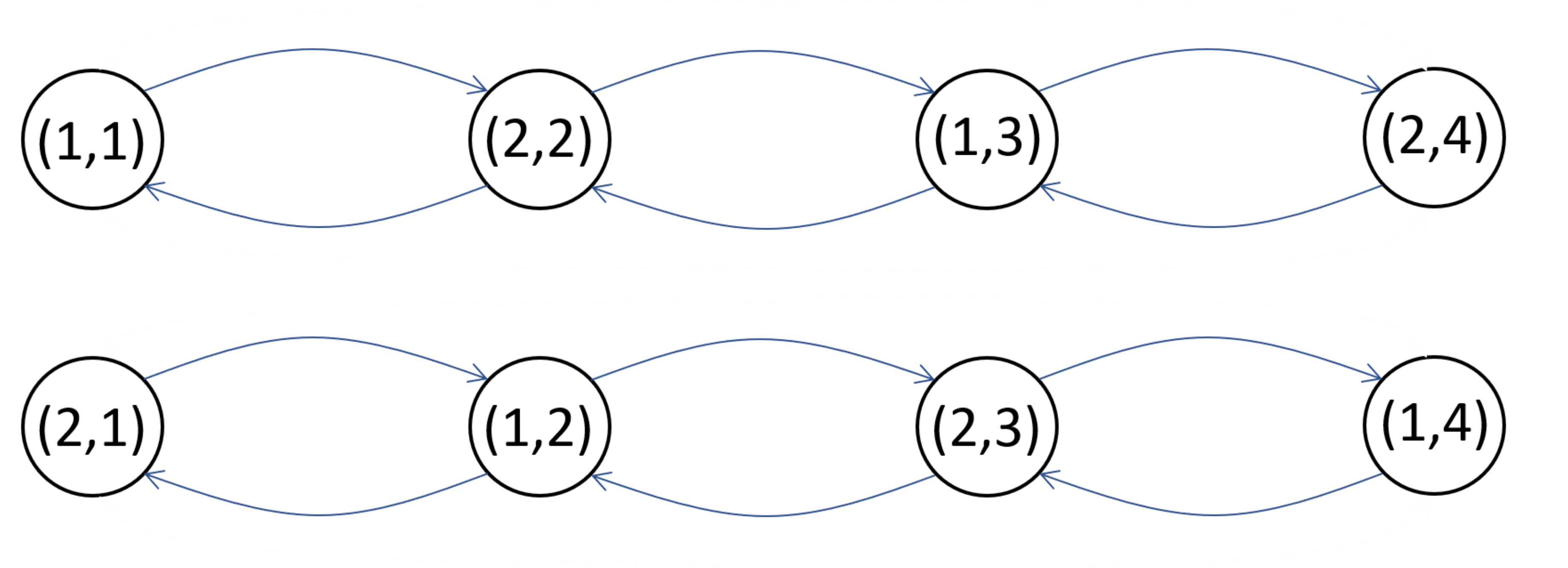}}
	\subfigure[The joint state transition graph after adding individual options]{
	\label{fig:-1(b)}
	\includegraphics[height=1.0in, width=3.4in]{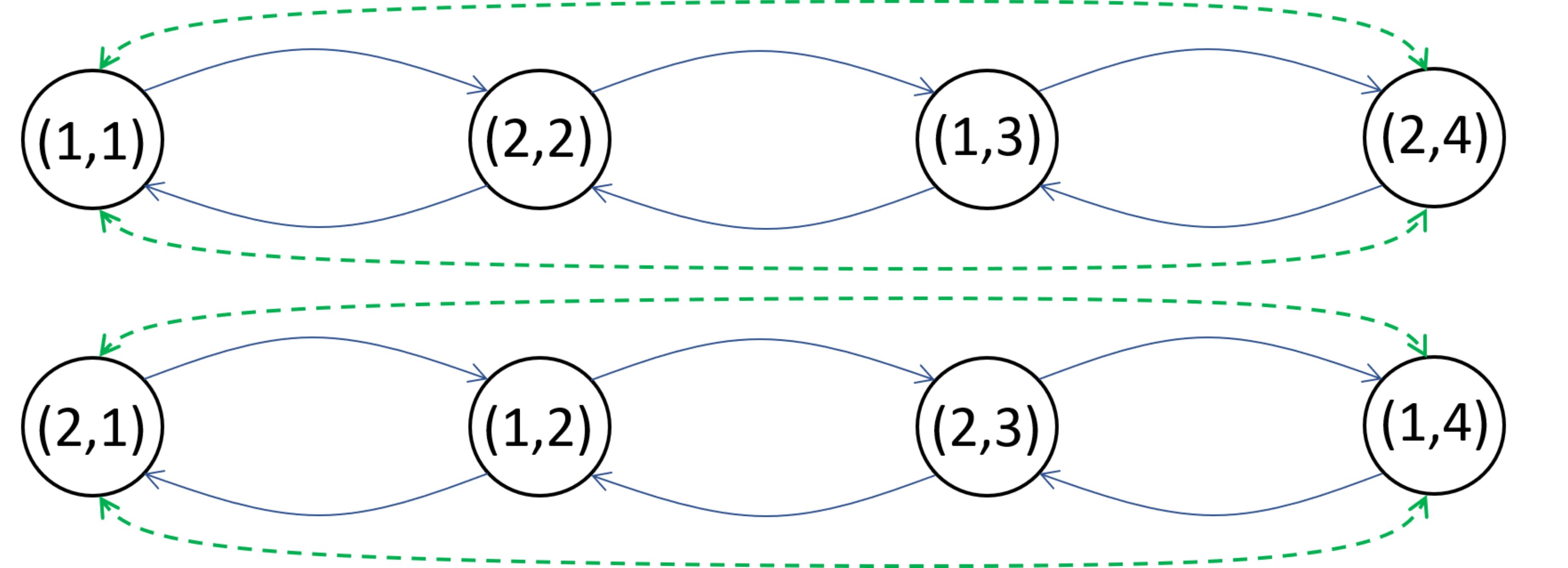}}
	\caption{An illustrative example showing the limitations of utilizing single-agent options alone for MARL.}
	\label{fig:-1}
\end{figure*}

\section{Background} \label{background}

\subsection{Basic Conceptions and Notations} \label{notation}
In this section, we will introduce the necessary conceptions and corresponding notations used in this paper.

\noindent \textbf{Markov Decision Process (MDP): }The RL problem can be described with an MDP, denoted by $\mathcal{M}=(\mathcal{S}, \mathcal{A}, \mathcal{P}, \mathcal{R}, \mathcal{\gamma})$, where $\mathcal{S}$ is the state space, $\mathcal{A}$ is the action space, $\mathcal{P}:\mathcal{S} \times \mathcal{A} \times \mathcal{S} \rightarrow [0,1]$ is the state transition function, $\mathcal{R}:\mathcal{S} \times \mathcal{A} \rightarrow R^{1}$ is the reward function, and $\mathcal{\gamma} \in (0,1]$ is the discount factor. 

\noindent \textbf{State transition graph in an MDP: }The state transitions in $\mathcal{M}$ can be modelled as a state transition graph $G=(V_G, E_G)$, where $V_G$ is a set of vertices representing the states in $\mathcal{S}$, and $E_G$ is a set of undirected edges representing state adjacency in $\mathcal{M}$. We note that:
\begin{remark} \label{rem:2}
There is an edge between state $s$ and $s'$ (i.e., $s$ and $s'$ are adjacent) if and only if $\exists \  a \in \mathcal{A},\  s.t.\  \mathcal{P}(s,a,s')>0\ \lor \ \mathcal{P}(s',a,s)>0$. 
\end{remark}

The adjacency matrix $A$ of $G$ is an $|\mathcal{S}| \times |\mathcal{S}|$ matrix whose $(i,j)$ entry is 1 when $s_{i}$ and $s_{j}$ are adjacent, and 0 otherwise. The degree matrix $D$ is a diagonal matrix whose entry $(i, i)$ equals the number of edges incident to $s_{i}$. The Laplacian matrix of $G$ is defined as $L=D-A$. Its second smallest eigenvalue $\lambda_{2}(L)$ is called the algebraic connectivity of the graph $G$, and the corresponding normalized eigenvector is called the Fiedler vector \cite{fiedler1973algebraic}. Last, the normalized Laplacian matrix is defined as $\mathcal{L}=D^{-\frac{1}{2}}LD^{-\frac{1}{2}}$.

\noindent \textbf{Kronecker product of graphs \cite{weichsel1962kronecker}: }Let $G_1=(V_{G_1}, E_{G_1})$ and $G_2=(V_{G_2}, E_{G_2})$ be two state transition graphs, corresponding to the individual state space $\mathcal{S}_{1}$ and $\mathcal{S}_{2}$ respectively. The Kronecker product of them denoted by $G_1 \otimes G_2$ is a graph defined on the set of vertices $V_{G_1} \times V_{G_2}$, such that:
\begin{remark} \label{rem:1}
Two vertices of $G_1 \otimes G_2$, namely $(g,h)$ and $(g',h')$, are adjacent if and only if $g$ and $g'$ are adjacent in $G_1$ and $h$ and $h'$ are adjacent in $G_2$.
\end{remark}
\noindent Thus, the Kronecker Product Graph can capture the joint transitions of the agents in their joint state space very well. In Section \ref{theory}, we propose to use the Kronecker Product Graph as an effective approximation of the joint state transition graph, so that we can discover the joint options based on the factor graphs. Further, $A_1 \otimes A_2$ is an $|\mathcal{S}_{1}||\mathcal{S}_{2}| \times |\mathcal{S}_{1}||\mathcal{S}_{2}|$ matrix with elements defined by $(A_1 \otimes A_2){(I,J)} =  A_1{(i,j)}A_2{(k,l)}$ with Equation (\ref{equ:-1}), where $A_1$ and $A_2$ are the adjacency matrices of $G_1$ and $G_2$, $A_1{(i,j)}$ is the element lies on the $i$-th row and $j$-th column of $A_1$ (indexed from 1).
\begin{equation} \label{equ:-1}
    \begin{aligned}
        I = (i-1) \times |\mathcal{S}_{2}| + k, \ \ 
        J = (j-1) \times |\mathcal{S}_{2}|+ l
    \end{aligned}
\end{equation}

\subsection{Covering Option Discovery} \label{option definition}

As defined in \cite{DBLP:journals/ai/SuttonPS99}, an option $\omega$ consists of three components: an intra-option policy $\pi_{\omega}: \mathcal{S} \text{ x } \mathcal{A} \rightarrow [0,1]$, a termination condition $ \beta_{\omega}: \mathcal{S} \rightarrow \{0,1\}$, and an initiation set $I_{\omega} \subseteq \mathcal{S}$. An option $<I_{\omega}, \pi_{\omega}, \beta_{\omega}>$ is available in state $s$ if and only if $s \in I_{\omega}$. If the option $\omega$ is taken, actions are selected according to $\pi_{\omega}$ until $\omega$ terminates stochastically according to $\beta_{\omega}$ (i.e., $\beta_{\omega}=1$). Therefore, in order to get an option, we need to train/define the intra-option policy, define the termination condition and initiation set.

The authors of \cite{DBLP:conf/icml/JinnaiPAK19} proposed \textit{Covering Option Discovery} -- discovering options by minimizing the upper bound of the expected cover time of the state space. First, they compute the Fiedler vector $F$ of the Laplacian matrix of the state transition graph. Then, they collect the states $s_{i}$ and $s_{j}$ with the largest $(F_i-F_j)^2$ ($F_i$ is the $i$-th element in $F$), based on which they construct two symmetric options: 
\begin{equation} \label{equ:-2}
    \begin{aligned}
        \omega_{ij}=\ <I_{\omega_{ij}}=\{s_{i}\},\  \pi_{\omega_{ij}}, \ \beta_{\omega_{ij}}=\{s_{j}\}> \\ 
        \omega_{ji}=\ <I_{\omega_{ji}}=\{s_{j}\},\  \pi_{\omega_{ji}}, \ \beta_{\omega_{ji}}=\{s_{i}\}>
    \end{aligned}
\end{equation}
to connect these two subgoal states bidirectionally, where $\pi_{\omega}$ is defined as the optimal path between the initiation and termination state. This whole process is repeated until they get the required number of options.

The intuition of this method is as follows. The authors of \cite{fast_graphs} prove that $(F_i-F_j)^2$ gives the first order approximation of the increase in $\lambda_{2}(L)$ (i.e., algebraic connectivity) by connecting $(s_i, s_j)$. Then, they propose a greedy heuristic to improve the algebraic connectivity of a graph: adding a certain number of edges one at a time, and each time connecting $(s_i, s_j)$ corresponding to the largest $(F_i-F_j)^2$. Further, in \cite{DBLP:conf/icml/JinnaiPAK19}, they prove that the larger the algebraic connectivity is, the smaller the upper bound of the expected cover time would be and the easier the exploration tends to be. Therefore, applying this greedy heuristic to the state transition graph can effectively improve the exploration in the state space.

\section{Proposed Algorithm} \label{alg}
\subsection{System Model} \label{example}

In this paper, we consider to compute covering options in multi-agent scenarios, with $n$ being the number of agents, $\widetilde{\mathcal{S}}=\mathcal{S}_1 \times \mathcal{S}_2 \times \cdots \times \mathcal{S}_n$ being the set of joint states, $\widetilde{\mathcal{A}}=\mathcal{A}_1 \times \mathcal{A}_2 \times \cdots \times \mathcal{A}_n$ being the set of joint actions, $\mathcal{S}_i$ and $\mathcal{A}_i$ being the individual state space and action space of agent $i$. Apparently, the size of the joint state space, i.e., $|\widetilde{\mathcal{S}}|=\prod_{i=1}^n |\mathcal{S}_i|$, grows exponentially with the number of agents. Thus, it is prohibitive to directly compute the covering options based on the joint state transition graph using the approach introduced in Section \ref{option definition} for a large $n$. 

A natural method to tackle this challenging problem is to compute the options for each individual agent by considering only its own state transitions, and then learn to collaboratively leverage these individual options. However, it fails to directly recognize joint options (i.e., multi-agent options) composed of multiple agents' temporal action sequences for encouraging the joint exploration of all the agents. In this case, the algebraic connectivity of the joint state space may not be improved with these single-agent options. We will illustrate this with a simple example.

\textbf{Illustrative example: }Figure \ref{fig:-1(a)} shows a joint state transition graph $\widetilde{G}$ of two agents, where agent 1 has two states denoted by $\mathcal{S}_1=\{1,2\}$ and agent 2 has four states denoted by $\mathcal{S}_2=\{1,2,3,4\}$. In order to compute the individual options, we can restrict our attention to the state transition graph of each agent, namely $G_1$ and $G_2$, with Laplacian given by $L_1$ and $L_2$,  respectively:
\begin{eqnarray}
L_1 = \left[ \begin{array}{cc} 1 &-1 \\ -1 & 1 \end{array} \right] \ \ {\rm ,} \ \ L_2 =\left[  \begin{array}{cccc} 1 &-1 & 0 & 0 \\ -1 & 2 & -1 & 0 \\ 0 & -1 & 2 & -1 \\ 0 & 0 & -1 & 1 \end{array}  \right].
\end{eqnarray}

To compute the options for each agent, we first compute the Fiedler vectors of $G_1$ and $G_2$ as:
\begin{eqnarray}
v^{G_1} = \frac{1}{\sqrt{2}} \left[ \begin{array}{c} -1 \\  1 \end{array} \right], \ \ v^{G_2} = \frac{1}{\sqrt{8-4\sqrt{2}}} \left[  \begin{array}{c} -1 \\  -\sqrt{2} + 1 \\ \sqrt{2} - 1 \\ 1  \end{array} \right].
\end{eqnarray}

\begin{figure*}[ht]
	\centering
	\subfigure[The joint state transition graph updated with option $\omega_1$]{
	\label{fig:-3(a)}
	\includegraphics[height=1.1in, width=3.4in]{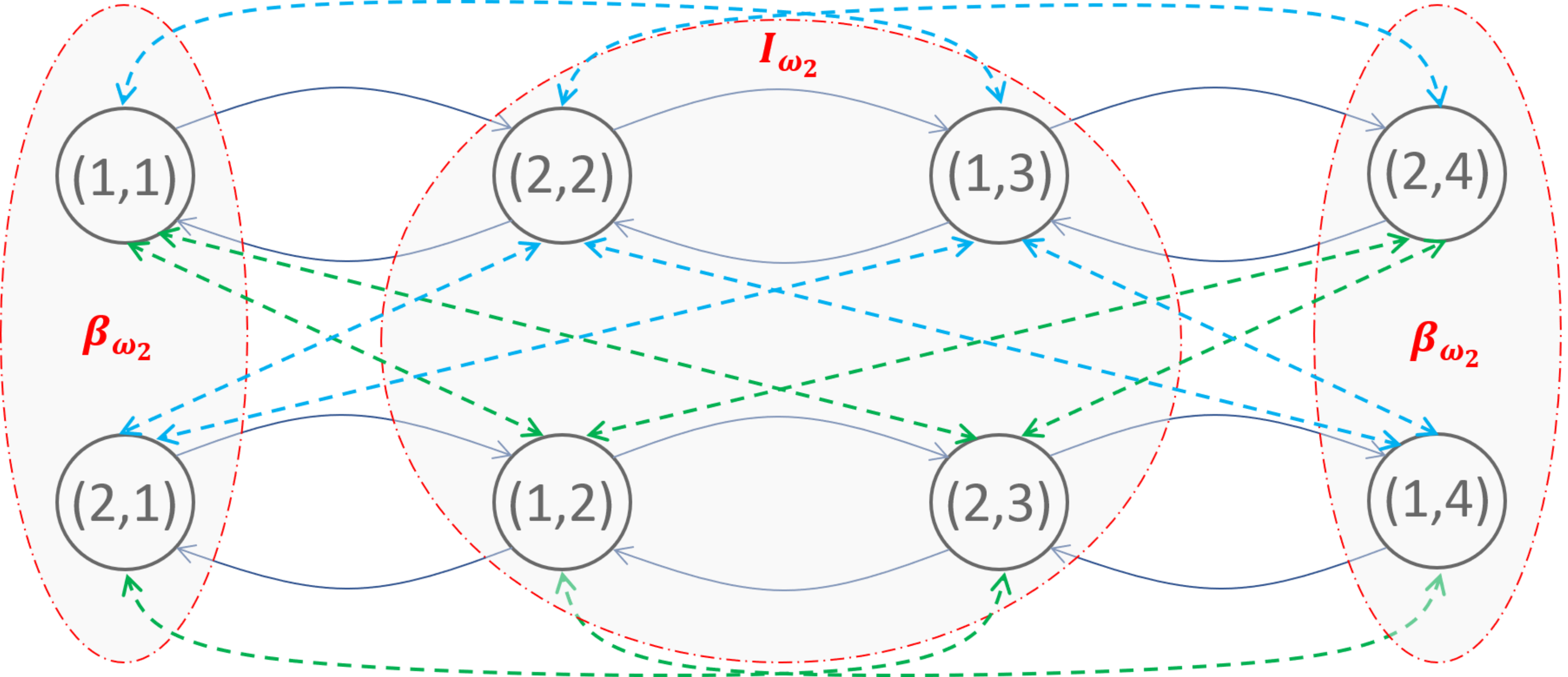}}
	\subfigure[The joint state transition graph updated with option $\omega_2$]{
	\label{fig:-3(b)}
	\includegraphics[height=1.1in, width=3.4in]{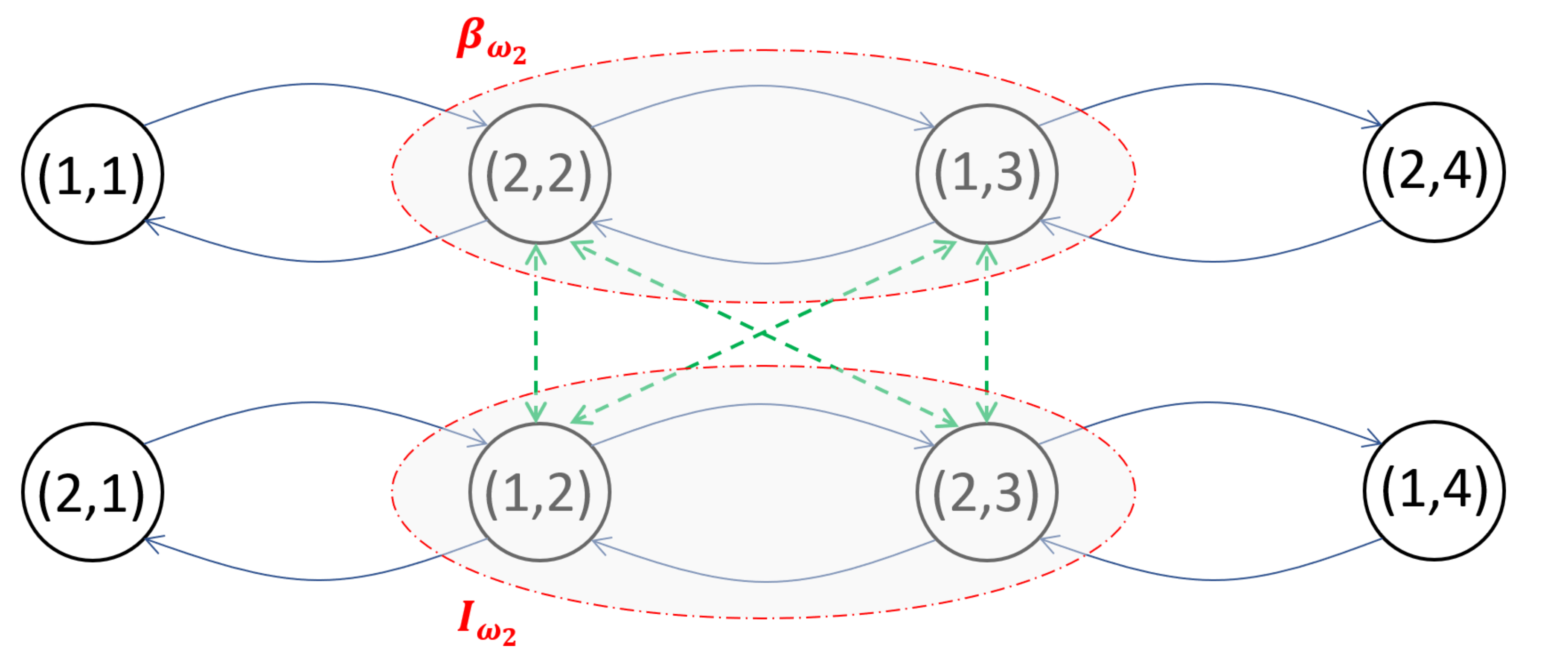}}
	\caption{The joint state transition graph updated with the detected multi-agent options}
	\label{fig:-3}

\end{figure*}

Then, according to the option discovery approach described in Section \ref{option definition}, we can get the individual options for agent 1 to connect its state 1 (minimum) and state 2 (maximum), and individual options for agent 2 to connect its state 1 (minimum) and state 4 (maximum). With these options, the updated joint state space would be like Figure \ref{fig:-1(b)}. Apparently, the straightforward decomposition of option discovery for MARL fails to create a connected graph. It implies that for the purpose of encouraging efficient exploration, utilizing the single-agent options alone may not be sufficient.

Therefore, we propose to build Multi-agent Covering Options to enhance the connectivity of the joint state space and accelerate the joint exploration of the agents within the scenario. Again, we can represent it as a tuple: $<I_{\omega}, \pi_{\omega}, \beta_{\omega}>$, where $I_{\omega} \subseteq \widetilde{\mathcal{S}}$ is the set of initiation joint states, $ \beta_{\omega}: \widetilde{\mathcal{S}} \rightarrow \{0,1\}$ indicates the joint states to terminate, $\pi_{\omega}=(\pi_{\omega}^{1}, \cdots, \pi_{\omega}^{n}) (\pi_{\omega}^{i}: \mathcal{S}_i \times \mathcal{A}_i \rightarrow [0,1]),$ is the joint intra-option policy that can lead the agents from the initiation states to the termination states. The key challenge is to calculate the Fiedler vector of the joint state space according to which we can define $<I_{\omega}, \pi_{\omega}, \beta_{\omega}>$ like Section \ref{option definition}. Given that $|\widetilde{\mathcal{S}}|$ grows exponentially with $n$, we propose to estimate the joint Fiedler vector based on the individual state spaces in the next section.

\subsection{Theory results} \label{theory}

This section shows the theoretical foundations of our approach. In order to discover the multi-agent options, we need to find the Fiedler vector of the joint state transition graph. Given that the size of the joint state space grows exponentially with the number of agents, we propose to use the Kronecker Product Graph to decompose the eigenfunction calculation to single-agent state spaces, which can make our approach more scalable. This decomposition is based on the facts: (1) the Kronecker product of individual state transition graphs $\otimes_{i=1}^{n}G_{i}=G_1\otimes\cdots\otimes G_n$ provides a good approximation of the joint state transition graph $\widetilde{G}$; (2) The Fielder vector of $\otimes_{i=1}^{n}G_{i}$ can be estimated with the eigenvectors of $G_i(i=1,\cdots,n)$.

First, we note that the Kronecker product of individual state transition graphs $\otimes_{i=1}^{n}G_{i}$ can be used as a simple yet powerful approximation of the joint state transition graph $\widetilde{G}$ for the purpose of computing multi-agent options. This approximation becomes exact (i.e., $\widetilde{G}=\otimes_{i=1}^{n}G_{i}$) if the state transitions of an agent would not be influenced by the other agents. 
Also, we show through numerical experiments in Section \ref{results} (Figure \ref{fig:5}) that such Kronecker product approximation can still effectively find the multi-agent options for problems where $\widetilde{G}=\otimes_{i=1}^{n}G_{i}$ is not exact. We note that to discover multi-agent options, we only need to identify the most distant states in the joint state space which are corresponding to the minimum and maximum of its Fiedler vector. Thus, we do not need to get the exact joint state transition graph $\widetilde{G}$ or the exact value of its Fiedler vector.

Next, we show how to effectively approximate the Fiedler vector of $\otimes_{i=1}^{n}G_{i}$ based on the Laplacian spectrum of the factor graphs, which enables an effective decomposition of multi-agent option discovery. Inspired by \cite{basic2021estimation} which proposed an estimation of the Laplacian spectrum of the Kronecker product of two factor graphs, we have the following THEOREM \ref{thm:2}.

\begin{theorem}
\label{thm:2}

For graph $\widetilde{G}=\otimes_{i=1}^{n}G_{i}$, we can approximate the eigenvalues $\mu$ and eigenvectors $v$ of its Laplacian $L$ by:
\begin{equation} \label{equ:mu}
    \setlength{\abovedisplayskip}{1pt}
    \setlength{\belowdisplayskip}{1pt}
    \begin{aligned}
    \mu_{k_1,\ldots,k_n} =\left\{\left[1-\prod_{i=1}^{n}(1-\lambda_{k_i}^{G_i})\right] \prod_{i=1}^{n}d_{k_i}^{G_i}\right\}
    \end{aligned}
\end{equation}

\begin{equation} \label{equ:v}
    \setlength{\abovedisplayskip}{1pt}
    \setlength{\belowdisplayskip}{1pt}
    v_{k_1,\ldots,k_n}=\otimes_{i=1}^{n}v_{k_i}^{G_i}
\end{equation}
where $\lambda_{k_i}^{G_i}$ and $v_{k_i}^{G_i}$ are the $k_i$-th smallest eigenvalue and corresponding eigenvector of $\mathcal{L}_{G_i}$ (normalized Laplacian matrix of $G_i$), and $d_{k_i}^{G_i}$ is the $k_i$-th smallest diagonal entry of $D_{G_{i}}$ (degree matrix of $G_i$).
\end{theorem}

The proof of THEOREM \ref{thm:2} is provided in Appendix \ref{thmproof}, which also contains the proof that the estimated eigenvalues through this theorem are non-negative. Note that if $G_i$ has $K_i$ ($i=1,\cdots,n$) eigenvalues, there would be $\prod_{i=1}^{n} K_i$ eigenvalues for $\widetilde{G}$. Through enumerating $(k_1, \cdots, k_n)$, we can collect the eigenvalues of $\otimes_{i=1}^{n}G_{i}$ by Equation (\ref{equ:mu}) and the corresponding eigenvectors by Equation (\ref{equ:v}). Then, the eigenvector $v_{\hat{k}_1,\cdots,\hat{k}_n}$ corresponding to the second smallest eigenvalue $\mu_{\hat{k}_1,\cdots,\hat{k}_n}$ is the estimated Fiedler vector of the joint state transition graph, namely $F_{\widetilde{G}}$. Based on it, we can define the joint states corresponding to the maximum or minimum in $F_{\widetilde{G}}$ as the initiation or termination joint states, which can be connected with joint options. As discussed in Section \ref{option definition}, connecting these two joint states with options can greedily improve the algebraic connectivity of the joint state space and accelerate the joint exploration within it.

\textbf{Illustrative example: }Now we consider again the example in Figure \ref{fig:-1(a)}, where $\widetilde{G} = G_{1} \otimes G_{2}$. We can approximate the Fiedler vector of $\widetilde{G}$ using THEOREM \ref{thm:2}. As a result, we get two approximations of the Fiedler vector (The computing details are shown in Appendix \ref{example2}):
\begin{equation} \label{v11}
v_{11}=\frac{1}{\sqrt{6}}\left[\frac{1}{\sqrt{2}},\ 1,\ 1,\ \frac{1}{\sqrt{2}},\ \frac{1}{\sqrt{2}},\ 1,\ 1,\ \frac{1}{\sqrt{2}}\right]^{T}
\end{equation}
\begin{equation}
v_{24}=\frac{1}{\sqrt{6}}\left[-\frac{1}{\sqrt{2}},\ 1,\ -1,\ \frac{1}{\sqrt{2}},\ \frac{1}{\sqrt{2}},\ -1,\ 1,\ -\frac{1}{\sqrt{2}}\right]^{T}
\end{equation}
Based on the two approximations and the indexing relationship between $\widetilde{G}$ and its factor graphs (Equation (\ref{equ:-1})), we can get two sets of multi-agent options: $\{I_{\omega_1}=\{(1,2),\ (1,3),\ (2,2),\\(2,3)\}, \beta_{\omega_1}=\{(1,1),\ (1,4),\ (2,1),\ (2,4)\}\}$ and $\{I_{\omega_2}=\{(1,2),\ (2,3)\}, \beta_{\omega_2}=\{(1,3),\ (2,2)\}\}$, where we set the joint states corresponding to the maximum and minimum as the initiation states and termination states respectively. For example, in $v_{11}$ (Equation (\ref{v11})), the $7$-th element (indexed from 1) is a maximum, so the $7$-th joint state is within the initiation set $I_{\omega_1}$ and denoted as $(2,3)$ according to Equation (\ref{equ:-1}). As shown in Figure \ref{fig:-3}, both of the two options can lead to a connected graph when applied to $\widetilde{G}$. Thus, the adoption of multi-agent options has the potential to encourage efficient exploration of the joint state space by improving its algebraic connectivity, and we can discover multi-agent options based on individual agents' state spaces, so that we can enjoy the ease of decomposition. 

In the following sections, we will formalize our algorithm -- Multi-agent Covering Option Discovery through Kronecker Product of Factor Graphs, and show empirically the significant performance improvement brought by integrating multi-agent options in MARL.

\begin{algorithm*}[t]
\caption{Multi-agent Covering Option Discovery}\label{alg:1}
{\begin{algorithmic}[1]
\State \textbf{Input}: number of agents $n$, list of adjacency matrices $A_{1:n}$, number of options to generate $tot\_num$
\State \textbf{Output}: list of multi-agent options $\Omega$
\State $\Omega \leftarrow \emptyset$, $cur\_num \leftarrow 0$
\While{$cur\_num < tot\_num$}
\State \textbf{Collect} the degree list of each individual state transition graph $D_{1:n}$ according to $A_{1:n}$
\State \textbf{Obtain} the list of normalized laplacian matrices $\mathcal{L}_{1:n}$ corresponding to $A_{1:n}$
\State \textbf{Calculate} the eigenvalues $U_{i}$ and corresponding eigenvectors $V_{i}$ for each $\mathcal{L}_{i}$ and collect them as $U_{1:n}$ and $V_{1:n}$
\State \textbf{Obtain} the Fielder vector $F$ of the joint state space using THEOREM \ref{thm:2} based on $D_{1:n}$, $U_{1:n}$ and $V_{1:n}$
\State \textbf{Collect} the list of joint states corresponding to the minimum or maximum in $F$, named $MIN$ and $MAX$ respectively
\State \textbf{Convert} each joint state $s_{joint}$ in $MIN$ and $MAX$ to $(s_{1}, \cdots, s_{n})$, where $s_{i}$ is the corresponding individual state of 
\Statex $\quad\ $ agent $i$, based on the equation: 
\Statex $\quad\quad\qquad$ $ind(s_{joint}) = ((ind(s_{1})*dim(A_{2})+ind(s_{2}))*dim(A_{3})+\cdots+ind(s_{n-1}))*dim(A_{n})+ind(s_{n})$
\Statex $\quad\ $ where $dim(A_{i})$ is the dimension of $A_{i}$, $ind(s_{i})$ is the index of $s_{i}$ (indexed from 0) in the state space of agent $i$
\State \textbf{Generate} a new list of options $\Omega'$ through Algorithm \ref{alg:2}
\State $\Omega \leftarrow \Omega \cup \Omega'$, $cur\_num \leftarrow cur\_num + len(\Omega')$
\State \textbf{Update} $A_{1:n}$ through Algorithm \ref{alg:3}

\EndWhile
\State \textbf{Return} $\Omega$

\end{algorithmic}}
\end{algorithm*}

\begin{algorithm*}[t]
\caption{Generate Multi-agent Options}\label{alg:2}
{\begin{algorithmic}[1]
\State \textbf{Input}: $MIN, MAX$: list of joint states corresponding to the minimum or maximum in the Fielder vector
\State \textbf{Output}: list of multi-agent options $\Omega'$
\State $\Omega' \leftarrow \emptyset$
\For{$s=(s_{1}, \cdots, s_{n})$ in ($MIN \cup MAX$)}
\State \textbf{Define} the initiation set $I_{\omega}$ as the joint states in the known region of the joint state space
\State \textbf{Define} the termination condition:
 $ \beta_{\omega}(s_{cur}) \leftarrow \left\{
                \begin{aligned}
                1 &  & if \ (s_{cur}==s) \ or\  (s_{cur}\ is\ unknown) \\
                0 &  & otherwise
                \end{aligned}
                \right.
$
\Statex $\quad\ $ where $s_{cur}$ is the current joint state
\State \textbf{Train} the intra-option policy $\pi_{\omega}=(\pi_{\omega}^{1}, \cdots, \pi_{\omega}^{n})$, where $\pi_{\omega}^{i}$ maps the individual state of agent $i$ to its action aiming at 
\Statex $\quad\ \ $leading agent $i$ from any state in its initiation set to its termination state $s_{i}$
\State $\Omega' \leftarrow \Omega' \cup \{<I_{\omega}, \pi_{\omega}, \beta_{\omega}>\}$
\EndFor
\State \textbf{Return} $\Omega'$

\end{algorithmic}}
\end{algorithm*}

\begin{algorithm}[t]
\caption{Update Adjacency Matrices}\label{alg:3}
{\begin{algorithmic}[1]
\State \textbf{Given}: list of adjacency matrices $A_{1:n}$, list of joint states corresponding to the minimum or maximum in the Fielder vector $MIN, MAX$
\For{$s_{min}=(s_{min}^{1}, \cdots, s_{min}^{n})$ in $MIN$}
    \For{$s_{max}=(s_{max}^{1}, \cdots, s_{max}^{n})$ in $MAX$}
        \For{$i=1$ to $n$}
        \State $A_{i}[ind(s_{min}^{i})][ind(s_{max}^{i})]=1$
        \State $A_{i}[ind(s_{max}^{i})][ind(s_{min}^{i})]=1$
        \EndFor
    \EndFor
\EndFor

\end{algorithmic}}
\end{algorithm}

\subsection{Multi-agent Covering Option Discovery}

In this paper, we adopt Algorithm \ref{alg:1} to construct multi-agent options, based on the individual state transition graphs of each agent which are represented as a list of adjacency matrices $A_{1:n}$. First, in Line 5-9 of Algorithm \ref{alg:1}, we acquire the estimation of the Fielder vector $F$ of the joint state space through THEOREM \ref{thm:2} based on $A_{1:n}$, so that we can collect the joint states corresponding to the minimum or maximum of $F$. Then, in Line 10 of Algorithm \ref{alg:1}, we split each joint state into a list of individual states. For example, after getting a pair of joint states ($s_{min}, s_{max}$), we convert them into ($(s_{min}^{1}, \cdots, s_{min}^{n})$, $(s_{max}^{1}, \cdots, s_{max}^{n})$), so that we can connect ($s_{min}, s_{max}$) in the joint state space by connecting each ($s_{min}^{i}, s_{max}^{i}$) in the corresponding individual state space. According to THEOREM \ref{thm:2}, we estimate $F$ as the Kronecker product of $n$ eigenvectors, where the $i$-th vector is an eigenvector of agent $i$'s normalized laplacian matrix. Therefore, we can get the relationship between the index of the joint state and the indexes of its corresponding individual states based on the definition of Kronecker product, which is shown in Line 10.

After decentralizing the joint states, we can generate the multi-agent options through Algorithm \ref{alg:2}. For each option $\omega$, we define its initiation set $I_{\omega}$ as the explored joint states, and its termination set $\beta_{\omega}$ as a joint state in $MIN \cup MAX$ or the unexplored area. As mentioned in Section \ref{option definition}, an option $\omega$ is available in state $s$ if and only if $s \in I_{\omega}$. Therefore, instead of constructing a point option between ($s_{min}, s_{max}$), i.e., setting \{$s_{min}$\} (\{$s_{max}$\}) as $I_{\omega}$ and \{$s_{max}$\} (\{$s_{min}$\}) as $\beta_{\omega}$, we extend $I_{\omega}$ to the known area to increase the accessibility of $\omega$ without loss of the connectivity of the joint state space. As for the multi-agent intra-option policy $\pi_{\omega}$ used for connecting the initiation joint state and termination joint state, we divide it into a list of single-agent policies $\pi_{\omega}^{i}\ (i=1,\cdots,n)$, where $\pi_{\omega}^{i}$ can be trained with any single-agent RL algorithm aiming at leading agent $i$ from its own initiation state to the termination state $s_{min}^{i}$ ($s_{max}^{i}$). At last, before entering the next loop, we adopt Algorithm \ref{alg:3} to update the individual state transition graphs with the newly-discovered options. This whole process (Line 5-13 in Algorithm \ref{alg:1}) is repeated until we get a certain number of options.  

To sum up, the proposed algorithm first discovers the joint states that need to be explored most, and then build multi-agent options to encourage agents to visit these sub-goals. More precisely, we project each sub-goal joint state into its corresponding individual state spaces and train the intra-option policy for each agent to visit the projection of the sub-goal state in its individual state space.

At last, we give out the computational complexity of our approach. Consider an MDP with $n$ agents and $m$ states for each agent. To compute the Fiedler vector directly from the joint state transition graph would require time complexity $\mathcal{O}(m^{3n})$, since there are in total $m^{n}$ joint states and the time complexity of eigenvalue decomposition (Line 7 in Algorithm \ref{alg:1}) is cubic with the size of the joint state space. While, with our Kronecker factor graph approach, we can decompose the original problem into computing eigenvectors of the individual state transition graphs, of which the overall time complexity is $\mathcal{O}(nm^{3})$. Thus, our solution significantly reduces the problem complexity from $\mathcal{O}(m^{3n})$ to $\mathcal{O}(nm^{3})$ for multi-agent problems. Also, we would like to point out that for problems with continuous or very large state space (i.e., $m$ is large), our approach could be directly integrated with sample-based techniques for eigenfunction estimation (Line 7 in Algorithm \ref{alg:1}), like \cite{DBLP:conf/iclr/WuTN19, DBLP:conf/icml/WangZZSHF21}. Hence, the bottleneck on computational complexity can be overcome. More precisely, the Laplacian spectrum of the factor graphs can be estimated using neural networks, and then leveraged by our proposed algorithm to find the Fiedler vector of the joint state transition graph, which will be considered as future work.

\subsection{Adopting Multi-agent Options in MARL} \label{framework}

\begin{figure}[t]
	\centering
	\includegraphics[height=2.0in, width=3.2in]{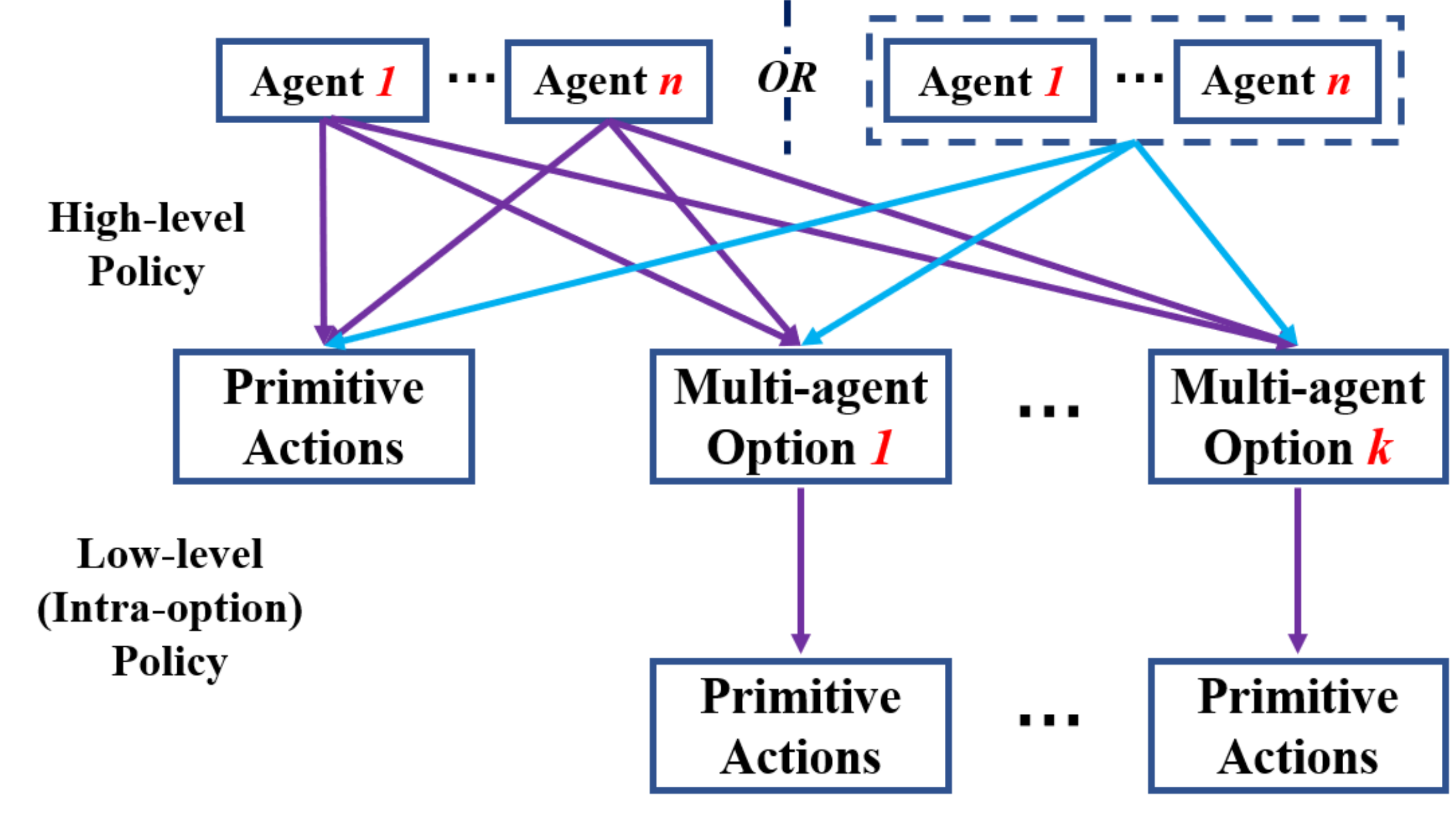}
	\caption{Hierarchical algorithm framework: When making decisions, the agent first decides on which option $\omega$ to use according to the high-level policy, and then decides on the primitive action to take based on the corresponding intra-option policy $\pi_{\omega}$. The agents can decide on their options independently (the left side) or jointly (the right side).}
	\label{fig:1}
\end{figure}

In order to take advantage of options in the learning process, we adopt a hierarchical algorithm framework, shown in Figure \ref{fig:1}. When making decisions, the RL agent first decides on which option $\omega$ to use according to the high-level policy (the primitive actions can be viewed as one-step options), and then decides on the action to take based on the corresponding intra-option policy $\pi_{\omega}$. Note that the agent does not decide on a new option with the high-level policy until the current option terminates.

For a multi-agent option $\omega:$ $<I_{\omega}=\{ the \ explored\ joint\ \\ states\},$ $\pi_{\omega}=(\pi_{\omega}^{1}, \cdots, \pi_{\omega}^{n}), \beta_{\omega}=\{(s_{1}, \cdots, s_{n})\}>$, it can be adopted either in a decentralized or centralized way. As shown by the purple arrows in Figure \ref{fig:1}, the agents choose their own options independently, and they may choose different options to execute in the meantime. In this case, if agent $i$ selects option $\omega$, it will execute $\pi_{\omega}^{i}$ until it reaches its termination state $s_{i}$ or an unknown individual state. On the other hand, we can force the agents to execute the same multi-agent option simultaneously. To realize this, as shown by the blue arrows in Figure \ref{fig:1}, we view the $n$ agents as a whole, which takes the joint state as the input and chooses primitive actions or the same multi-agent option to execute at a time. Once a multi-agent option $\omega$ is chosen, agent $1:n$ will execute $\pi_{\omega}^{1:n}$ until they reach the termination joint state $(s_{1}, \cdots, s_{n})$ or an unexplored joint state. We note that if there are $m$ primitive actions and $k$ multi-agent options, the size of the search space would be $(m+k)^{n}$ for the decentralized approach and $m^n+k$ for the centralized approach. Therefore, the decentralized way is more flexible but has a larger search space. While, the centralized way fails to consider all the possible solutions but makes it easier for the agents to visit the sub-goal joint states, since the agents simultaneously select the same joint option which will not terminate until the agents arrive at a sub-goal state. In this paper, we use Independent Q-Learning \cite{DBLP:conf/icml/Tan93} (adopting Q-Learning \cite{watkins1992q} to each individual agent) to train the decentralized high-level policy, and Centralised Q-Learning (viewing the $n$ agents as a whole and adopting Q-Learning to this joint agent) to train the centralised high-level policy. We will present the comparisons between these two ways in Section \ref{evaluation}.

Further, we note that the centralized high-level policy may not be applicable when the number of agents $n$ is large, since both the input space and output space will grow exponentially with $n$. Thus, we propose to partition the agents into some sub-groups first, and then define/learn the joint options within each sub-group. The intuition behind this is as follows. In practice, a multi-agent task can usually be divided into some sub-tasks, each of which can be completed by a sub-group of the agents. For each sub-group, we can learn a list of multi-agent options, and then the agents within this group can make use of these options in a decentralized or centralized way as mentioned above. Further, if there is no way to divide the (identical) agents based on sub-tasks, we can still group them randomly to a list of two-agent or three-agent sub-groups. Agents within the same sub-group will co-explore their joint state space using the algorithm framework shown as Figure \ref{fig:1}. In Section \ref{evaluation}, we show that the adoption of grouping techniques can not only accelerate the exploration but also can greatly improve the scalability of our algorithm.

\section{Evaluation and Results} \label{evaluation}

\begin{figure}[t]
\centering
\subfigure[four-room]{
\label{fig:2(b)} 
\includegraphics[width=1.6in, height=1.4in]{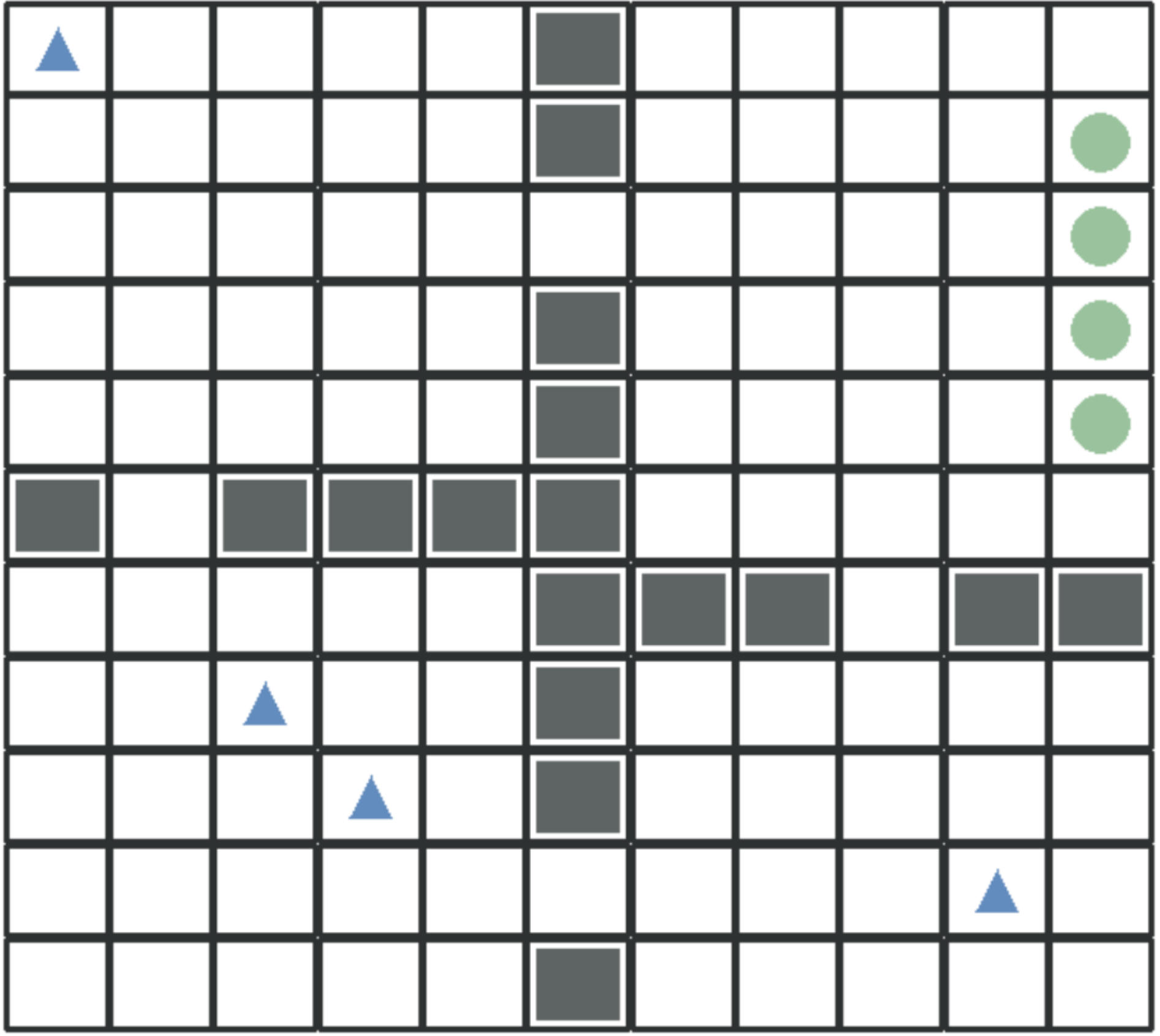}}
\subfigure[four-room with grouping]{
\label{fig:2(c)} 
\includegraphics[width=1.6in, height=1.4in]{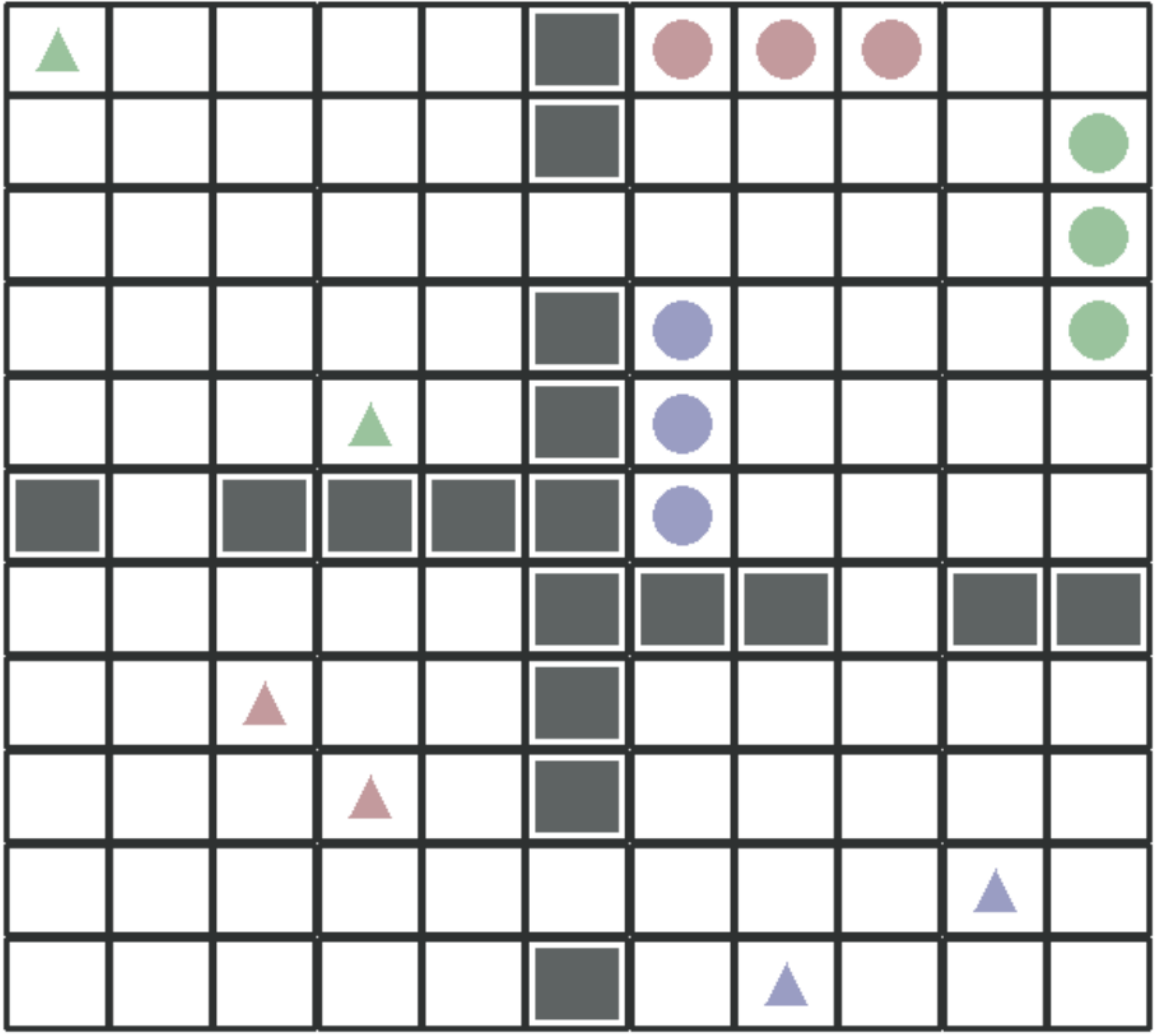}}
\subfigure[maze]{
\label{fig:2(d)} 
\includegraphics[width=1.6in, height=1.4in]{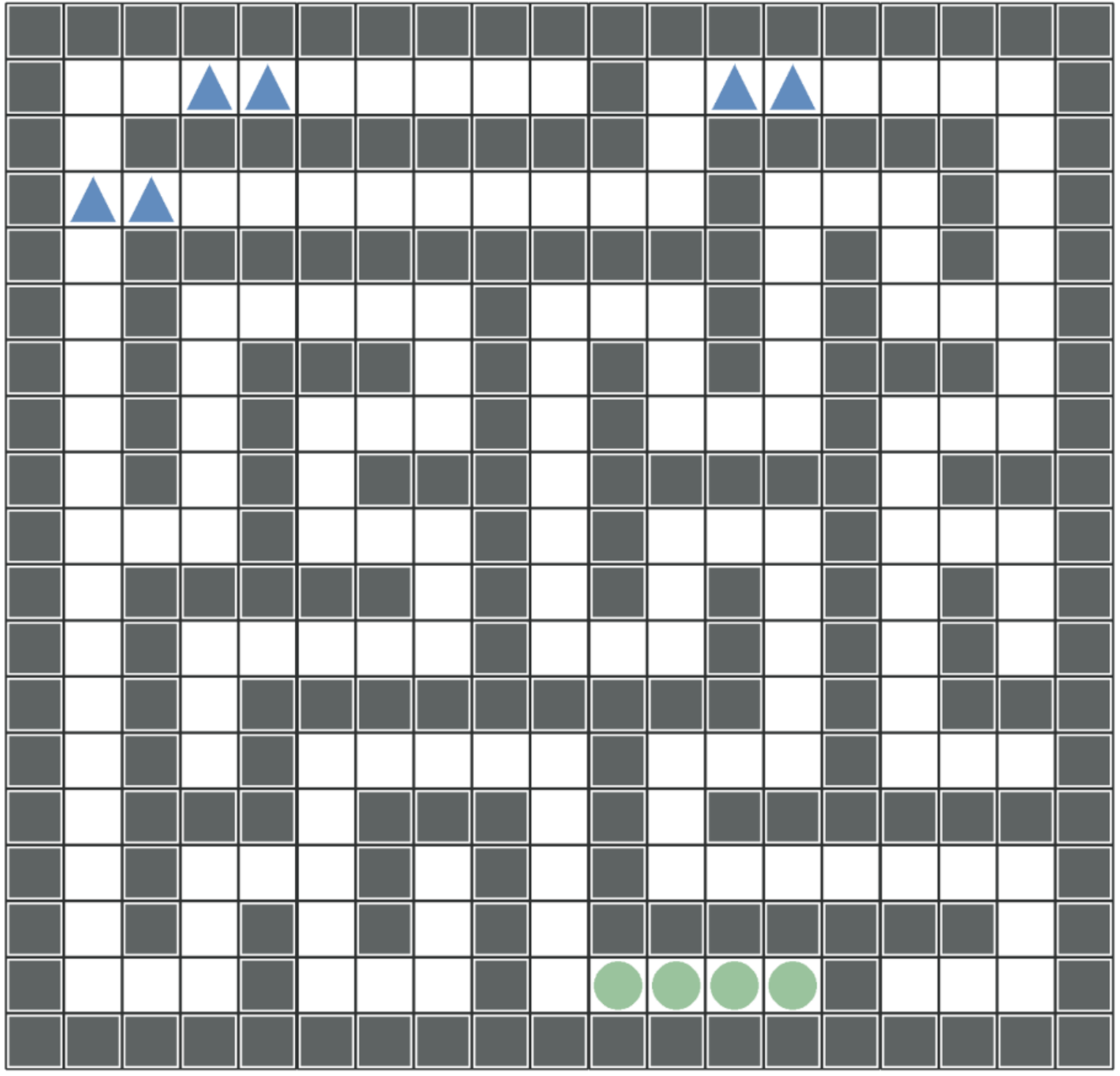}}
\subfigure[maze with grouping]{
\label{fig:2(e)} 
\includegraphics[width=1.6in, height=1.4in]{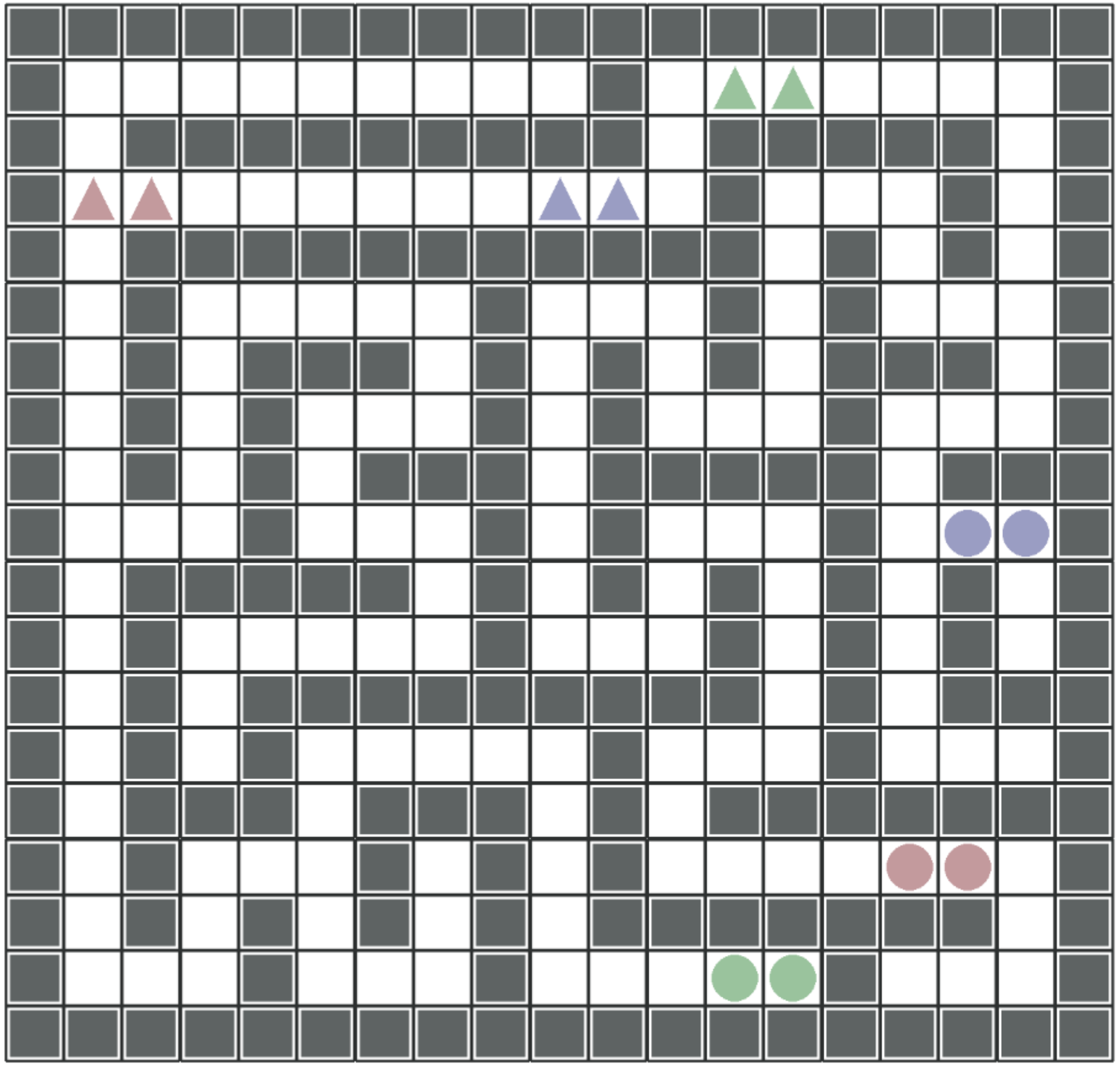}}
\caption{Simulators for Evaluation: All the agents (triangles) must reach the goal area (circles) simultaneously to complete the task, based on only their current locations. In (b) and (d), agents are assigned with different goals; the agents and their corresponding goals are labeled with the same color.}
\label{fig:2} 
\end{figure}

\begin{table*}[t] 
\centering
    \begin{tabular}{ |p{5cm}||p{3cm}|p{3cm}|p{5cm}| }
 \hline
 \makecell[c]{Algorithm} & \makecell[c]{Input} & \makecell[c]{Output} & \makecell[c]{How to utilize multi-agent options}\\
 \hline
\makecell[c]{Random} & \makecell[c]{--} & \makecell[c]{Individual action} &  \makecell[c]{Decentralized}\\
\makecell[c]{Independent Q-Learning \cite{DBLP:conf/icml/Tan93}} & \makecell[c]{Individual state} & \makecell[c]{Individual action} & \makecell[c]{Decentralized}\\
\makecell[c]{Distributed Q-Learning \cite{DBLP:conf/icml/LauerR00}} & \makecell[c]{Joint state} & \makecell[c]{Individual action} & \makecell[c]{Decentralized}\\
\makecell[c]{Centralized Q-Learning} & \makecell[c]{Joint state} & \makecell[c]{Joint action} & \makecell[c]{Decentralized}\\
\makecell[c]{Centralized Q-Learning + Force} & \makecell[c]{Joint state} & \makecell[c]{Joint action} & \makecell[c]{Centralized}\\
\hline
\end{tabular}
\caption{Comparisons among different high-level policy algorithms}
\label{tab:1}
\end{table*}

\begin{figure*}[htbp]
\centering
\subfigure[Random]{
\label{fig:8(a)} 
\includegraphics[width=2.2in, height=1.2in]{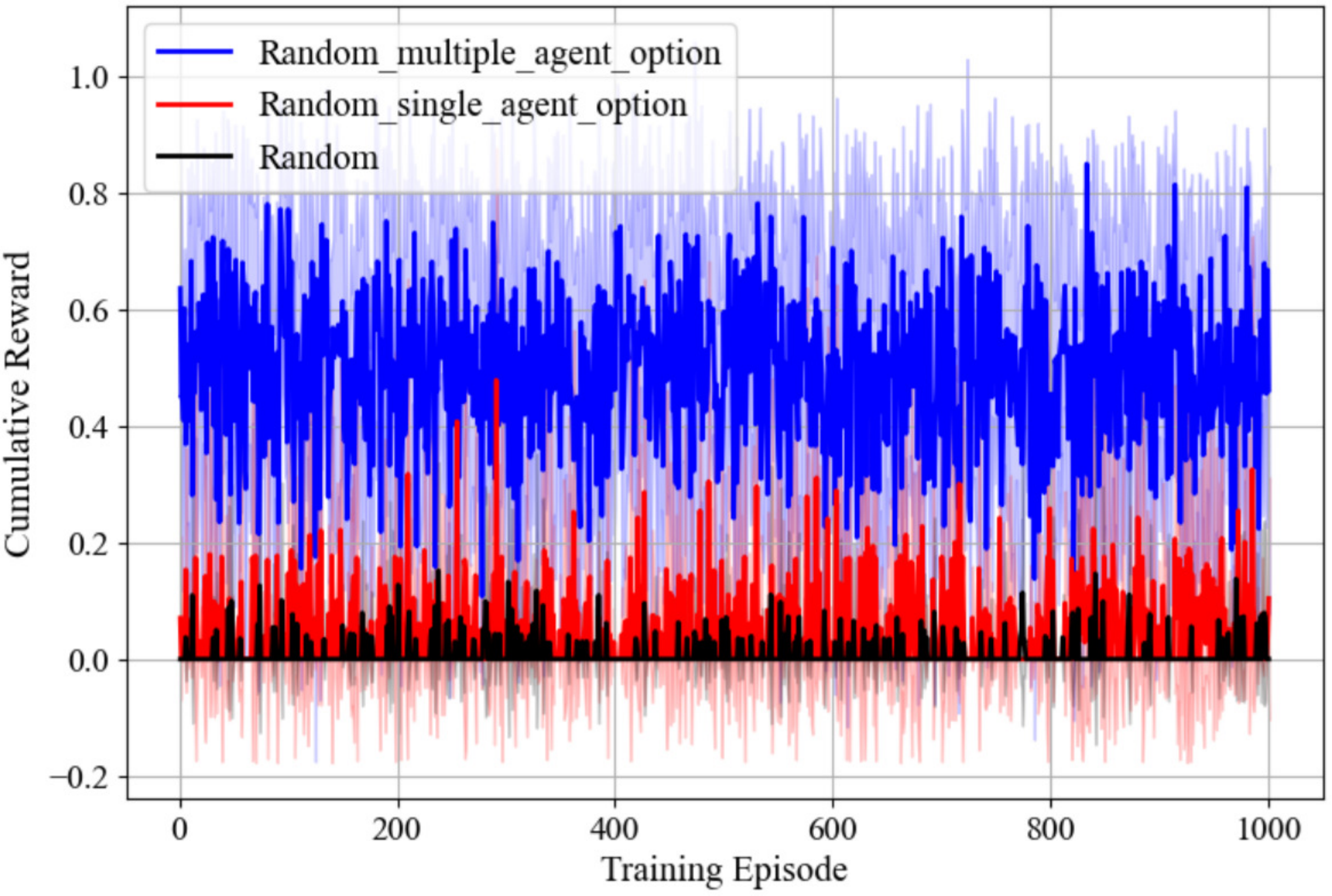}}
\subfigure[Independent Q-Learning]{
\label{fig:8(b)} 
\includegraphics[width=2.2in, height=1.2in]{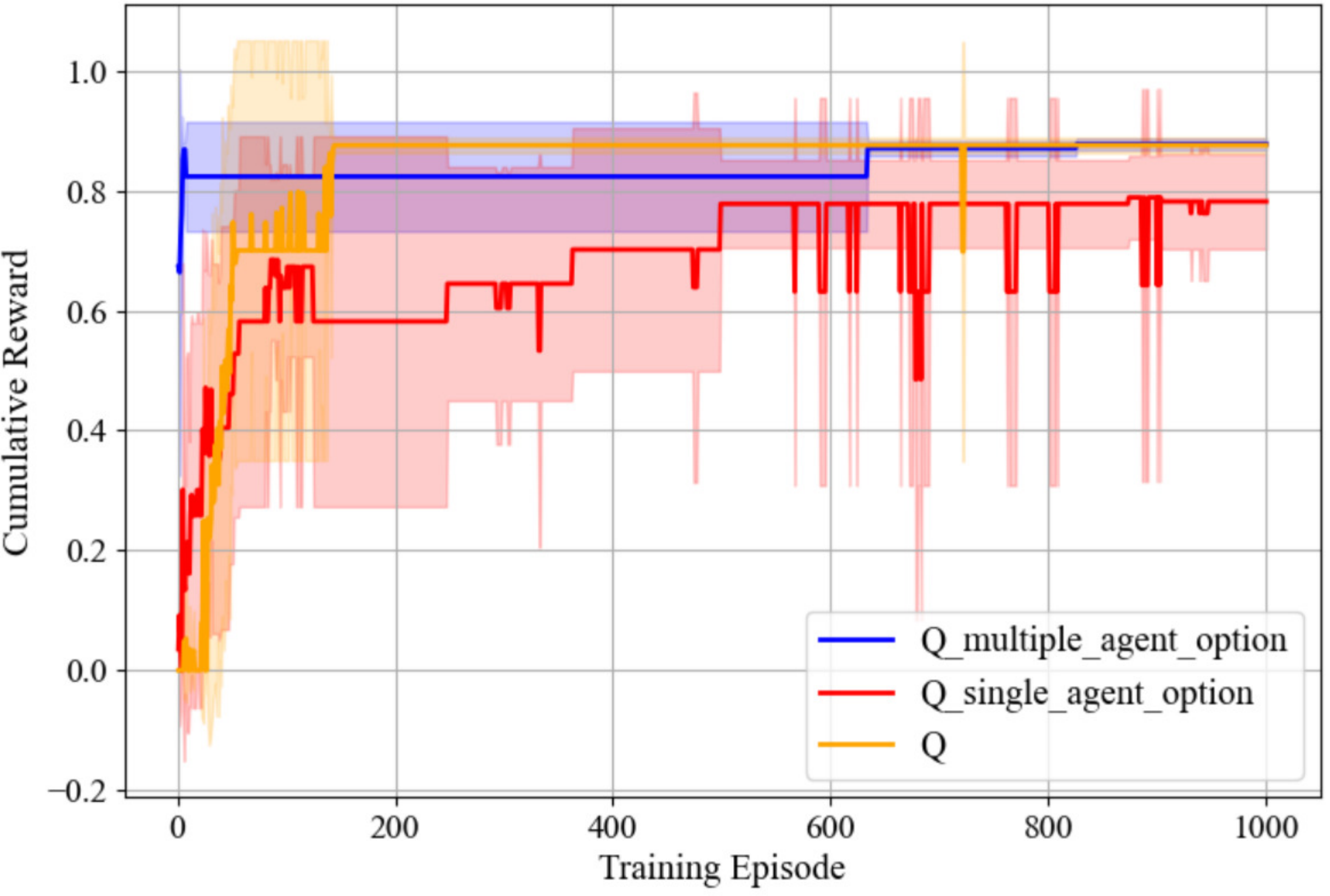}}
\subfigure[Distributed Q-Learning]{
\label{fig:8(c)} 
\includegraphics[width=2.2in, height=1.2in]{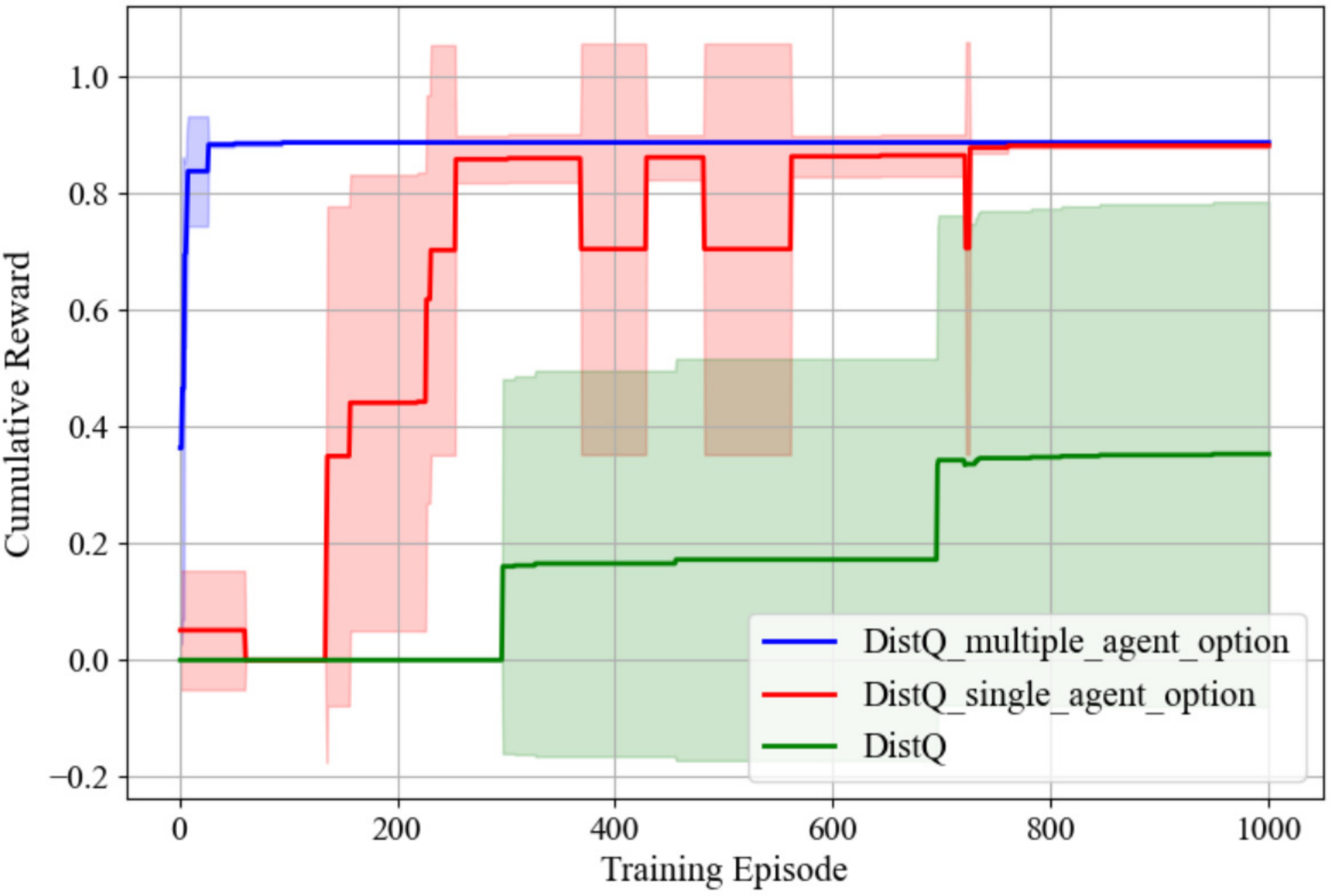}}

\subfigure[Centralized Q-Learning]{
\label{fig:8(d)} 
\includegraphics[width=2.2in,height=1.2in]{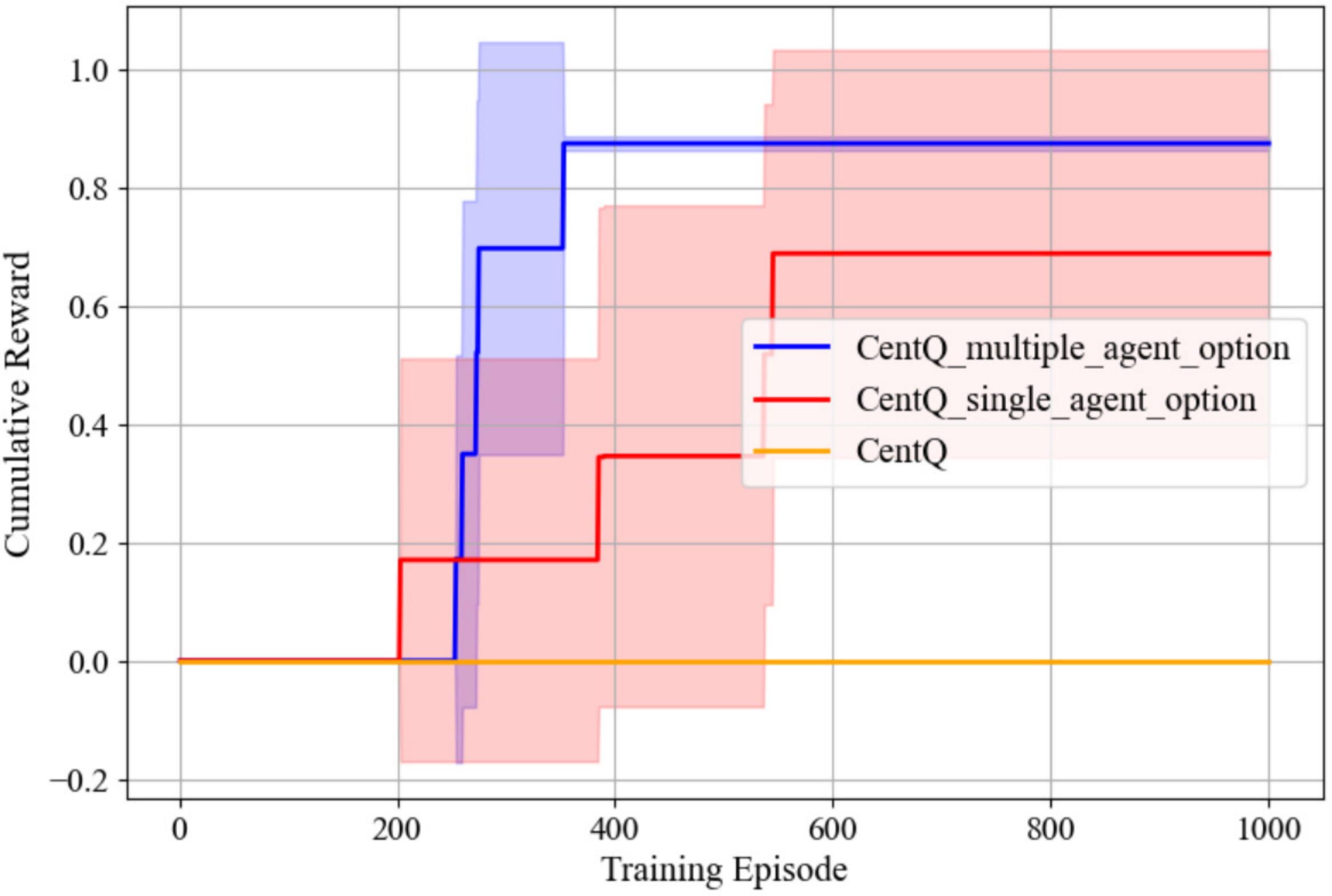}}
\subfigure[Centralized Q-Learning + Force]{
\label{fig:8(e)} 
\includegraphics[width=2.2in, height=1.2in]{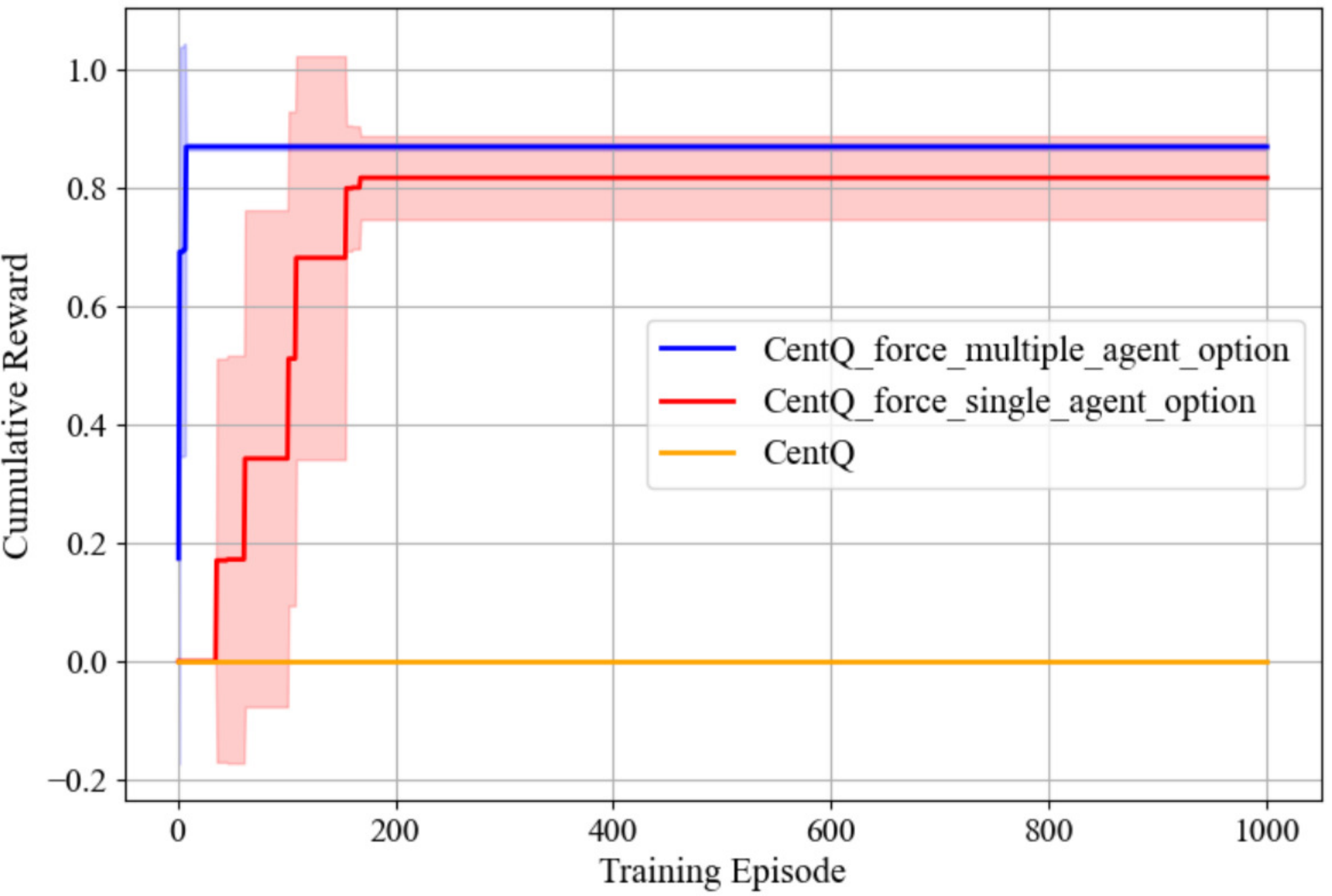}}

\caption{Comparisons on the two-agent four-room task: (a)-(e) show the results of using different algorithms as the high-level policy. No matter which algorithm we adopt, agents with multi-agent options can converge faster than the baselines. Also, shown as (a)(d)(e), our approach converges to a higher cumulative reward.}

\label{fig:8} 
\end{figure*}

\begin{figure*}[t]
\centering

\subfigure[four-room with 3 agents]{
\label{fig:3(b)} 
\includegraphics[width=2.3in, height=1.25in]{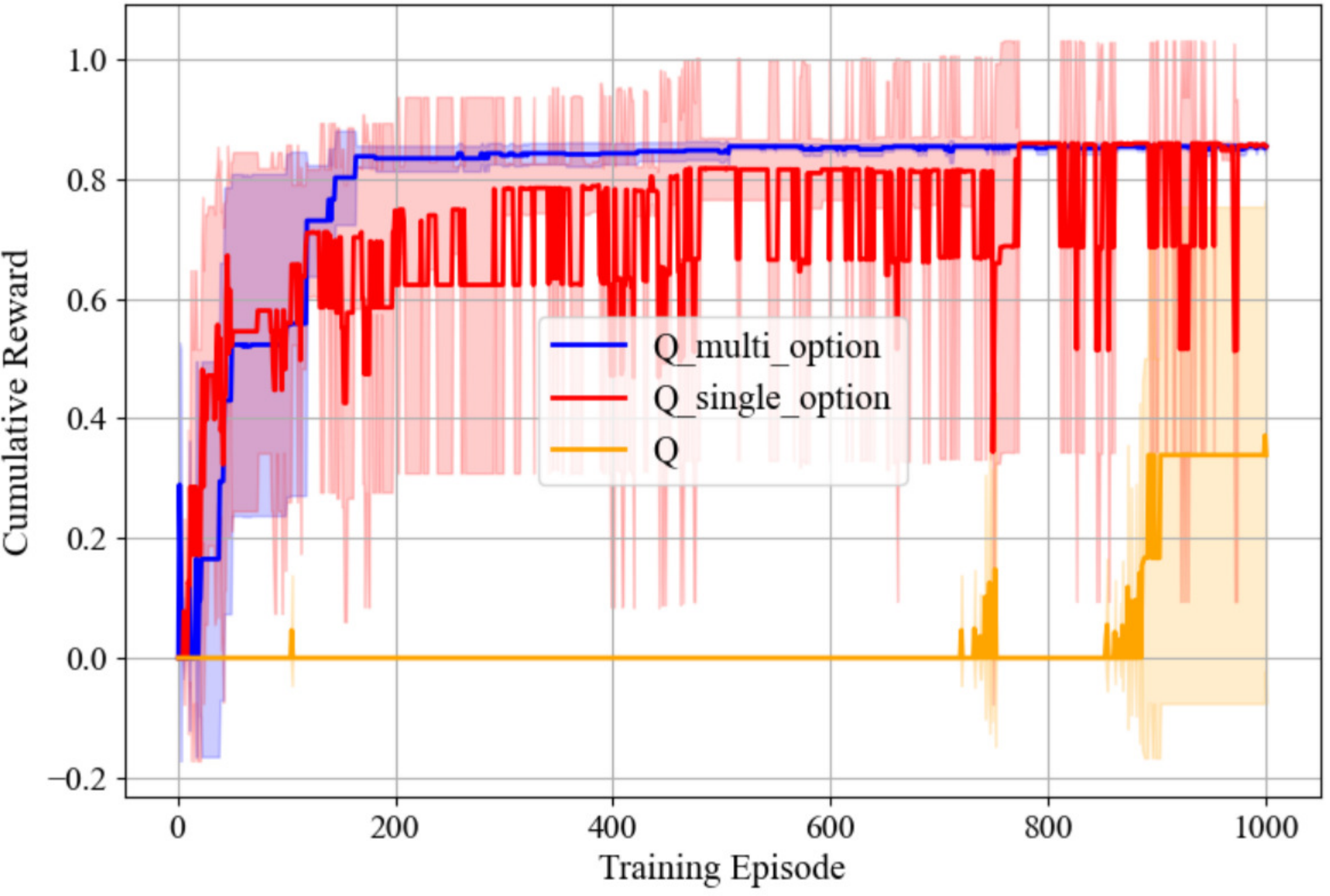}}
\subfigure[four-room with 4 agents]{
\label{fig:3(c)} 
\includegraphics[width=2.3in, height=1.25in]{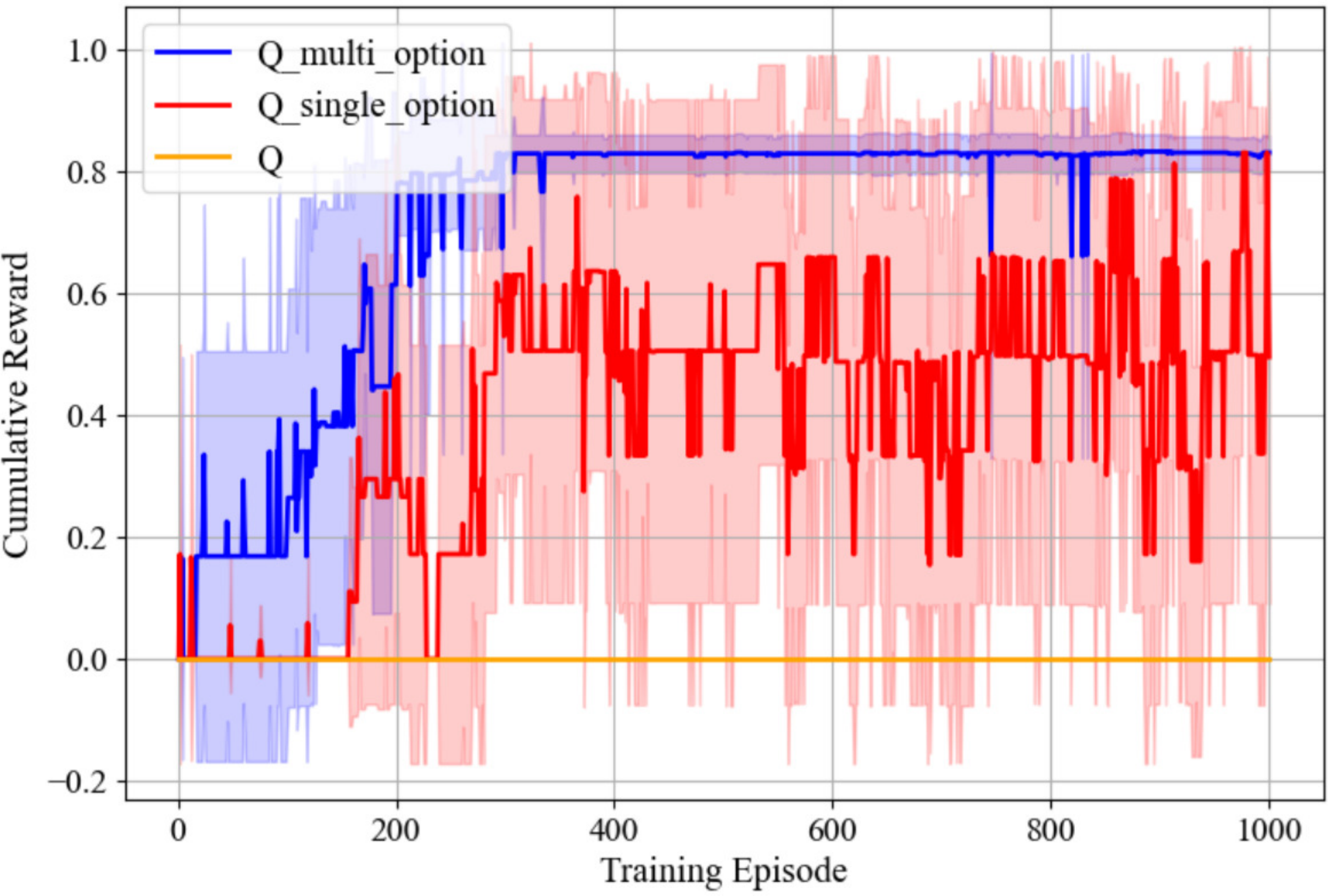}}
\subfigure[four-room with 5 agents]{
\label{fig:3(d)} 
\includegraphics[width=2.3in, height=1.25in]{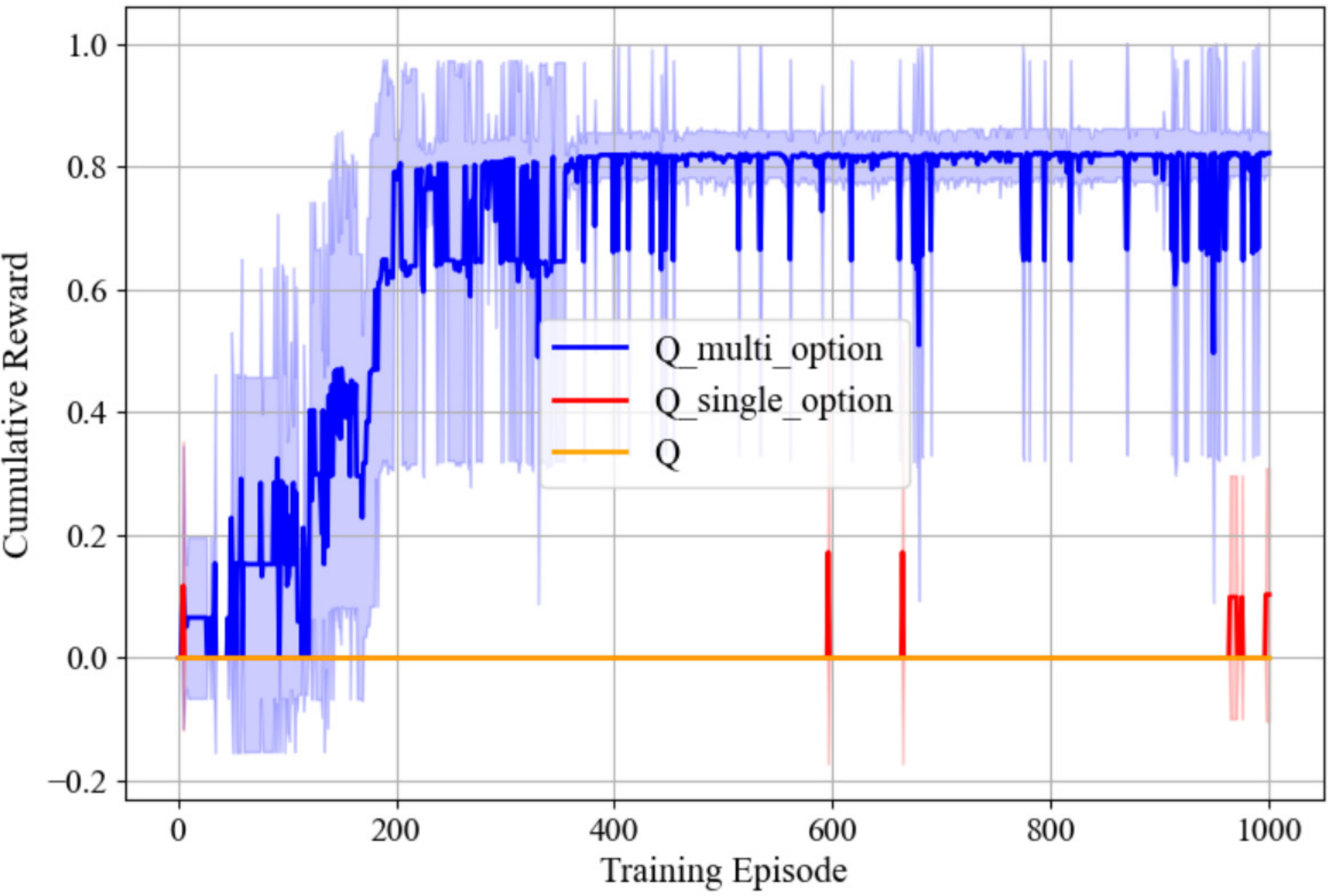}}

\subfigure[four-room with 3 agents]{
\label{fig:3(e)} 
\includegraphics[width=2.3in, height=1.25in]{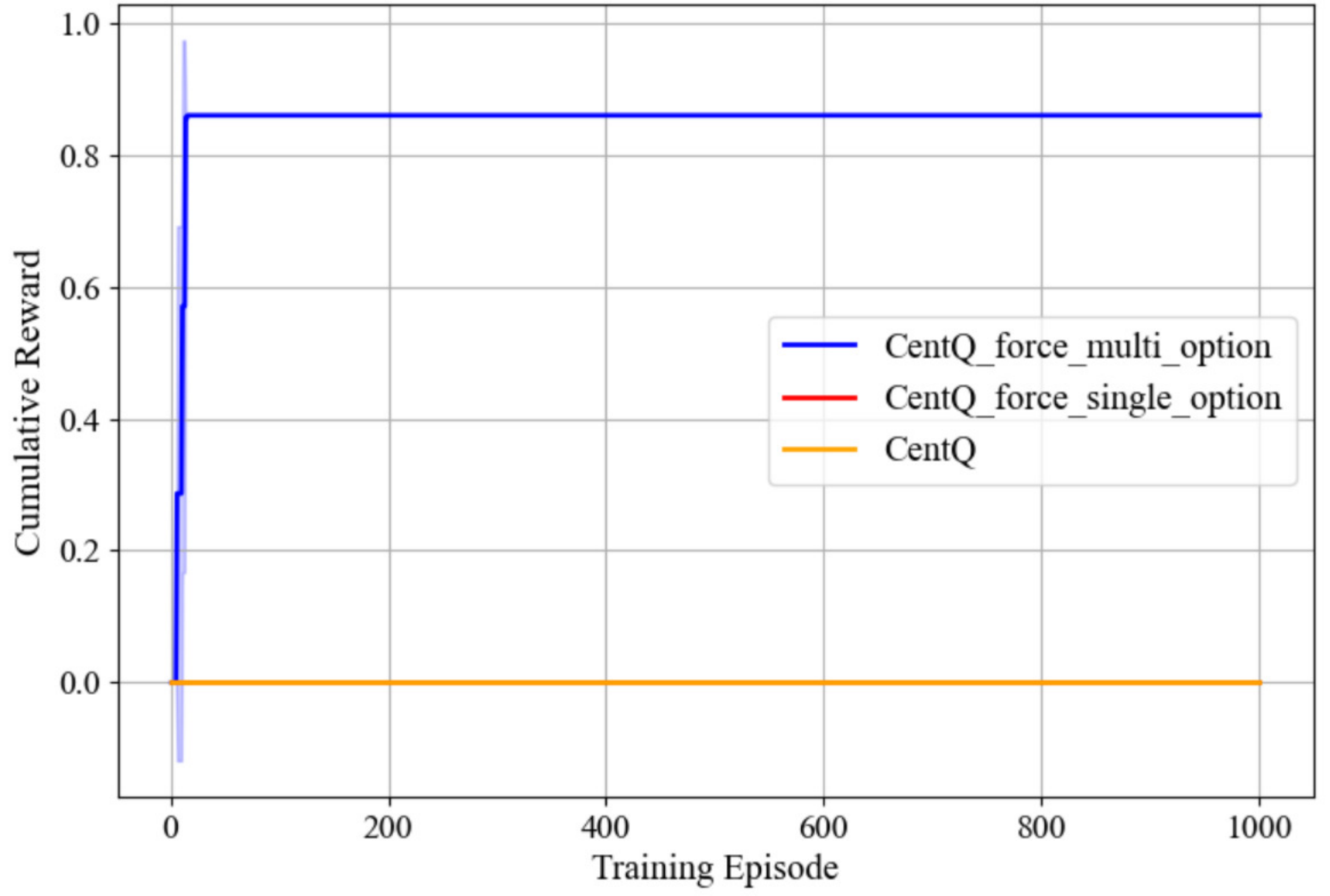}}
\subfigure[four-room with 4 agents]{
\label{fig:3(f)} 
\includegraphics[width=2.3in, height=1.25in]{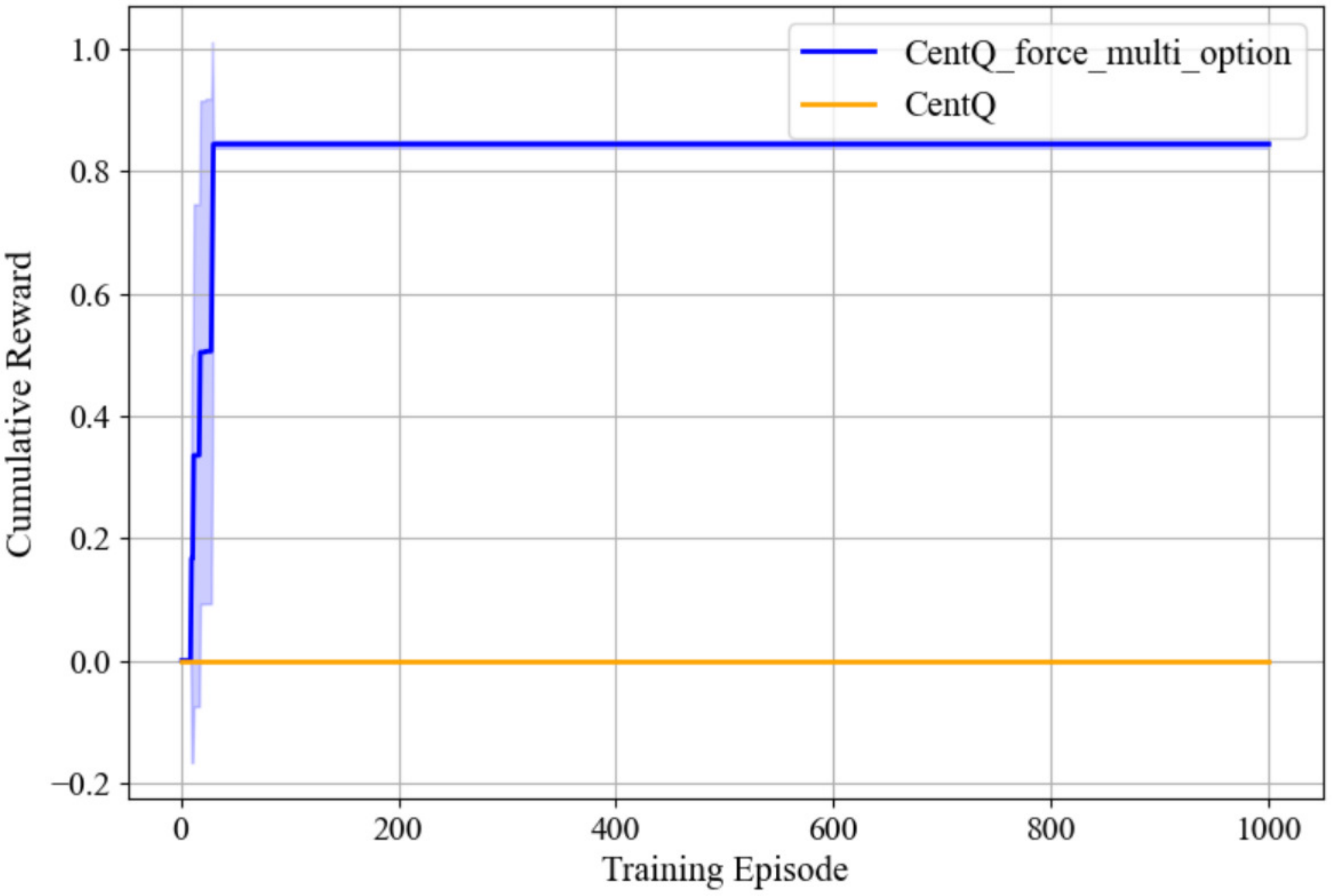}}
\subfigure[four-room with 5 agents]{
\label{fig:3(g)} 
\includegraphics[width=2.3in, height=1.25in]{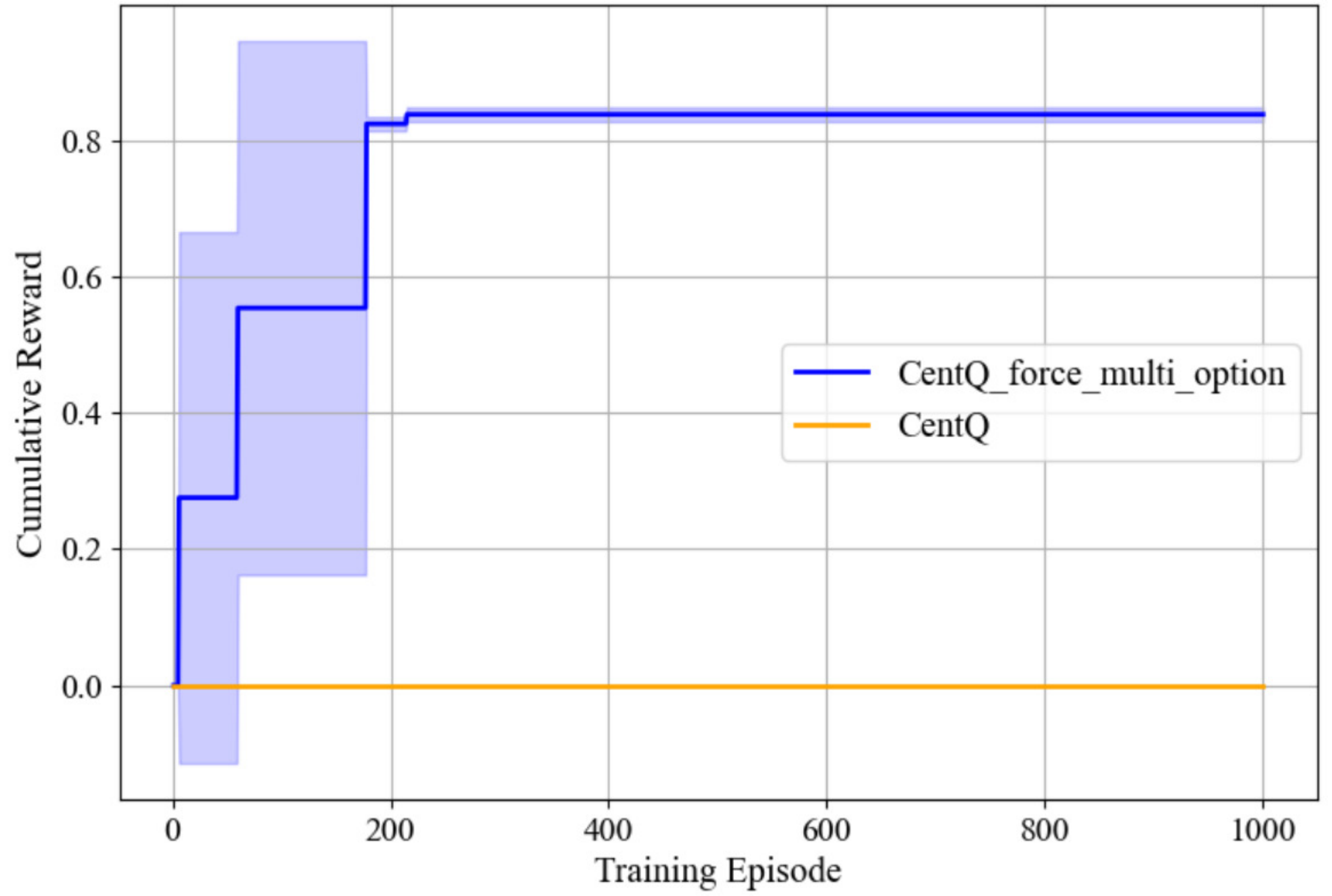}}

\caption{Evaluation on $n$-agent four-room tasks: (a)-(c) using Independent Q-Learning as the high-level policy. The performance improvement of our approach are more and more significant as the number of agents increases. (d)-(f) using Centralized Q-Learning + Force as the high-level policy. Agents with single-agent options start to fail since the 3-agent case. Also, it can be observed that the centralized way to utilize the $n$-agent options leads to faster convergence.}
\label{fig:3} 
\end{figure*}

\begin{figure}[t]
\centering
\subfigure[four-room with 2$\times$2 agents]{
\label{fig:6(a)} 
\includegraphics[width=1.66in, height=1.25in]{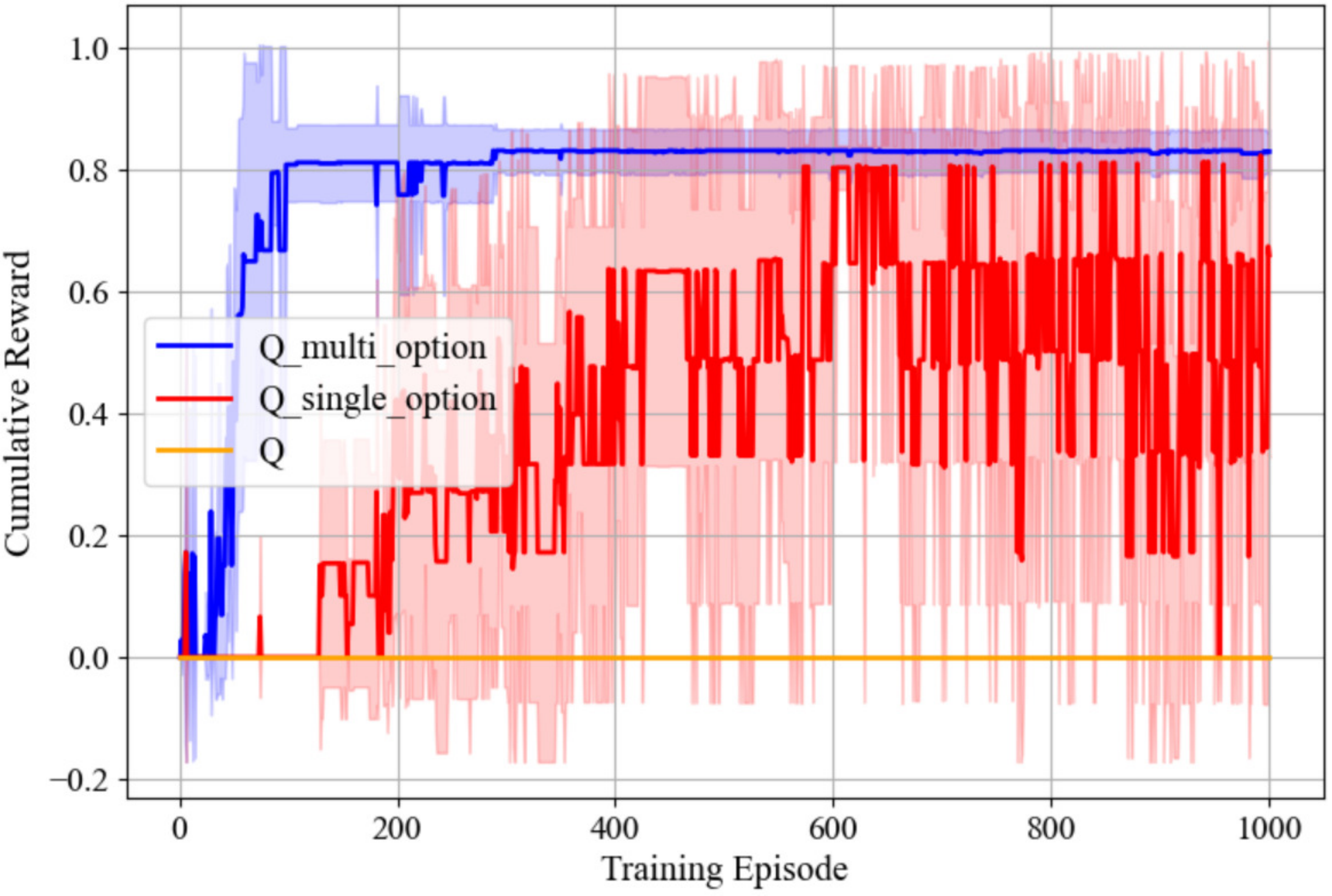}}
\subfigure[four-room with 3$\times$2 agents]{
\label{fig:6(b)} 
\includegraphics[width=1.66in, height=1.25in]{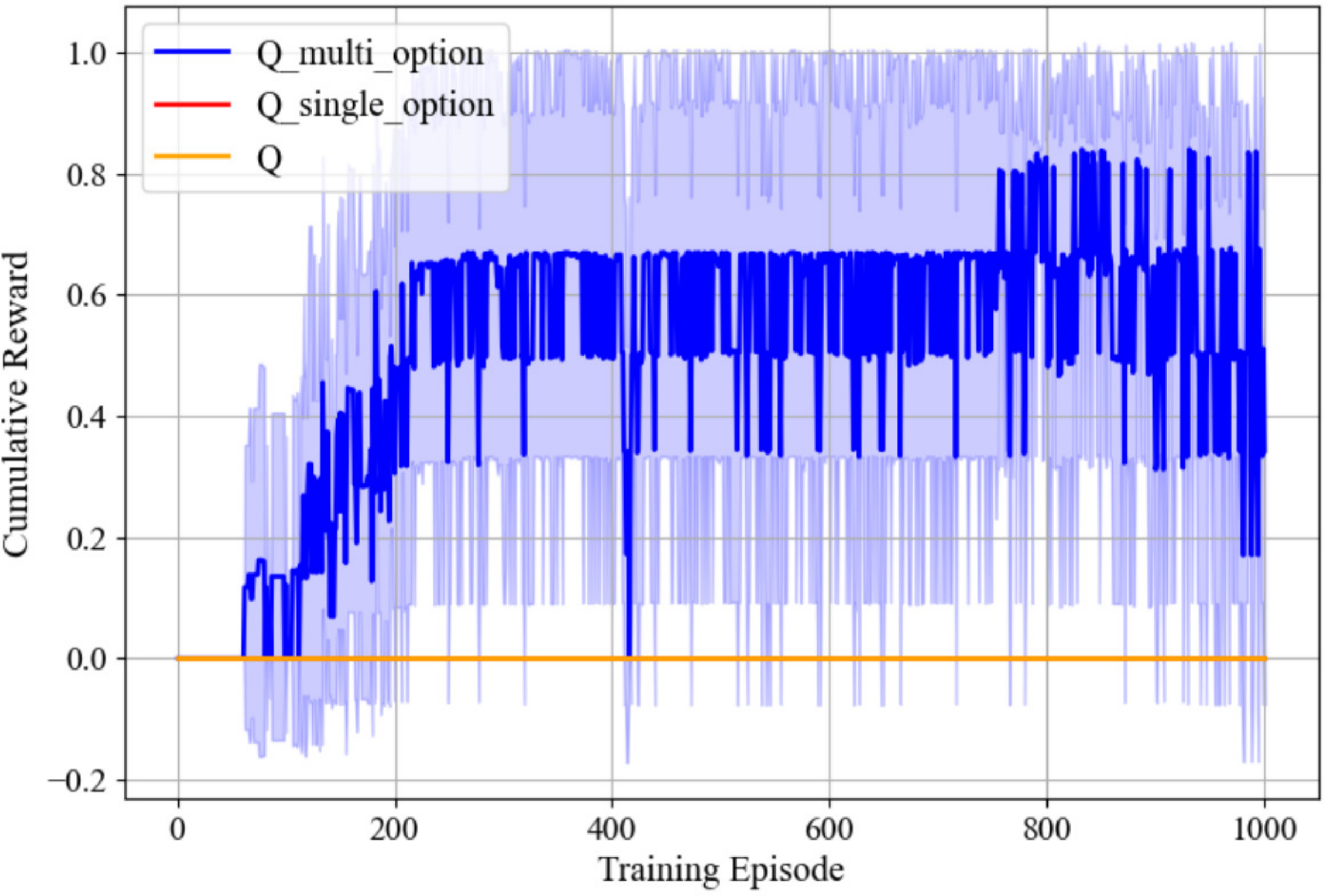}}
\subfigure[four-room with 2$\times$2 agents]{
\label{fig:6(c)} 
\includegraphics[width=1.66in, height=1.25in]{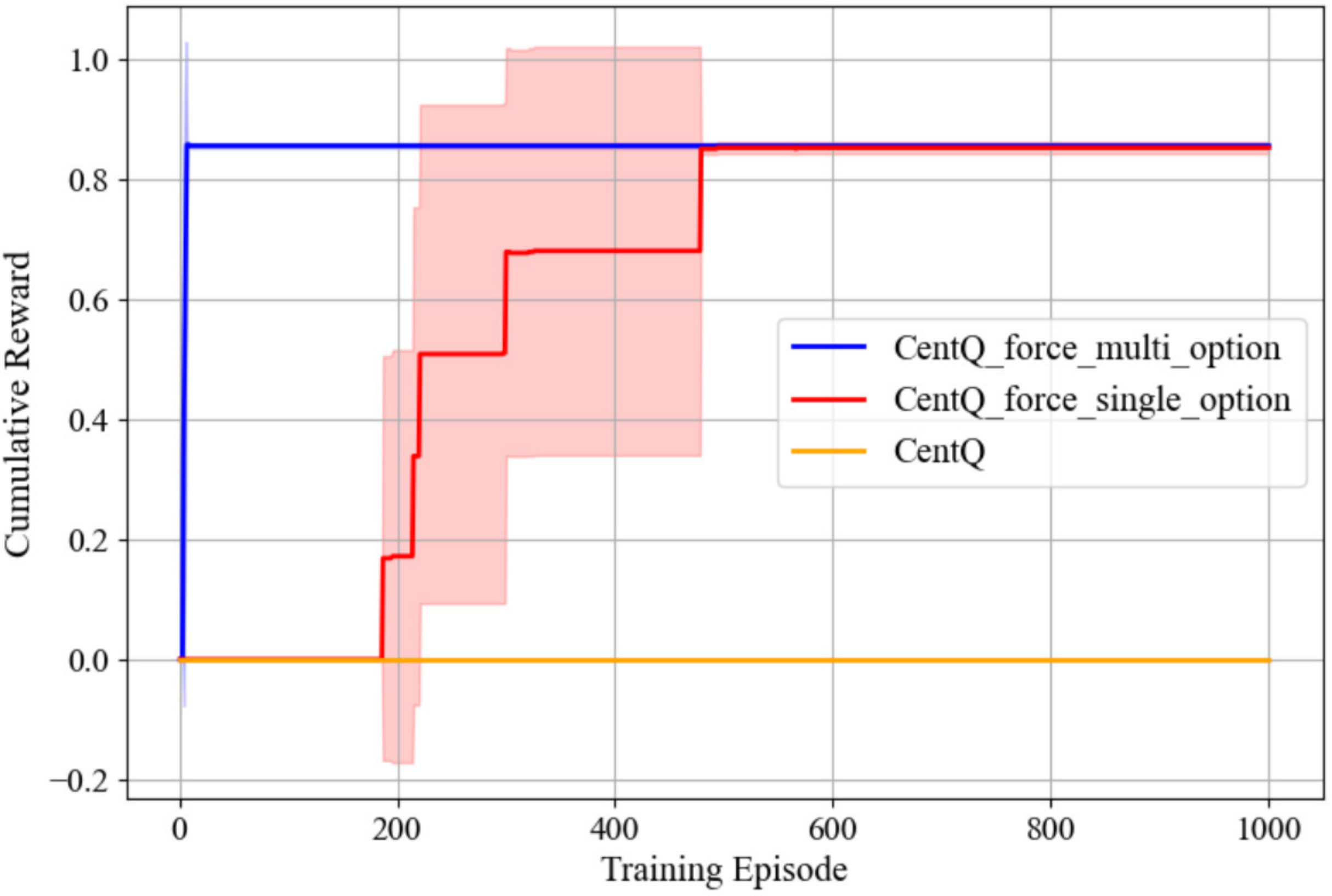}}
\subfigure[four-room with 3$\times$2 agents]{
\label{fig:6(d)} 
\includegraphics[width=1.66in, height=1.25in]{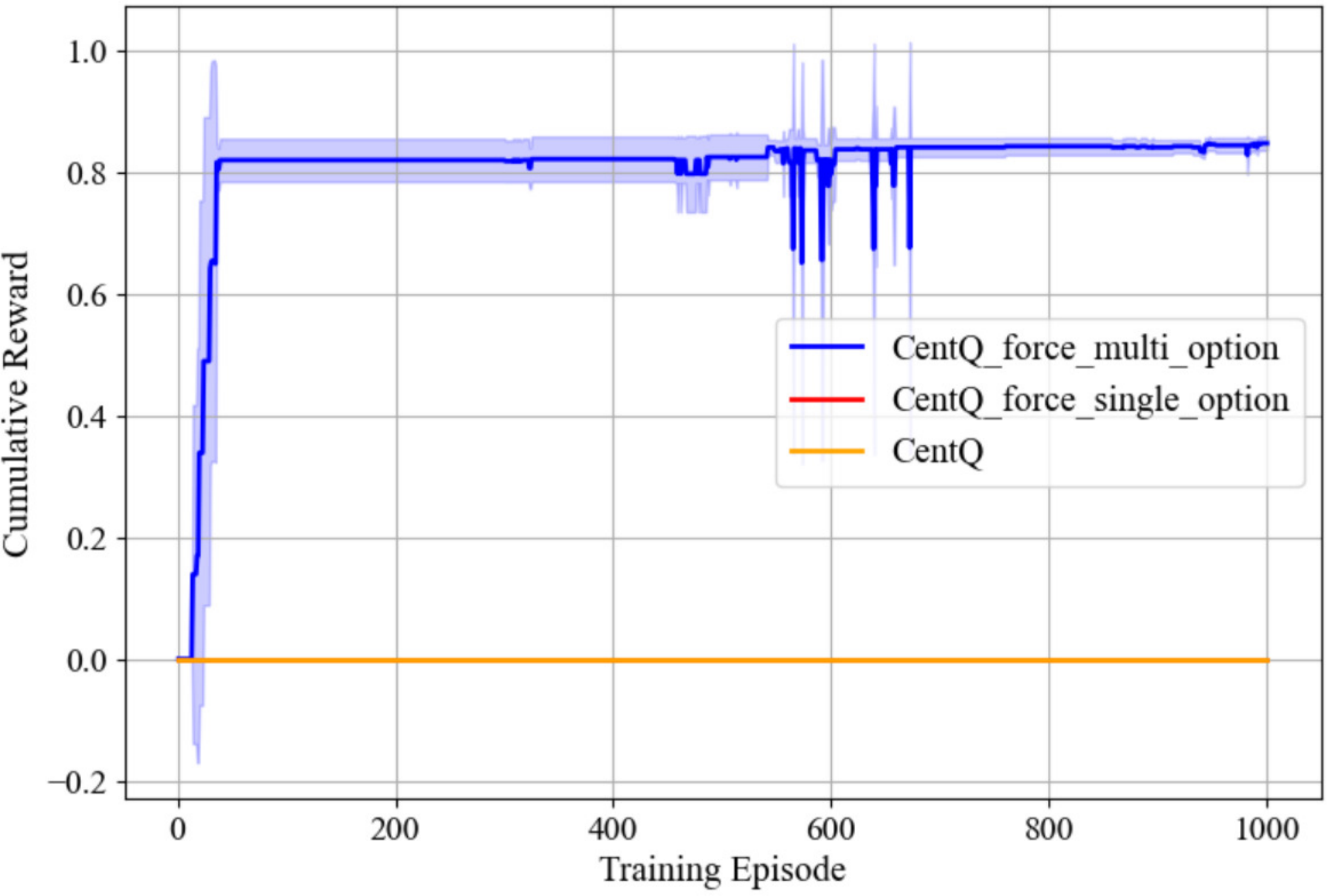}}
\caption{Comparisons on the $m \times n$ four-room tasks: (a)-(b) Independent Q-Learning; (c)-(d) Centralized Q-Learning + Force. Agents with pairwise options can learn these tasks much faster than the baselines, even when both the baselines fail on the $3\times2$ four-room task. Also, agents trained with Centralized Q-Learning + Force have faster convergence speed and higher convergence value than those trained with Independent Q-Learning.}
\label{fig:6} 
\end{figure}

\subsection{Simulation Setup}

As shown in Figure \ref{fig:2}, the proposed approach is evaluated on four multi-agent goal-achieving tasks. (1) For tasks shown as Figure \ref{fig:2(b)} and \ref{fig:2(d)}, $n$ (2$\sim$8) agents agents (triangles) must reach the goal area (circles) at the same time to complete this task, without going through the walls (squares). (2) For tasks shown as Figure \ref{fig:2(c)} and \ref{fig:2(e)}, there are $m$ groups of agents, and each group contains $n$ agents. Each group of agents has a special goal area labeled with the same color. Note that all the $m\times n$ agents should get to their related goal areas at the same time to complete this task, and the agents don't know which goal area is related to them at first. For all the four tasks, different agents can share the same grid, and only when the agents complete the task can they receive a reward signal $r=1.0$, which is shared by all the agents; otherwise, they will receive $r=0.0$. In the following experiments, we use the episodic cumulative reward as the metric, which is defined as: $\sum_{i=0}^{l}\mathcal{\lambda}^{i}r$, where $\mathcal{\lambda}=0.99$, and $l\leq200$ is the length of each episode. 

\begin{figure*}[t]
\centering
\subfigure[four-room with 4 agents]{
\label{fig:4(a)} 
\includegraphics[width=2.3in, height=1.25in]{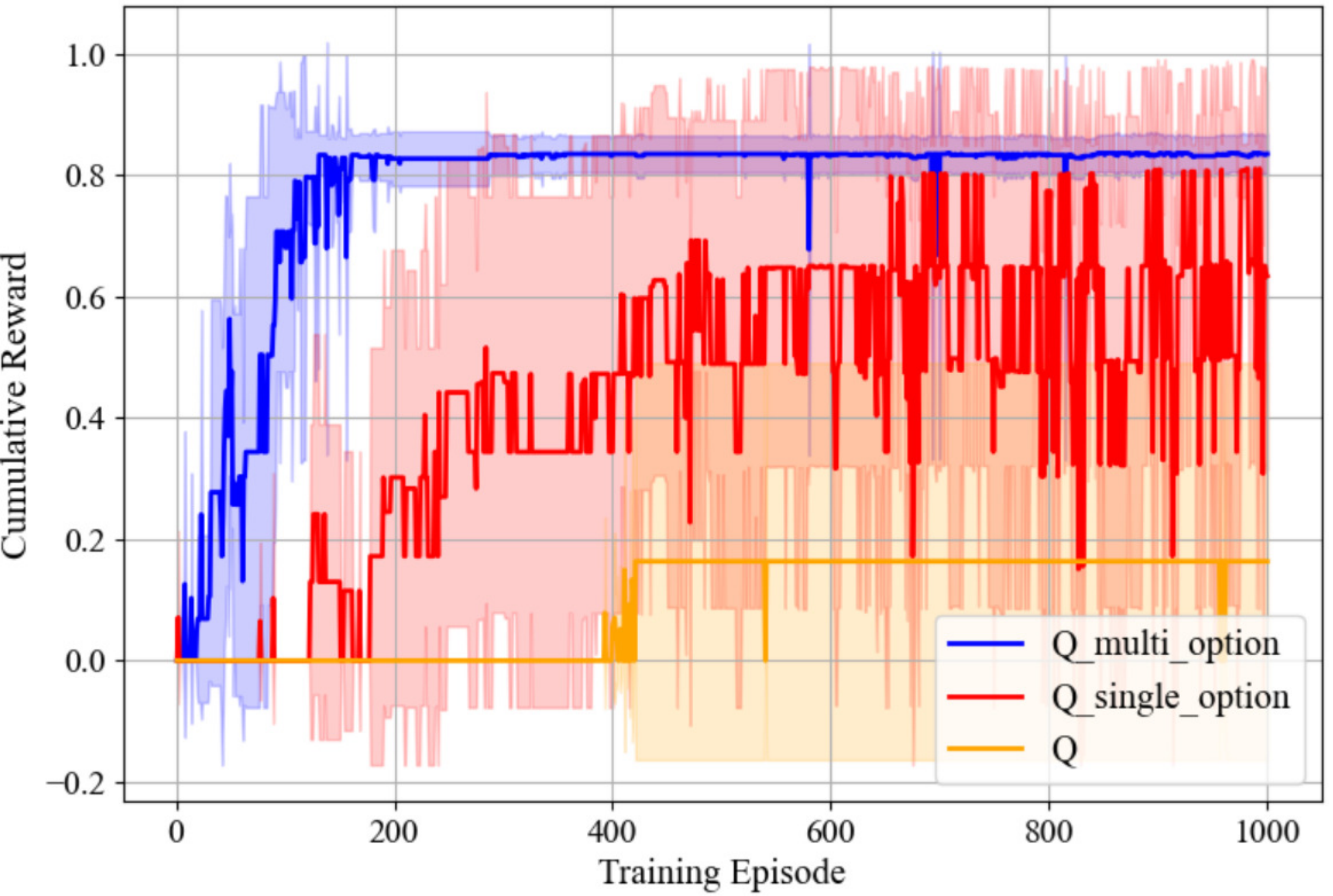}}
\subfigure[four-room with 6 agents]{
\label{fig:4(b)} 
\includegraphics[width=2.3in, height=1.25in]{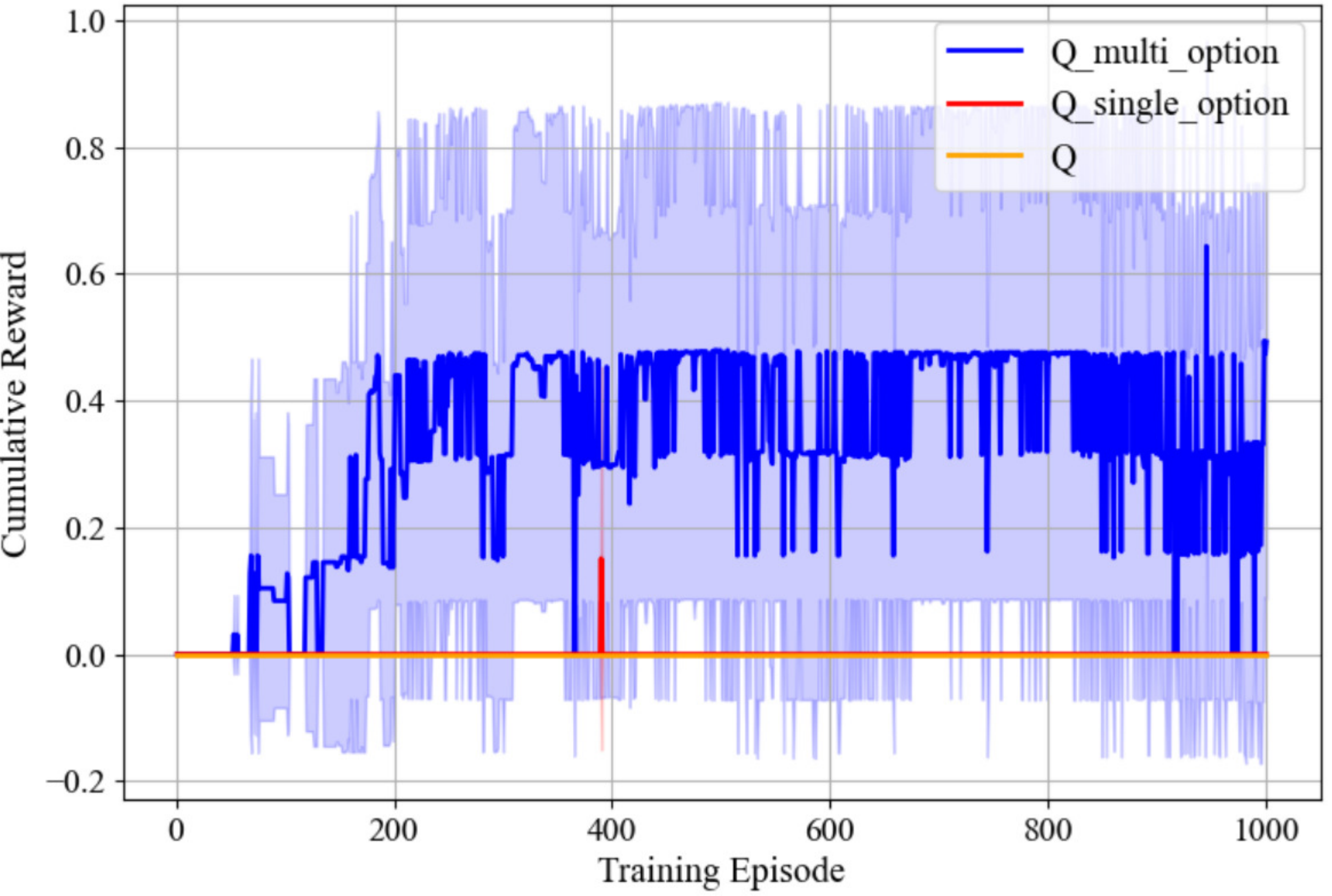}}
\subfigure[four-room with 8 agents]{
\label{fig:4(c)} 
\includegraphics[width=2.3in, height=1.25in]{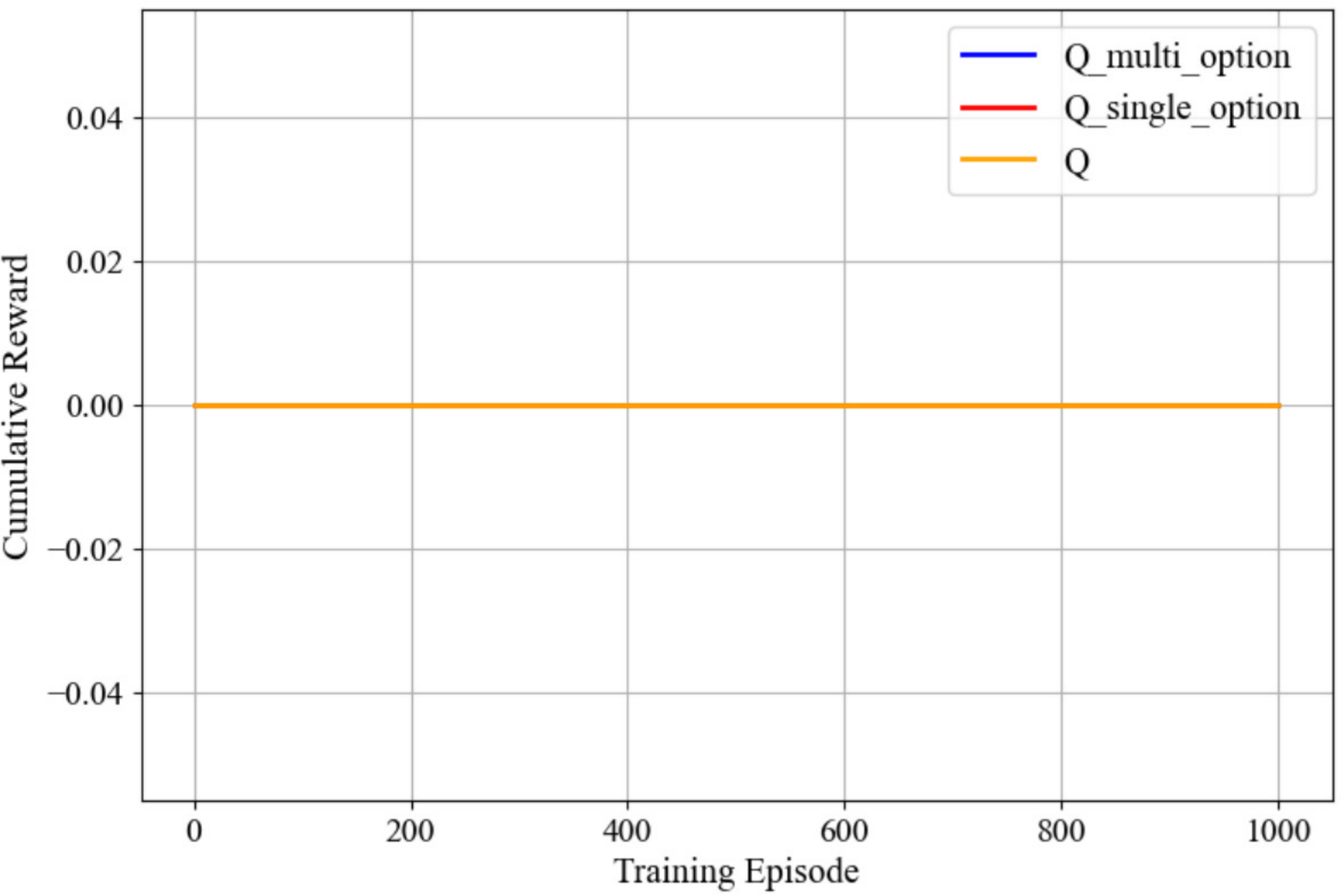}}

\subfigure[four-room with 4 agents]{
\label{fig:4(d)} 
\includegraphics[width=2.3in,height=1.25in]{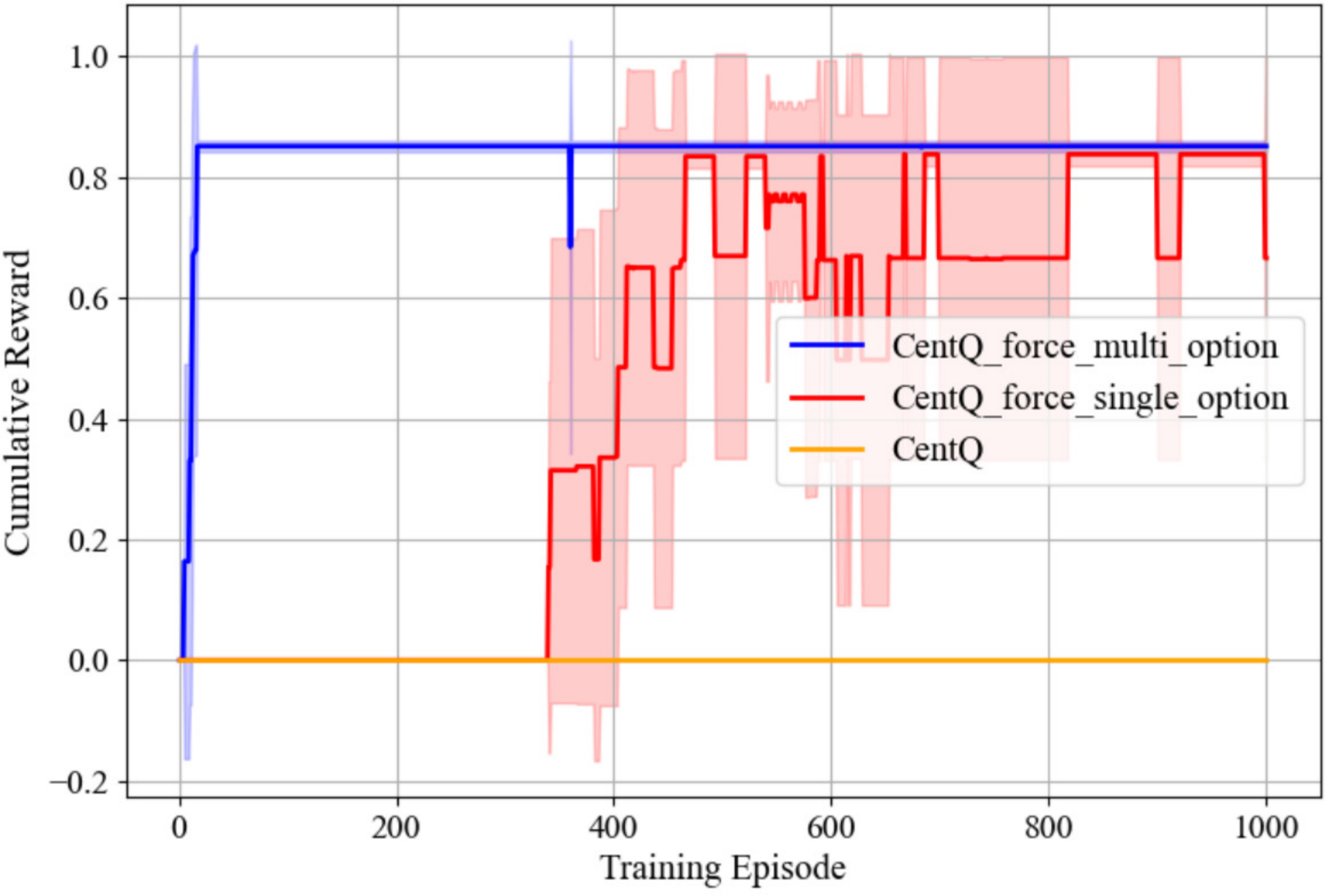}}
\subfigure[four-room with 6 agents]{
\label{fig:4(e)} 
\includegraphics[width=2.3in, height=1.25in]{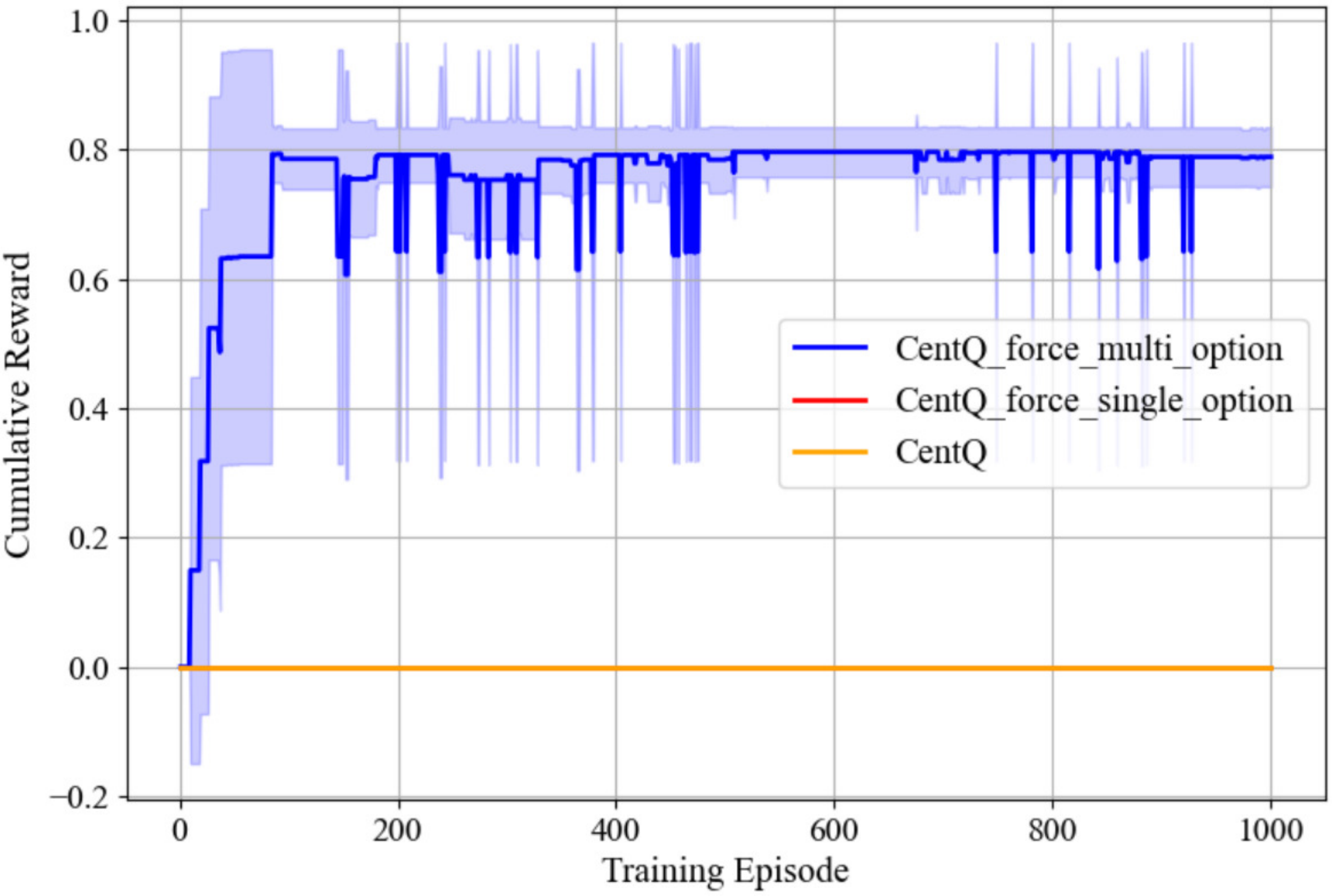}}
\subfigure[four-room with 8 agents]{
\label{fig:4(f)} 
\includegraphics[width=2.3in, height=1.25in]{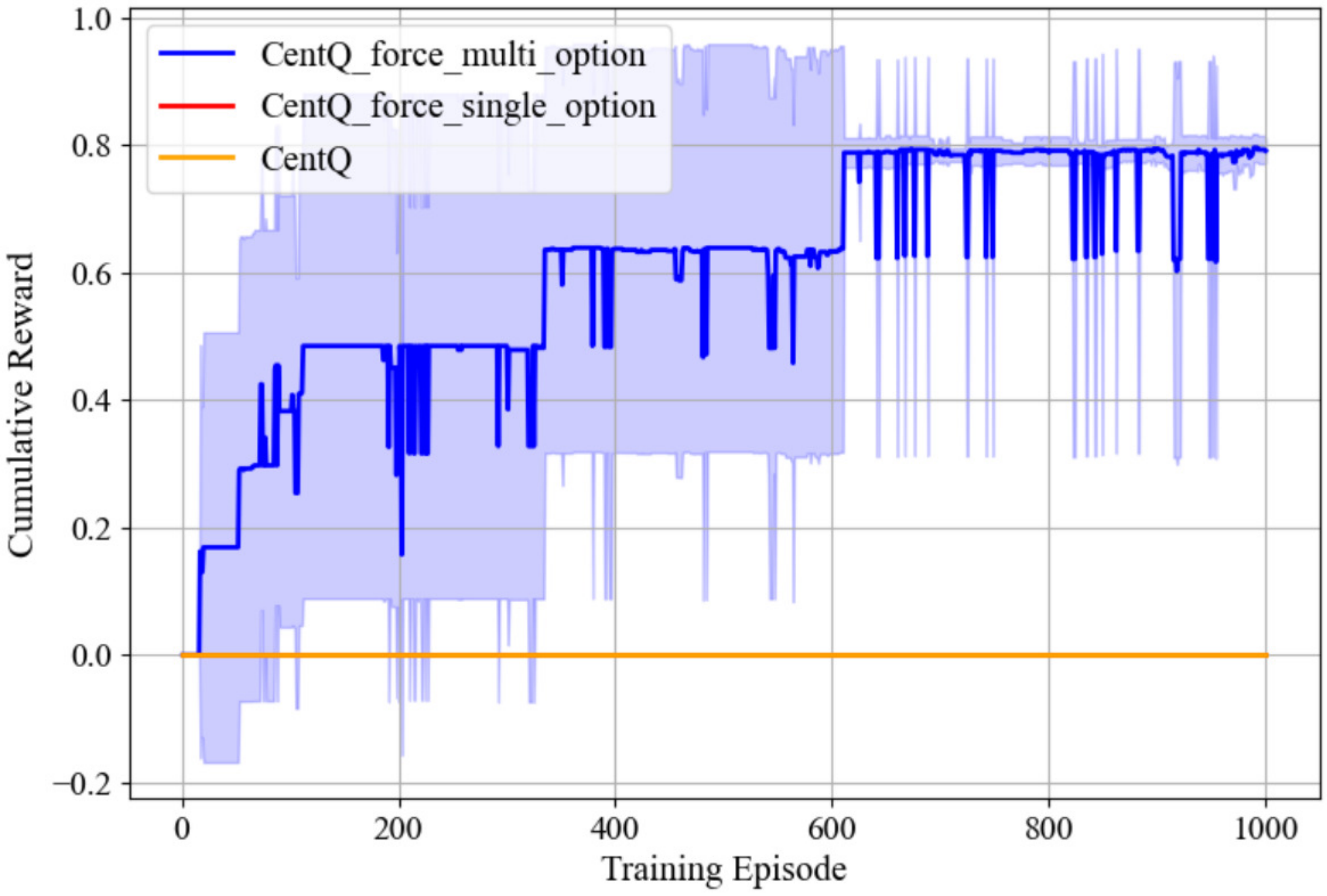}}
\caption{Comparisons on the $n$-agent four-room tasks: (a)-(c) Independent Q-Learning; (d)-(f) Centralized Q-Learning + Force. When $n$-agent options are not available, we can still get a significant performance improvement with only pairwise options. Adopting Centralized Q-Learning + Force within these sub-groups can further improve the convergence speed and value, compared with adopting Independent Q-Learning, e.g., when $n=6$, the cumulative reward is improved by $\mathbf{\sim 100 \%}$.}
\label{fig:4} 
\end{figure*}

Note that these evaluation tasks are quite challenging. On the one hand, the agents need to make decisions based on only their current locations (i.e., without knowing where the goal area is).  On the other hand, the reward space is very sparse: for example, in the eight-agent four-room task, the ratio of the rewarding joint states is $(4/121)^8\approx1.4\times10^{-12}$,  which is also the probability that they can complete this task through the random walk \cite{pearson1905problem}. Hence, agents without highly-efficient exploration strategies cannot complete these tasks. In Section \ref{results}, we conduct experiments with tasks of increasing complexity (e.g., Figure \ref{fig:3} and \ref{fig:4}), showing that the more difficult the task is, the more advantageous our approach becomes.

We compare our approach -- agents with multi-agent options, with two baselines: (1) Agents without options: the high-level policy is directly used to choose primitive actions, rather than choosing the option first and then choosing the primitive action according to the corresponding intra-option policy. Comparisons with this baseline can show the effectiveness of using options to aid the exploration. (2) Agents with single-agent options: we first construct covering options for each agent based on their individual state spaces, and then utilize these options in MARL, like what they do in \cite{DBLP:conf/atal/AmatoKK14, amato2019modeling, shen2006multi, DBLP:conf/atal/ChakravortyWRCB20, DBLP:conf/iclr/LeeYL20}. As for the option discovery method, we adopt the state-of-the-art algorithm proposed in \cite{DBLP:conf/icml/JinnaiPAK19}, which claims to outperform previous option discovery algorithms for sparse reward scenarios, like \cite{DBLP:conf/nips/SimsekB08} and \cite{DBLP:journals/corr/MachadoBB17}. Comparisons with this baseline can show the superiority of our approach to directly identify and adopt joint options in multi-agent scenarios. Note that, in order to confirm the fairness, we set the number of single-agent options and multi-agent options for each agent to select as the same. Also, we extend the initiation set of each single-agent option to the known area to increase their accessibility, like what we do with multi-agent options.

There are two kinds of policies in Figure \ref{fig:1}: the high-level policy for selecting among options, and the low-level policy for selecting among primitive actions. In the following experiments, we evaluate the performance of agents with five different algorithms as the high-level policy: random policy, Independent Q-Learning \cite{DBLP:conf/icml/Tan93}, Distributed Q-Learning \cite{DBLP:conf/icml/LauerR00} (each agent decides on their own option based on the joint state), Centralized Q-Learning and Centralized Q-Learning + Force, to make sure that the performance improvement is not specific to a certain algorithm.  Table \ref{tab:1} shows the comparisons among these algorithms: (1) If adopting ``Independent Q-Learning" as the high-level policy, agents need to make decisions based on only their local states; otherwise, agents within the same sub-group can share their views and make decisions based on their joint states. (2) For ``Centralized Q-Learning + Force", agents are forced to choose the same multi-agent option at a time (centralized); while, for the others, agents can choose different options to execute simultaneously (decentralized). As for the low-level policy, we adopt Value Iteration \cite{sutton2018reinforcement} to find the optimal path between each pair of initiation and termination state for each agent $i$ as $\pi_{\omega}^{i}$. Compared with Baseline (2), our approach does not require extra cost for learning the low-level policy, since the number of single-agent options and multi-agent options are the same for each agent.

\begin{figure*}[htbp]
\centering

\subfigure[Independent Q-Learning]{
\label{fig:10(c)} 
\includegraphics[width=2.4in, height=1.25in]{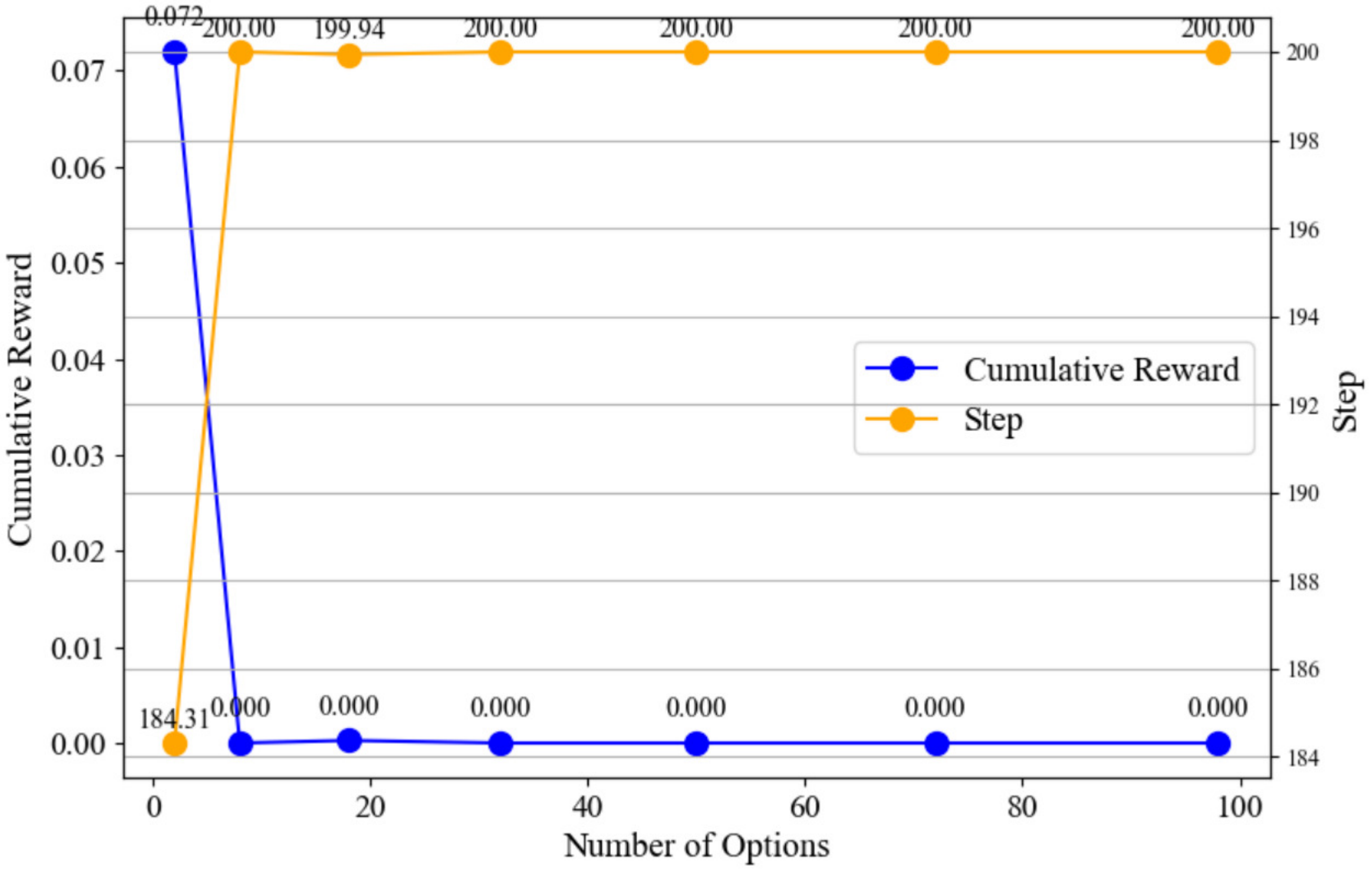}}
\subfigure[Centralized Q-Learning + Force]{
\label{fig:10(d)} 
\includegraphics[width=2.4in, height=1.25in]{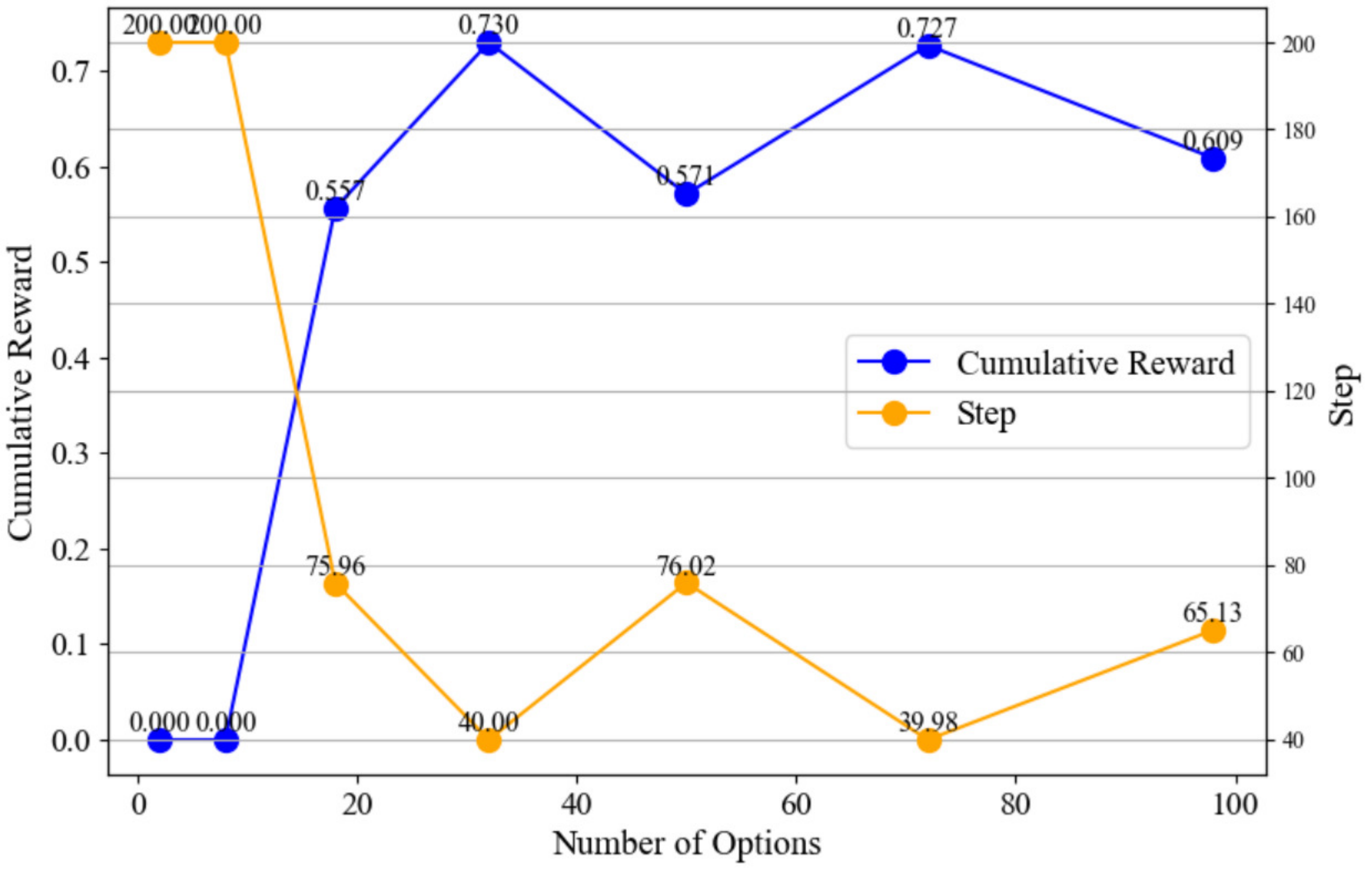}}

\caption{Performance change of the agents with pairwise options as the number of options increase, evaluated on the eight-agent four-room task. As the number of options increases, the performance of agents with Centralized Q-Learning + Force as the high-level policy can be improved further. While, if using the Independent Q-Learning as the high-level policy, the agents' performance would go worse.}

\label{fig:10} 
\end{figure*}

\begin{figure*}[t]
\centering
\subfigure[four-room with 4 agents]{
\label{fig:5(a)} 
\includegraphics[width=2.3in, height=1.25in]{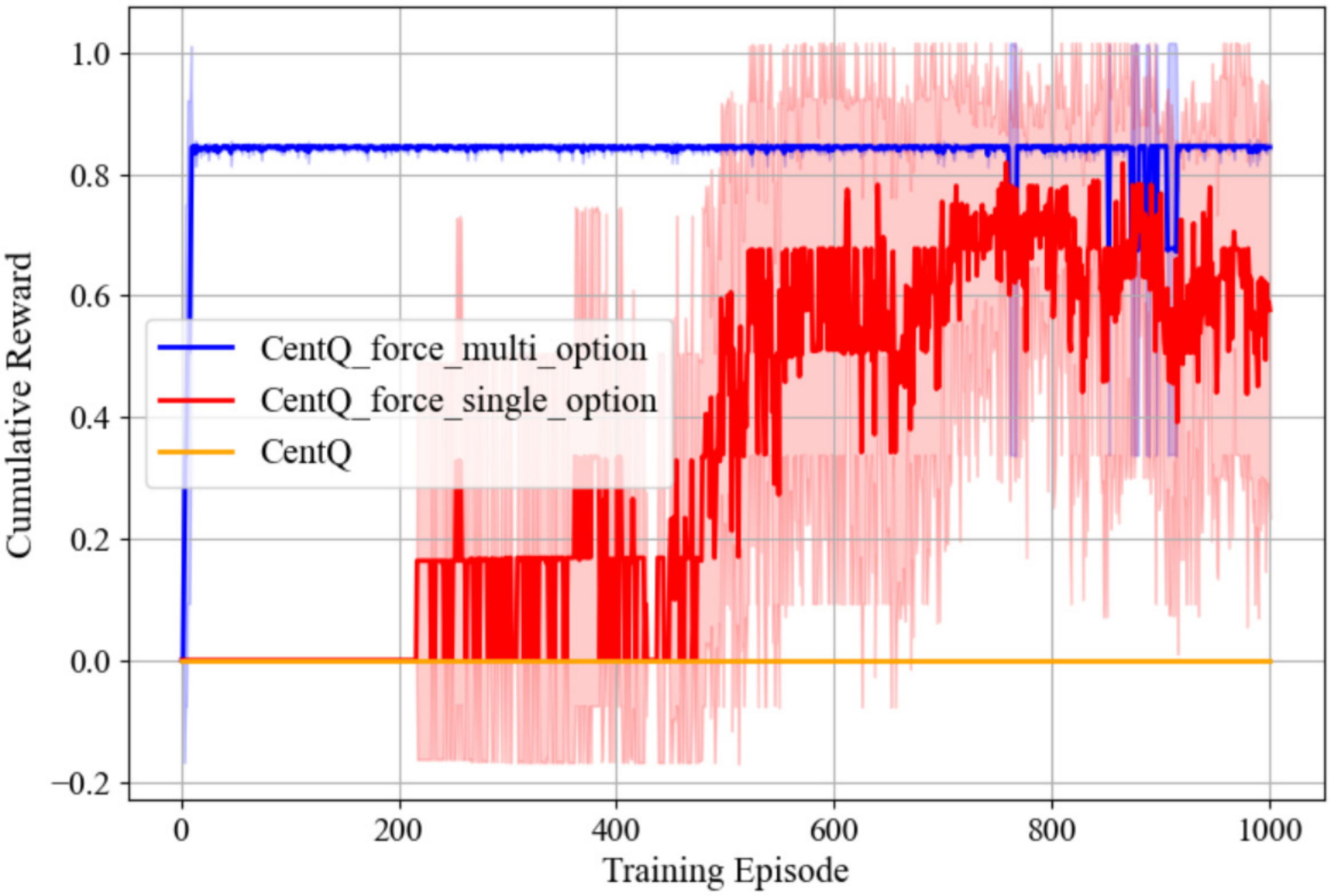}}
\subfigure[four-room with 6 agents]{
\label{fig:5(b)} 
\includegraphics[width=2.3in, height=1.25in]{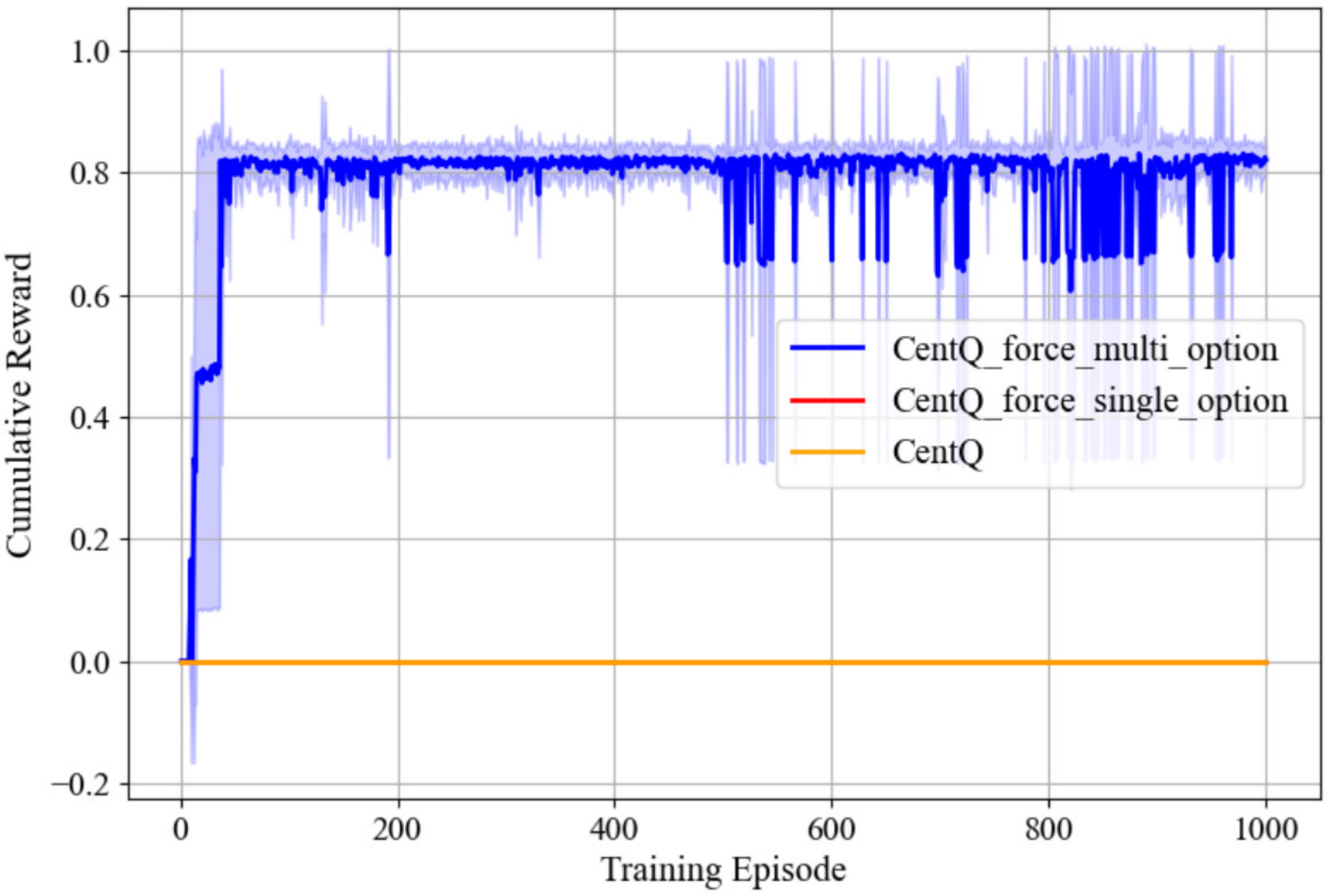}}
\subfigure[four-room with 8 agents]{
\label{fig:5(c)} 
\includegraphics[width=2.3in, height=1.25in]{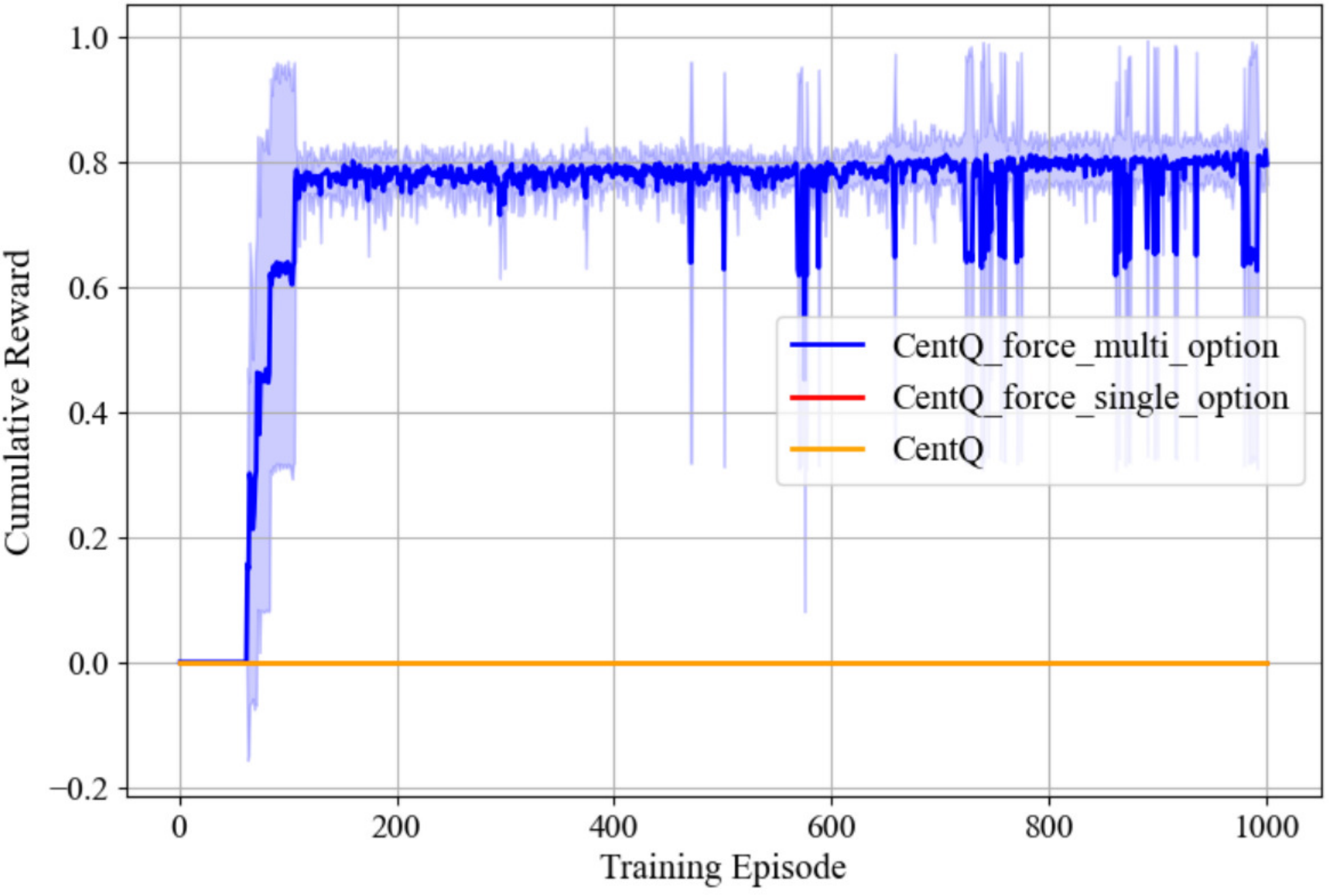}}
\caption{Comparisons on the $n$-agent four-room tasks where agent's state transitions can be influenced by the others, using Centralized Q-Learning + Force as the high-level policy. On this setting, we can still obtain good approximations of the multi-agent options based on the theory introduced in Section \ref{theory} and use them to get superior performance.}
\label{fig:5} 
\end{figure*}

\begin{figure*}[htbp]
\centering
\subfigure[maze with 6 agents]{
\label{fig:11(a)} 
\includegraphics[width=2.3in, height=1.25in]{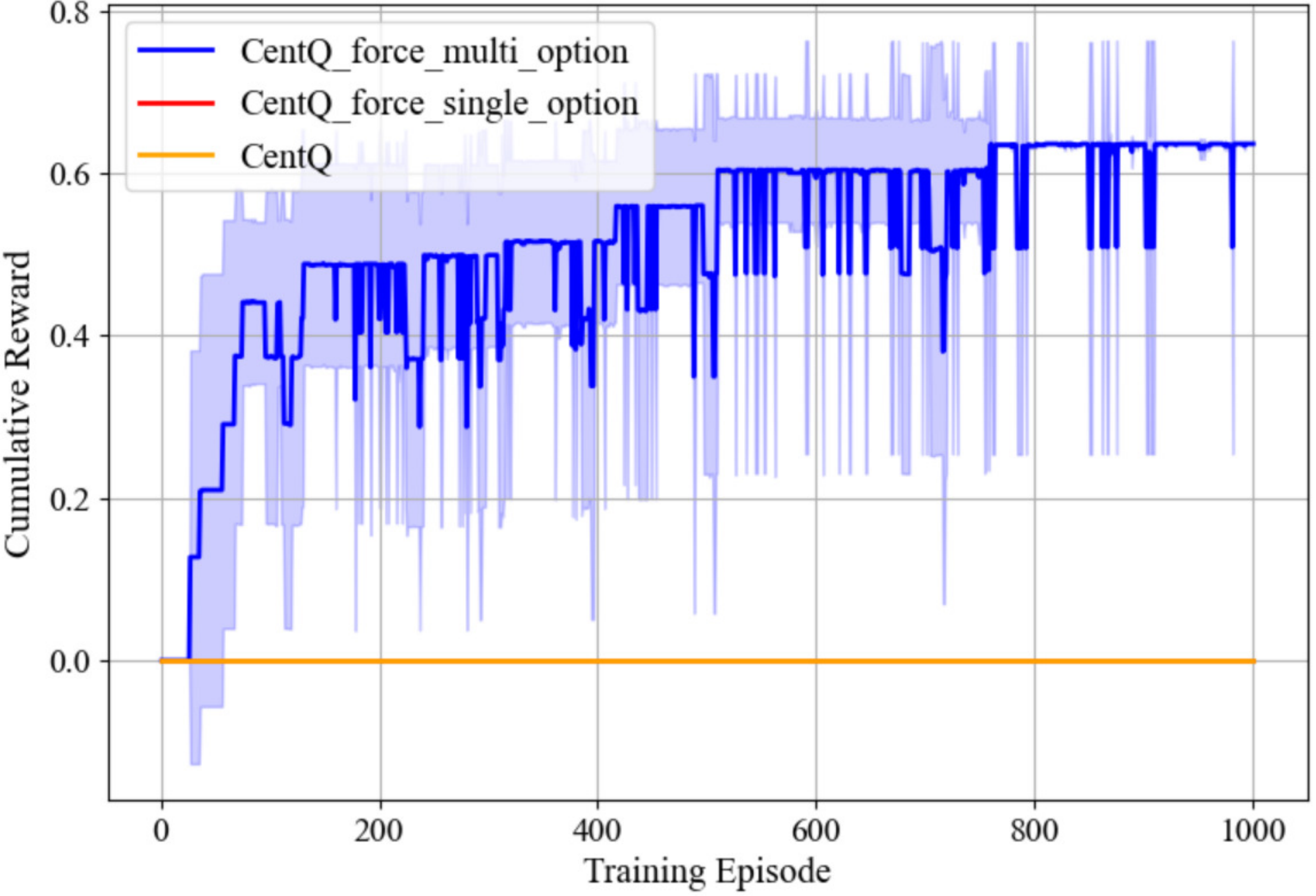}}
\subfigure[maze with 3$\times$2 agents]{
\label{fig:11(b)} 
\includegraphics[width=2.4in, height=1.25in]{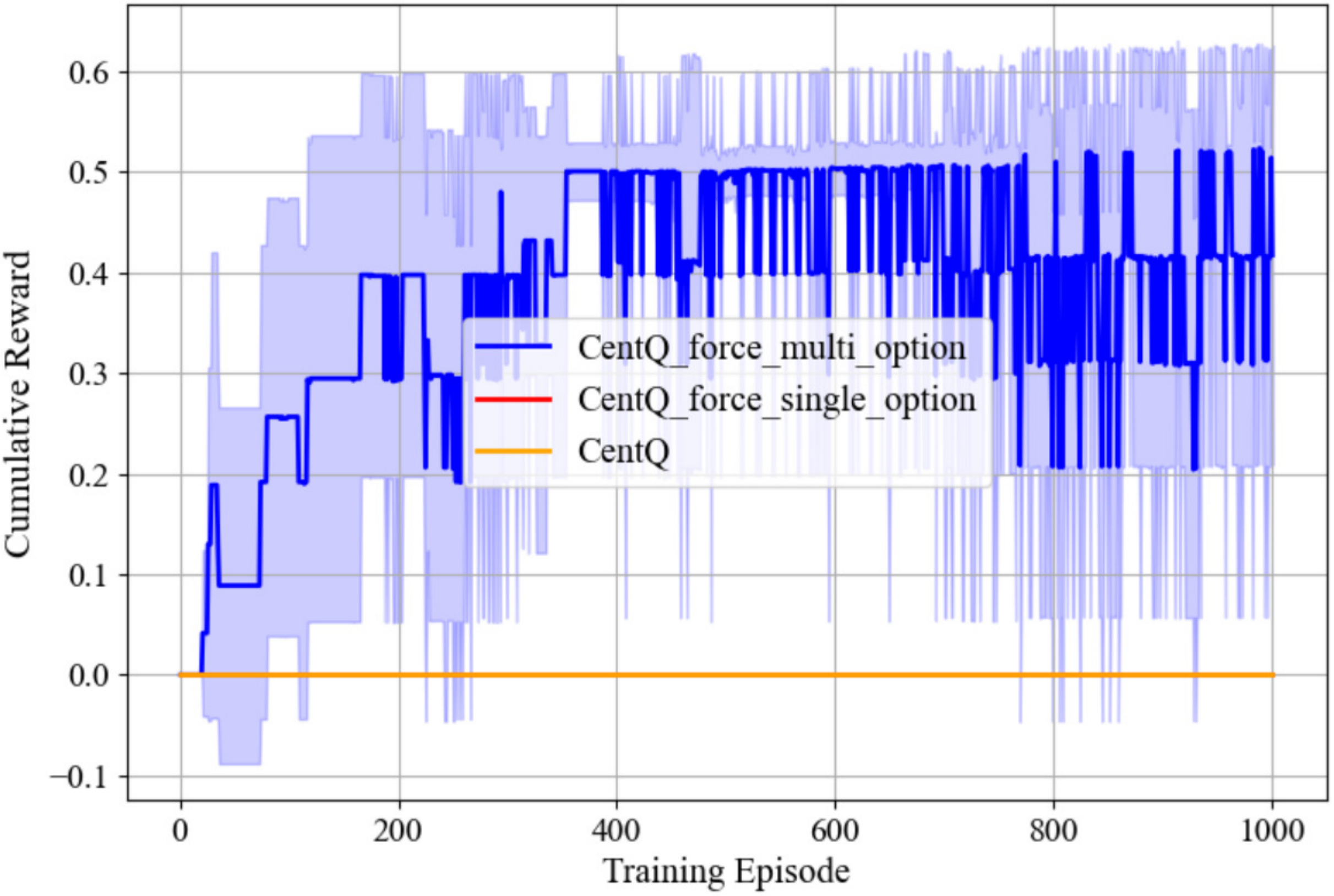}}
\caption{Comparisons on the more challenging maze tasks using Centralized Q-Learning + Force as the high-level policy, where (a) and (b) are corresponding to Figure \ref{fig:2(d)} and \ref{fig:2(e)} respectively. Although both baselines fail to complete the tasks, our approach can converge within 500 episodes with a fairly high cumulative reward.}

\label{fig:11} 
\end{figure*}

\subsection{Main Results} \label{results}

For each experiment in this section, we present comparisons among the performance of agents with multi-agent options (the blue line), agents with single-agent options (the red line) and agents without options. We run each experiment for five times and plot the change of the mean value and standard deviation of the episodic cumulative reward during the training process (1000 episodes).

\textbf{Two-agent four-room task:} As shown in Figure \ref{fig:8}, we present comparisons on the two-agent four-room task, with different algorithms (listed in Table \ref{tab:1}) as the high-level policy. It can be observed that no matter which algorithm we adopt as the high-level policy, agents with multi-agent options can converge faster than the baselines. However, when using Independent Q-Learning to train the high-level policy, the performance of our approach and the baselines are very close. Thus, in the follow-up experiments, we compare these approaches on more challenging tasks with Independent Q-Learning as the high-level policy to see if there will be more significant performance increase for our proposed method. Also, we will adopt Centralized Q-Learning + Force to train the high-level policy in the following experiments, to compare the two manners (decentralized or centralized) to utilize the multi-agent options.

\textbf{$N$-agent four-room task:} In Figure \ref{fig:3(b)}-\ref{fig:3(d)}, we test these methods on $n$-agent four-room tasks ($n=3\sim5$), using Independent Q-learning as the high-level policy. We can observe that the performance improvement brought by our approach are more and more significant as the number of agents increases. When $n=5$, both the baselines fail to complete the task, while agents with five-agent options can converge within $\sim200$ episodes. On the other hand, in Figure \ref{fig:3(e)}-\ref{fig:3(g)}, we show the results of using Centralized Q-Learning + Force as the high-level policy on the same tasks. We can see that the centralized way to utilize the $n$-agent options leads to faster convergence, since the joint output space of the agents is pruned (shown as Figure \ref{fig:1}). Note that when the number of agents is three, the agents with single-agent options already fail to complete the four-room task. We don't include the results of agents with single-agent options in Figure \ref{fig:3(f)}-\ref{fig:3(g)}, because it takes a tremendously long time to run those experiments and it can be predicted that the results will be the same as Figure \ref{fig:3(e)}. 

\textbf{Four-room task with sub-task grouping:} The size of the joint state space and output space grows exponentially with the number of agents, making it infeasible to directly construct $n$-agent options and adopt Centralized Q-Learning for a large $n$. However, in real-life scenarios, a multi-agent task can usually be divided into sub-tasks, and the agents can be divided into sub-groups based on the sub-tasks they are responsible for. Thus, we test our proposed method on the $m \times n$ four-room tasks shown as Figure \ref{fig:2(c)}, where we divide the agents into $m$ sub-groups, each of which contains $n$ agents with the same goal area. 
Figure \ref{fig:6} shows comparisons between our method and the baselines on $m \times n$ four-room tasks. Note that, in the $2\times2$ ($3\times2$) four-room task, we use two-agent (pairwise) options rather than four-agent (six-agent) options, and when using Centralized Q-Learning + Force, we only use the joint state space of the two agents as input to decide on their joint option choice. We can see that agents with pairwise options can learn to complete the tasks much faster than the baselines (e.g., improved by about two orders of magnitude in the $2\times2$ four-room task), even when both the baselines fail to complete the $3\times2$ four-room task. Also, we see that agents trained with Centralized Q-Learning + Force (Figure \ref{fig:6(c)}-\ref{fig:6(d)}) have faster convergence speed and higher convergence value compared to training with Independent Q-Learning (Figure \ref{fig:6(a)}-\ref{fig:6(b)}).

\textbf{Four-room task with random grouping:} Further, we note that our method also works with random grouping when sub-task grouping may not work. The intuition is that adopting two-agent or three-agent options can encourage the joint exploration of the agents in small sub-groups, which can increase the overall performance compared with only utilizing single-agent explorations. As shown in Figure \ref{fig:4}, we compare the performance of agents with pairwise options, single-agent options and no options on the $n$-agent four-room tasks ($n=4,6,8$).  We can observe that when $n=6\ or\ 8$, agents with single-agent options or no options can't complete this task, while we can get a significant performance improvement with only pairwise options. On the other hand, agents with pairwise options can't complete the most challenging eight-agent four-room task, if we use Independent Q-Learning to train the high-level policy, shown as Figure \ref{fig:4(c)}. However, if we adopt Centralized Q-Learning + Force, agents with pairwise options can still complete this challenging task with satisfaction, shown as Figure \ref{fig:4(f)}. 

Further, in Figure \ref{fig:10}, we show how the performance of agents with pairwise options would change with the number of options, based on the eight-agent four-room tasks (the orange line: number of steps to complete the task; the blue line: episodic cumulative reward). When increasing the number of options, the performance of agents with pairwise options and using Centralized Q-Learning + Force as the high-level policy can be improved further. While, if using the Independent Q-Learning as the high-level policy, the agents' performance would go worse as the number of options increase. The reason is that, as mentioned in Section \ref{framework}, the joint output space of the agents will grow exponentially with the number of options if we utilize the multi-agent options in a decentralized way. In contrast, the size of the joint output space is linear with the number of options when we use the multi-agent options in a centralized manner.

\textbf{Four-room task with random grouping and dynamic influences among agents:} Further, we show that even if in environments where an agent's state transitions can be strongly influenced by the others, we can still obtain good approximations of the multi-agent options to encourage joint exploration using THEOREM \ref{thm:2}. For this new setting, we make some modifications based on the $n$-agent four-room task (Figure \ref{fig:2(b)}) -- different agents cannot share the same grid so that an agent may be blocked by others when moving ahead, and this influence is highly dynamic. We use the Centralized Q-Learning + Force as the high-level policy, of which the results are shown as Figure \ref{fig:5}. We can see that although this modification affects the performance of agents with single-agent options, we can still get significant performance improvement with the pairwise options.  

\textbf{Maze task with random grouping or sub-task grouping:} Finally, in order to show the effectiveness of our approach on more challenging tasks, we compare it with the baselines on the maze tasks shown as Figure \ref{fig:2(d)} and \ref{fig:2(e)}, of which the results are shown in Figure \ref{fig:11(a)} and \ref{fig:11(b)} respectively. Compared with the four-room task, the state space of the maze task is larger and the path finding toward the goal-area is much more difficult. Again, both baselines fail to complete the tasks, while our approach can converge within 500 episodes with a fairly high cumulative reward. Note that, for both tasks, we first group the agents based on sub-tasks (Figure \ref{fig:2(e)}) or randomly (Figure \ref{fig:2(d)}), then learn the pairwise options for each sub-group and utilize these options in a centralized manner to aid the exploration.

We present the numeric results of all the experiments in tabular in Appendix \ref{numeric results}, which show that agents with multi-agent options outperform the baselines in \textbf{ALL} the multi-agent tasks in terms of faster exploration and higher cumulative rewards.

\section{Conclusion} \label{conc}

In this paper, we propose to approximate the joint state space in MARL as a Kronecker graph and estimate its Fiedler vector using the Laplacian spectrum of the individual agents' state transition graphs. Based on the approximation of the Fiedler vector, multi-agent covering options are constructed, containing multiple agents’ temporal action sequence towards the sub-goal joint states which are usually infrequently visited, so as to accelerate the joint exploration in the environment. Further, we propose algorithms to adopt these multi-agent options in MARL, using centralized, decentralized, and group-based strategies, respectively. We show through evaluation results that agents with multi-agent options have significantly superior performance than agents relying on single-agent options or no options. 

As a future direction, it will be of great use to scale our algorithm for real-life applications with SOTA representation learning and deep learning techniques, like \cite{DBLP:conf/iclr/WuTN19, DBLP:conf/icml/WangZZSHF21}. On the other hand, there will be non-negligible differences between $\otimes_{i=1}^{n}G_{i}$ and the joint state transition graph $\widetilde{G}$, if the state transitions of an agent are hugely influenced by the other ones. Therefore, mechanisms to detect these situations in a task scenario and integrate them with  $\otimes_{i=1}^{n}G_{i}$ for a better approximation of $\widetilde{G}$ will also be an interesting future direction.


\bibliographystyle{IEEEtran}
 \bibliography{references}

\begin{thebibliography}{10}
\providecommand{\url}[1]{#1}
\csname url@samestyle\endcsname
\providecommand{\newblock}{\relax}
\providecommand{\bibinfo}[2]{#2}
\providecommand{\BIBentrySTDinterwordspacing}{\spaceskip=0pt\relax}
\providecommand{\BIBentryALTinterwordstretchfactor}{4}
\providecommand{\BIBentryALTinterwordspacing}{\spaceskip=\fontdimen2\font plus
\BIBentryALTinterwordstretchfactor\fontdimen3\font minus
  \fontdimen4\font\relax}
\providecommand{\BIBforeignlanguage}[2]{{%
\expandafter\ifx\csname l@#1\endcsname\relax
\typeout{** WARNING: IEEEtran.bst: No hyphenation pattern has been}%
\typeout{** loaded for the language `#1'. Using the pattern for}%
\typeout{** the default language instead.}%
\else
\language=\csname l@#1\endcsname
\fi
#2}}
\providecommand{\BIBdecl}{\relax}
\BIBdecl

\bibitem{ebert2018visual}
F.~Ebert, C.~Finn, S.~Dasari, A.~Xie, A.~X. Lee, and S.~Levine, ``Visual
  foresight: Model-based deep reinforcement learning for vision-based robotic
  control,'' \emph{CoRR}, vol. abs/1812.00568, 2018.

\bibitem{lillicrap2015continuous}
T.~P. Lillicrap, J.~J. Hunt, A.~Pritzel, N.~Heess, T.~Erez, Y.~Tassa,
  D.~Silver, and D.~Wierstra, ``Continuous control with deep reinforcement
  learning,'' in \emph{Proceedings of the 4th International Conference on
  Learning Representations, {ICLR} 2016}, 2016.

\bibitem{brown2019superhuman}
N.~Brown and T.~Sandholm, ``Superhuman ai for multiplayer poker,''
  \emph{Science}, vol. 365, pp. 885--890, 2019.

\bibitem{DBLP:journals/nature/SilverHMGSDSAPL16}
D.~Silver, A.~Huang, C.~J. Maddison, A.~Guez, L.~Sifre, G.~van~den Driessche,
  J.~Schrittwieser, I.~Antonoglou, V.~Panneershelvam, M.~Lanctot, S.~Dieleman,
  D.~Grewe, J.~Nham, N.~Kalchbrenner, I.~Sutskever, T.~P. Lillicrap, M.~Leach,
  K.~Kavukcuoglu, T.~Graepel, and D.~Hassabis, ``Mastering the game of go with
  deep neural networks and tree search,'' \emph{Nature}, vol. 529, pp.
  484--489, 2016.

\bibitem{DBLP:journals/corr/HosuR16}
I.~Hosu and T.~Rebedea, ``Playing atari games with deep reinforcement learning
  and human checkpoint replay,'' \emph{CoRR}, vol. abs/1607.05077, 2016.

\bibitem{DBLP:journals/ai/SuttonPS99}
R.~S. Sutton, D.~Precup, and S.~P. Singh, ``Between mdps and semi-mdps: {A}
  framework for temporal abstraction in reinforcement learning,''
  \emph{Artificial Intelligence}, vol. 112, pp. 181--211, 1999.

\bibitem{DBLP:conf/icml/JinnaiAHLK19}
Y.~Jinnai, D.~Abel, D.~E. Hershkowitz, M.~L. Littman, and G.~D. Konidaris,
  ``Finding options that minimize planning time,'' in \emph{Proceedings of the
  36th International Conference on Machine Learning, {ICML} 2019}, vol.~97,
  2019, pp. 3120--3129.

\bibitem{DBLP:conf/icml/JinnaiPAK19}
Y.~Jinnai, J.~W. Park, D.~Abel, and G.~D. Konidaris, ``Discovering options for
  exploration by minimizing cover time,'' in \emph{Proceedings of the 36th
  International Conference on Machine Learning, {ICML} 2019}, vol.~97, 2019,
  pp. 3130--3139.

\bibitem{fiedler1973algebraic}
M.~Fiedler, ``Algebraic connectivity of graphs,'' \emph{Czechoslovak
  Mathematical Journal}, vol.~23, pp. 298--305, 1973.

\bibitem{fast_graphs}
A.~Ghosh and S.~P. Boyd, ``Growing well-connected graphs,'' in \emph{45th
  {IEEE} Conference on Decision and Control, {CDC} 2006}.\hskip 1em plus 0.5em
  minus 0.4em\relax {IEEE}, 2006, pp. 6605--6611.

\bibitem{DBLP:conf/atal/AmatoKK14}
C.~Amato, G.~D. Konidaris, and L.~P. Kaelbling, ``Planning with macro-actions
  in decentralized pomdps,'' in \emph{International conference on Autonomous
  Agents and Multi-Agent Systems, {AAMAS} 2014}.\hskip 1em plus 0.5em minus
  0.4em\relax {IFAAMAS/ACM}, 2014, pp. 1273--1280.

\bibitem{amato2019modeling}
C.~Amato, G.~Konidaris, L.~P. Kaelbling, and J.~P. How, ``Modeling and planning
  with macro-actions in decentralized pomdps,'' \emph{Journal of Artificial
  Intelligence Research}, vol.~64, pp. 817--859, 2019.

\bibitem{shen2006multi}
J.~Shen, G.~Gu, and H.~Liu, ``Multi-agent hierarchical reinforcement learning
  by integrating options into maxq,'' in \emph{Proceedings of the 1st
  International Multi-Symposiums on Computer and Computational Sciences,
  {IMSCCS} 2006}, vol.~1, 2006, pp. 676--682.

\bibitem{DBLP:conf/atal/ChakravortyWRCB20}
J.~Chakravorty, P.~N. Ward, J.~Roy, M.~Chevalier{-}Boisvert, S.~Basu, A.~Lupu,
  and D.~Precup, ``Option-critic in cooperative multi-agent systems,'' in
  \emph{Proceedings of the 19th International Conference on Autonomous Agents
  and Multiagent Systems, {AAMAS} 2020}.\hskip 1em plus 0.5em minus 0.4em\relax
  {IFAAMAS/ACM}, 2020, pp. 1792--1794.

\bibitem{DBLP:conf/iclr/LeeYL20}
Y.~Lee, J.~Yang, and J.~J. Lim, ``Learning to coordinate manipulation skills
  via skill behavior diversification,'' in \emph{8th International Conference
  on Learning Representations, {ICLR} 2020}, 2020.

\bibitem{DBLP:books/wi/Puterman94}
M.~L. Puterman, \emph{Markov Decision Processes: Discrete Stochastic Dynamic
  Programming}.\hskip 1em plus 0.5em minus 0.4em\relax Wiley, 1994.

\bibitem{DBLP:conf/aistats/FruitL17}
R.~Fruit and A.~Lazaric, ``Exploration-exploitation in mdps with options,'' in
  \emph{Proceedings of the 20th International Conference on Artificial
  Intelligence and Statistics, {AISTATS} 2017}, vol.~54, 2017, pp. 576--584.

\bibitem{DBLP:conf/icml/VezhnevetsOSHJS17}
A.~S. Vezhnevets, S.~Osindero, T.~Schaul, N.~Heess, M.~Jaderberg, D.~Silver,
  and K.~Kavukcuoglu, ``Feudal networks for hierarchical reinforcement
  learning,'' in \emph{Proceedings of the 34th International Conference on
  Machine Learning, {ICML} 2017}, ser. Proceedings of Machine Learning
  Research, vol.~70, 2017, pp. 3540--3549.

\bibitem{DBLP:journals/csur/PateriaSTQ21}
S.~Pateria, B.~Subagdja, A.~Tan, and C.~Quek, ``Hierarchical reinforcement
  learning: {A} comprehensive survey,'' \emph{{ACM} Computing Surveys},
  vol.~54, pp. 109:1--109:35, 2021.

\bibitem{DBLP:conf/icml/McGovernB01}
A.~McGovern and A.~G. Barto, ``Automatic discovery of subgoals in reinforcement
  learning using diverse density,'' in \emph{Proceedings of the 8th
  International Conference on Machine Learning {ICML} 2001}, 2001, pp.
  361--368.

\bibitem{Menache02q-cut-}
I.~Menache, S.~Mannor, and N.~Shimkin, ``Q-cut - dynamic discovery of sub-goals
  in reinforcement learning,'' in \emph{Proceedings of the 13th European
  Conference on Machine Learning, {ECML} 2002}.\hskip 1em plus 0.5em minus
  0.4em\relax {Springer}, 2002, pp. 295--306.

\bibitem{DBLP:conf/nips/MankowitzMM16}
D.~J. Mankowitz, T.~A. Mann, and S.~Mannor, ``Adaptive skills adaptive
  partitions {(ASAP)},'' in \emph{Advances of the 30th Conference on Neural
  Information Processing Systems, {NIPS} 2016}, vol.~29, 2016, pp. 1588--1596.

\bibitem{Harb2018WhenWI}
J.~Harb, P.~Bacon, M.~Klissarov, and D.~Precup, ``When waiting is not an
  option: Learning options with a deliberation cost,'' in \emph{Proceedings of
  the 32nd {AAAI} Conference on Artificial Intelligence, {AAAI} 2018}.\hskip
  1em plus 0.5em minus 0.4em\relax {AAAI} Press, 2018, pp. 3165--3172.

\bibitem{stolle2002learning}
M.~Stolle and D.~Precup, ``Learning options in reinforcement learning,'' in
  \emph{Proceedings of the 5th International Symposium on Abstraction,
  Reformulation and Approximation, {SARA} 2002}, vol. 2371.\hskip 1em plus
  0.5em minus 0.4em\relax Springer, 2002, pp. 212--223.

\bibitem{DBLP:conf/icml/SimsekWB05}
{\"{O}}.~Simsek, A.~P. Wolfe, and A.~G. Barto, ``Identifying useful subgoals in
  reinforcement learning by local graph partitioning,'' in \emph{Proceedings of
  the 22nd International Conference on Machine Learning, {ICML} 2005}, vol.
  119, 2005, pp. 816--823.

\bibitem{DBLP:conf/nips/SimsekB08}
{\"{O}}.~Simsek and A.~G. Barto, ``Skill characterization based on
  betweenness,'' in \emph{Proceedings of the 22nd Conference on Neural
  Information Processing Systems, {NIPS} 2008}, vol.~21, 2008, pp. 1497--1504.

\bibitem{DBLP:journals/corr/MachadoBB17}
M.~C. Machado, M.~G. Bellemare, and M.~H. Bowling, ``A laplacian framework for
  option discovery in reinforcement learning,'' in \emph{Proceedings of the
  34th International Conference on Machine Learning, {ICML} 2017}, vol.~70,
  2017, pp. 2295--2304.

\bibitem{DBLP:conf/iclr/MachadoRGLTC18}
M.~C. Machado, C.~Rosenbaum, X.~Guo, M.~Liu, G.~Tesauro, and M.~Campbell,
  ``Eigenoption discovery through the deep successor representation,'' in
  \emph{Proceedings of the 6th International Conference on Learning
  Representations, {ICLR} 2018}, 2018.

\bibitem{DBLP:conf/atal/YangBZ20}
J.~Yang, I.~Borovikov, and H.~Zha, ``Hierarchical cooperative multi-agent
  reinforcement learning with skill discovery,'' in \emph{Proceedings of the
  19th International Conference on Autonomous Agents and Multiagent Systems,
  {AAMAS} 2020}.\hskip 1em plus 0.5em minus 0.4em\relax {IFAAMAS/ACM}, 2020,
  pp. 1566--1574.

\bibitem{DBLP:conf/iclr/EysenbachGIL19}
B.~Eysenbach, A.~Gupta, J.~Ibarz, and S.~Levine, ``Diversity is all you need:
  Learning skills without a reward function,'' in \emph{Proceedings of the 7th
  International Conference on Learning Representations, {ICLR} 2019}, 2019.

\bibitem{weichsel1962kronecker}
P.~M. Weichsel, ``The kronecker product of graphs,'' in \emph{Proceedings of
  the 13th American Mathematical Society, {AMS} 1962}, vol.~13.\hskip 1em plus
  0.5em minus 0.4em\relax JSTOR, 1962, pp. 47--52.

\bibitem{basic2021estimation}
M.~Basic, B.~Arsic, and Z.~Obradovic, ``Another estimation of laplacian
  spectrum of the kronecker product of graphs,'' \emph{CoRR}, vol.
  abs/2102.02924, 2021.

\bibitem{DBLP:conf/iclr/WuTN19}
Y.~Wu, G.~Tucker, and O.~Nachum, ``The laplacian in {RL:} learning
  representations with efficient approximations,'' in \emph{Proceedings of the
  7th International Conference on Learning Representations, {ICLR} 2019}, 2019.

\bibitem{DBLP:conf/icml/WangZZSHF21}
K.~Wang, K.~Zhou, Q.~Zhang, J.~Shao, B.~Hooi, and J.~Feng, ``Towards better
  laplacian representation in reinforcement learning with generalized graph
  drawing,'' in \emph{Proceedings of the 38th International Conference on
  Machine Learning, {ICML} 2021}, vol. 139, 2021, pp. 11\,003--11\,012.

\bibitem{DBLP:conf/icml/Tan93}
M.~Tan, ``Multi-agent reinforcement learning: Independent versus cooperative
  agents,'' in \emph{Proceedings of the 10th International Conference on
  Machine Learning, {ICML} 1993}, 1993, pp. 330--337.

\bibitem{watkins1992q}
C.~J. Watkins and P.~Dayan, ``Q-learning,'' \emph{Machine learning}, vol.~8,
  pp. 279--292, 1992.

\bibitem{DBLP:conf/icml/LauerR00}
M.~Lauer and M.~A. Riedmiller, ``An algorithm for distributed reinforcement
  learning in cooperative multi-agent systems,'' in \emph{Proceedings of the
  17th International Conference on Machine Learning, {ICML} 2000}, 2000, pp.
  535--542.

\bibitem{pearson1905problem}
K.~Pearson, ``The problem of the random walk,'' \emph{Nature}, vol.~72, pp.
  342--342, 1905.

\bibitem{sutton2018reinforcement}
R.~S. Sutton and A.~G. Barto, \emph{Reinforcement learning - an introduction},
  ser. Adaptive computation and machine learning.\hskip 1em plus 0.5em minus
  0.4em\relax {MIT} Press, 1998.

\bibitem{DBLP:conf/aaai/FoersterFANW18}
J.~N. Foerster, G.~Farquhar, T.~Afouras, N.~Nardelli, and S.~Whiteson,
  ``Counterfactual multi-agent policy gradients,'' in \emph{Proceedings of the
  32nd {AAAI} Conference on Artificial Intelligence, {AAAI} 2018}.\hskip 1em
  plus 0.5em minus 0.4em\relax {AAAI} Press, 2018, pp. 2974--2982.

\bibitem{DBLP:conf/nips/RashidFPW20}
T.~Rashid, G.~Farquhar, B.~Peng, and S.~Whiteson, ``Weighted {QMIX:} expanding
  monotonic value function factorisation for deep multi-agent reinforcement
  learning,'' in \emph{Proceedings of the 34th Conference on Neural Information
  Processing Systems, {NIPS} 2020}, vol.~33, 2020.

\bibitem{DBLP:conf/nips/MahajanRSW19}
A.~Mahajan, T.~Rashid, M.~Samvelyan, and S.~Whiteson, ``{MAVEN:} multi-agent
  variational exploration,'' in \emph{Proceedings of the 33th Conference on
  Neural Information Processing Systems, {NIPS} 2019}, vol.~32, 2019, pp.
  7611--7622.

\bibitem{github}
J.~Chen, J.~Chen, T.~Lan, and V.~Aggarwal, ``Multi-agent option discovery based
  on kronecker product,''
  \url{https://github.itap.purdue.edu/Clan-labs/MAOD_via_KP}, 2022.

\bibitem{DBLP:journals/dam/Sayama16}
H.~Sayama, ``Estimation of laplacian spectra of direct and strong product
  graphs,'' \emph{Discrete Applied Mathematics}, vol. 205, pp. 160--170, 2016.

\bibitem{west2001introduction}
D.~B. West \emph{et~al.}, \emph{Introduction to graph theory}.\hskip 1em plus
  0.5em minus 0.4em\relax Prentice hall Upper Saddle River, 2001, vol.~2.

\end{thebibliography}

\clearpage
\setcounter{page}{1}
\onecolumn


\onecolumn
\appendices

\section{Proof of THEOREM \ref{thm:2}} \label{thmproof}

For convenience, we use $G_i$ to represent the $i$-th factor graph and its adjacency matrix. Also, we denote the number of nodes in $G_i$ as $K_i$ and an identity matrix with $K_i$ diagonal elements as $I_{K_i}$. 
\begin{proof}
The normalized laplacian matrix of the Kronecker product of $n$ factor graphs $\otimes_{i=1}^{n}G_{i}$ can be written as:
\begin{eqnarray}
\mathcal{L}_{\otimes_{i=1}^{n}G_{i}} =& \otimes_{i=1}^{n}I_{K_i}-(\otimes_{i=1}^{n}D_{G_i}^{-\frac{1}{2}}) (\otimes_{i=1}^{n}G_{i}) (\otimes_{i=1}^{n} D_{G_i}^{-\frac{1}{2}}).
\end{eqnarray}
Using the property of the Kronecker product of matrices, $(A\otimes B)(C\otimes D)=AC\otimes BD$, we can obtain that:
\begin{equation}
    \begin{aligned}
    \mathcal{L}_{\otimes_{i=1}^{n}G_{i}} &=\otimes_{i=1}^{n}I_{K_i}-\otimes_{i=1}^{n}(D_{G_i}^{-\frac{1}{2}} G_i D_{G_i}^{-\frac{1}{2}})\\
    &=\otimes_{i=1}^{n}I_{K_i}-\otimes_{i=1}^{n}(I_{K_i}-\mathcal{L}_{G_i}).
    \end{aligned}
\end{equation}

Let $\{\lambda_{k_1}^{G_1}\}, \{\lambda_{k_2}^{G_2}\}, \ldots,  \{\lambda_{k_n}^{G_n}\}$ be the eigenvalues of matrices $\mathcal{L}_{G_1}, \mathcal{L}_{G_2}$ $,\ldots,\mathcal{L}_{G_n}$, with the corresponding orthonormal eigenvectors $\{v_{k_1}^{G_1}\}, \{v_{k_2}^{G_2}\},$ $\ldots,\{v_{k_n}^{G_n}\}$, where $k_i=1,2,\ldots,K_i$. Also, denote the diagonal matrices, whose diagonal elements are the values $\{1-\lambda_{k_1}^{G_1}\}, \{1-\lambda_{k_2}^{G_2}\},\ldots,\{1-\lambda_{k_n}^{G_n}\}$, as $\Lambda_{G_1},\Lambda_{G_2},\ldots,\Lambda_{G_n}$, and the square matrices containing the eigenvectors $\{v_{k_1}^{G_1}\}, \{v_{k_2}^{G_2}\},\ldots, \{v_{k_n}^{G_n}\}$ as the column vectors as $V_{G_1}, V_{G_2},\ldots,V_{G_n}$. Using the spectral decomposition of the matrix $I_{K_i}-\mathcal{L}_{G_i}\ (i=1,\ldots,n)$, we can obtain that:

\begin{equation}
    \begin{aligned}
    \mathcal{L}_{\otimes_{i=1}^{n}G_{i}}&=\otimes_{i=1}^{n}I_{K_i}-\otimes_{i=1}^{n}(V_{G_i}\Lambda_{G_i}V_{G_i}^T)\\
    &=\otimes_{i=1}^{n}I_{K_i}-(\otimes_{i=1}^{n}V_{G_i})(\otimes_{i=1}^{n}\Lambda_{G_i})(\otimes_{i=1}^{n}V_{G_i})^T\\
    &=(\otimes_{i=1}^{n}V_{G_i})(\otimes_{i=1}^{n}I_{K_i}-\otimes_{i=1}^{n}\Lambda_{G_i})(\otimes_{i=1}^{n}V_{G_i})^T,
    \end{aligned}
\end{equation}
since $\otimes_{i=1}^{n}I_{K_i} = \otimes_{i=1}^{n}[(V_{G_i})(V_{G_i})^T]= (\otimes_{i=1}^{n}V_{G_i})(\otimes_{i=1}^{n}V_{G_i})^T$. This implies that $\mathcal{L}_{\otimes_{i=1}^{n}G_{i}}$ has eigenvalues $\{[1-\prod_{i=1}^{n}(1-\lambda_{k_i}^{G_i})]\} $ and corresponding eigenvectors $\{\otimes_{i=1}^{n}v_{k_i}^{G_i}\}$.

Then, we let $\Lambda=\otimes_{i=1}^{n}I_{K_i}-\otimes_{i=1}^{n}\Lambda_{G_i}$ and $D=\otimes_{i=1}^{n}D_{G_i}$. Since the normalized Laplacian could be expressed in terms of Laplacian matrix as $\mathcal{L}=D^{-\frac{1}{2}}LD^{-\frac{1}{2}}$, we can get $L_{\otimes_{i=1}^{n}G_{i}}(\otimes_{i=1}^{n}V_{G_i})=D^{\frac{1}{2}}\mathcal{L}_{\otimes_{i=1}^{n}G_{i}}D^{\frac{1}{2}}(\otimes_{i=1}^{n}V_{G_i})$. By making assumption (used and testified in \cite{basic2021estimation, DBLP:journals/dam/Sayama16}) that $D^{\frac{1}{2}}_{G_i}V_{G_i}\approx V_{G_i}D^{\frac{1}{2}}_{G_i}$, for $i=1,2,\ldots,n$, we can derive that:

\begin{equation}
    \begin{aligned}
    L_{\otimes_{i=1}^{n}G_{i}}(\otimes_{i=1}^{n}V_{G_i})& \approx D^{\frac{1}{2}}\mathcal{L}_{\otimes_{i=1}^{n}G_{i}}(\otimes_{i=1}^{n}V_{G_i})D^{\frac{1}{2}}\\
    &=D^{\frac{1}{2}}\Lambda (\otimes_{i=1}^{n}V_{G_i}) D^{\frac{1}{2}}.
    \end{aligned}
\end{equation}
After applying the same assumption again, we finally obtain that:
\begin{equation} \label{equ:final}
    \begin{aligned}
    L_{\otimes_{i=1}^{n}G_{i}}(\otimes_{i=1}^{n}V_{G_i})&\approx (D \Lambda)(\otimes_{i=1}^{n}V_{G_i}). 
    \end{aligned}
\end{equation}

Based on Equation (\ref{equ:final}), we can get an approximation of the Laplacian spectrum, including the eigenvalues and corresponding eigenvectors, of the Kronecker product of $n$ factor graphs, shown as THEOREM \ref{thm:2}.

Next, we will prove that the estimated eigenvalues $\mu_{k_1k_2,\ldots,k_n}$ in THEOREM \ref{thm:2} are non-negative. It is obvious that $d_{k_i}^{G_i}$ and $\prod_{i=1}^{n}d_{k_i}^{G_i}$ are non-negative. Then, we need to prove $[1-\prod_{i=1}^{n}(1-\lambda_{k_i}^{G_i})]$ is non-negative. We know that if $\lambda$ is an eigenvalue of a normalized Laplacian matrix, we can get $0 \leq \lambda \leq 2 $. Hence, $-1 \le 1-\lambda_{k_i}^{G_i} \le 1$, for $i=1,2,\ldots,n$. Based on this, we can get that $\left|\prod_{i=1}^{n}(1-\lambda_{k_i}^{G_i})\right| \leq 1$ and thus $[1-\prod_{i=1}^{n}(1-\lambda_{k_i}^{G_i})]$ is non-negative. 
\end{proof}
\clearpage

\section{Finding the Fiedler vector for the illustrative example shown in Figure \ref{fig:-1(a)}} \label{example2}

\noindent(1) Compute the normalized Laplacian matrix of $G_1$ and $G_2$, namely $\mathcal{L}_{1}$ and $\mathcal{L}_{2}$:
\begin{eqnarray}
\mathcal{L}_{1} = \left[ \begin{array}{cc} 1 &-1 \\ -1 & 1 \end{array} \right] \ \ {\rm ,} \ \ \mathcal{L}_{2} =\left[  \begin{array}{cccc} 1 & -\frac{1}{\sqrt{2}} & 0 & 0 \\ -\frac{1}{\sqrt{2}} & 1 & -\frac{1}{2} & 0 \\ 0 & -\frac{1}{2} & 1 & -\frac{1}{\sqrt{2}} \\ 0 & 0 & -\frac{1}{\sqrt{2}} & 1 \end{array}  \right].
\end{eqnarray}
(2) Compute the eigenvalues and eigenvectors of $\mathcal{L}_{1}$ and $\mathcal{L}_{2}$:
\begin{eqnarray}
\lambda_1^{G_1} = 0,  \ \  \lambda_2^{G_1} = 2, \ \ v_{1:2}^{G_1} = \frac{1}{\sqrt{2}} \left[\left[\begin{array}{c} 1 \\ 1 \end{array} \right], \left[\begin{array}{c} -1 \\ 1 \end{array} \right]\right].  
\end{eqnarray}

\begin{eqnarray}
 \lambda_1^{G_2} = 0, \ \ \lambda_2^{G_2} = 0.5, \ \  \lambda_3^{G_2}  = 1.5, \ \  \lambda_4^{G_2}  = 2, \ \  v_{1:4}^{G_2}=  \frac{1}{\sqrt{3}}\left[\left[\begin{array}{c} \frac{1}{\sqrt{2}} \\ 1 \\ 1 \\ \frac{1}{\sqrt{2}} \end{array} \right], \left[  \begin{array}{c} -1 \\  -\frac{1}{\sqrt{2}} \\ \frac{1}{\sqrt{2}} \\ 1 \end{array} \right], \left[\begin{array}{c} 1 \\ -\frac{1}{\sqrt{2}} \\ -\frac{1}{\sqrt{2}} \\ 1 \end{array} \right], \left[  \begin{array}{c} \frac{1}{\sqrt{2}} \\  -1 \\ 1\\ -\frac{1}{\sqrt{2}}  \end{array} \right] \right].
\end{eqnarray}

\noindent(3) Compute the degree list of $G_1$ and $G_2$ (sorted in ascending order), namely $d^{G_{1}}$ and $d^{G_{2}}$:

\begin{eqnarray}
d^{G_{1}} = [1,\ 1]^{T},\ d^{G_{2}} = [1,\ 1,\ 2,\ 2]^{T}.
\end{eqnarray}
(4) According to THEOREM \ref{thm:2}, we can get two approximations of the Fiedler vector:
\begin{eqnarray}
v_{11}=v_{1}^{G_{1}} \otimes v_{2}^{G_{2}}=\frac{1}{\sqrt{6}}\left[\frac{1}{\sqrt{2}},\ 1,\ 1,\ \frac{1}{\sqrt{2}},\ \frac{1}{\sqrt{2}},\ 1,\ 1,\ \frac{1}{\sqrt{2}}\right]^{T},
\\
v_{24}=v_{2}^{G_{1}} \otimes v_{4}^{G_{2}}=\frac{1}{\sqrt{6}}\left[-\frac{1}{\sqrt{2}},\ 1,\ -1,\ \frac{1}{\sqrt{2}},\ \frac{1}{\sqrt{2}},\ -1,\ 1,\ -\frac{1}{\sqrt{2}}\right]^{T}.
\end{eqnarray}


\section{Numeric results of the experiments} \label{numeric results}

In this section, we show the numeric results of our experiments in tabular with the following notations.  IQL: Independent Q-Learning, DistQ: Distributed Q-Learning, CentQ: Centralized Q-Learning, CentQ+Force: Centralized Q-Learning + Force, Value: the mean of the episodic cumulative reward during the training process, Step: the mean of the number of steps to complete the task during the training process, Multiple: multi-agent options, Single: single-agent options.

\begin{table}[htbp]
\centering
\begin{tabular}{ |p{2cm}|p{1.0cm}|p{1.0cm}||p{2cm}|p{1.0cm}|p{1.0cm}||p{2cm}|p{1.0cm}|p{1.0cm}|}
\hline
\multicolumn{3}{|c||}{Three-agent four-room task} & \multicolumn{3}{c||}{Four-agent four-room task} & \multicolumn{3}{c|}{Five-agent four-room task}\\
\hline \hline
 \cellcolor{gray} \makecell[c]{IQL} & \makecell[c]{Value} & \makecell[c]{Step} & \cellcolor{gray} \makecell[c]{IQL} & \makecell[c]{Value} & \makecell[c]{Step} & \cellcolor{gray} \makecell[c]{IQL} & \makecell[c]{Value} & \makecell[c]{Step}\\
\hline
\makecell[c]{Multiple} & \makecell[c]{\textbf{0.790}} & \makecell[c]{\textbf{27.96}} &  \makecell[c]{Multiple} & \makecell[c]{\textbf{0.716}} & \makecell[c]{\textbf{42.95}} & \makecell[c]{Multiple} & \makecell[c]{\textbf{0.674}} & \makecell[c]{\textbf{50.83}}\\
\hline
\makecell[c]{Single} & \makecell[c]{0.705} & \makecell[c]{43.94} & \makecell[c]{Single} & \makecell[c]{0.386} & \makecell[c]{115.1} & \makecell[c]{Single} & \makecell[c]{0.002} & \makecell[c]{199.6}\\
\hline
\makecell[c]{No options} & \makecell[c]{0.038} & \makecell[c]{191.7} & \makecell[c]{No options} & \makecell[c]{0.0} & \makecell[c]{200.0} & \makecell[c]{No options} & \makecell[c]{0.0} & \makecell[c]{200.0}\\
\hline \hline
\cellcolor{gray} \makecell[c]{CentQ+Force} & \makecell[c]{Value} & \makecell[c]{Step} & \cellcolor{gray} \makecell[c]{CentQ+Force} & \makecell[c]{Value} & \makecell[c]{Step} & \cellcolor{gray} \makecell[c]{CentQ+Force} & \makecell[c]{Value} & \makecell[c]{Step}\\
\hline
\makecell[c]{Multiple} & \makecell[c]{\textbf{0.852}} & \makecell[c]{\textbf{16.73}} &  \makecell[c]{Multiple} & \makecell[c]{\textbf{0.829}} & \makecell[c]{\textbf{20.25}} & \makecell[c]{Multiple} & \makecell[c]{\textbf{0.769}} & \makecell[c]{\textbf{32.47}}\\
\hline
\makecell[c]{Single} & \makecell[c]{0.0} & \makecell[c]{200.0} & \makecell[c]{Single} & \makecell[c]{--} & \makecell[c]{--} & \makecell[c]{Single} & \makecell[c]{--} & \makecell[c]{--}\\
\hline
\makecell[c]{No options} & \makecell[c]{0.0} & \makecell[c]{200.0} & \makecell[c]{No options} & \makecell[c]{0.0} & \makecell[c]{200.0} & \makecell[c]{No options} & \makecell[c]{0.0} & \makecell[c]{200.0}\\
\hline
\end{tabular}
\caption{Numeric results on the $n$-agent four-room tasks}
\end{table}

\begin{table}[htbp]
\centering
\begin{tabular}{ |p{2cm}|p{1.0cm}|p{1.0cm}||p{2cm}|p{1.0cm}|p{1.0cm}|}
\hline
\multicolumn{3}{|c||}{$2\times2$ four-room task} & \multicolumn{3}{c|}{$3\times2$ four-room task} \\
\hline \hline
 \cellcolor{gray} \makecell[c]{IQL} & \makecell[c]{Value} & \makecell[c]{Step} & \cellcolor{gray} \makecell[c]{IQL} & \makecell[c]{Value} & \makecell[c]{Step}\\
\hline
\makecell[c]{Multiple} & \makecell[c]{\textbf{0.782}} & \makecell[c]{\textbf{28.66}} &  \makecell[c]{Multiple} & \makecell[c]{\textbf{0.504}} & \makecell[c]{\textbf{89.34}}\\
\hline
\makecell[c]{Single} & \makecell[c]{0.398} & \makecell[c]{111.2} & \makecell[c]{Single} & \makecell[c]{0.0} & \makecell[c]{200.0}\\
\hline
\makecell[c]{No options} & \makecell[c]{0.0} & \makecell[c]{200.0} & \makecell[c]{No options} & \makecell[c]{0.0} & \makecell[c]{200.0}\\
\hline \hline
\cellcolor{gray} \makecell[c]{CentQ+Force} & \makecell[c]{Value} & \makecell[c]{Step} & \cellcolor{gray} \makecell[c]{CentQ+Force} & \makecell[c]{Value} & \makecell[c]{Step} \\
\hline
\makecell[c]{Multiple} & \makecell[c]{\textbf{0.853}} & \makecell[c]{\textbf{16.17}} &  \makecell[c]{Multiple} & \makecell[c]{\textbf{0.809}} & \makecell[c]{\textbf{23.25}}\\
\hline
\makecell[c]{Single} & \makecell[c]{0.614} & \makecell[c]{67.33} & \makecell[c]{Single} & \makecell[c]{0.0} & \makecell[c]{200.0}\\
\hline
\makecell[c]{No options} & \makecell[c]{0.0} & \makecell[c]{200.0} & \makecell[c]{No options} & \makecell[c]{0.0} & \makecell[c]{200.0}\\
\hline
\end{tabular}
\caption{Numeric results on the $m \times n$ four-room tasks}
\end{table}

\begin{table}[htbp]
\centering
\begin{tabular}{ |p{2cm}|p{1.0cm}|p{1.0cm}||p{2cm}|p{1.0cm}|p{1.0cm}|}
\hline
\multicolumn{3}{|c||}{Two-agent two-room task} & \multicolumn{3}{c|}{Two-agent four-room task} \\
\hline \hline
 \cellcolor{gray} \makecell[c]{Random} & \makecell[c]{Value} & \makecell[c]{Step} & \cellcolor{gray} \makecell[c]{Random} & \makecell[c]{Value} & \makecell[c]{Step}\\
\hline
\makecell[c]{Multiple} & \makecell[c]{\textbf{0.609}} & \makecell[c]{\textbf{57.58}} &  \makecell[c]{Multiple} & \makecell[c]{\textbf{0.496}} & \makecell[c]{\textbf{83.98}}\\
\hline
\makecell[c]{Single} & \makecell[c]{0.344} & \makecell[c]{119.1} & \makecell[c]{Single} & \makecell[c]{0.045} & \makecell[c]{189.5}\\
\hline
\makecell[c]{No options} & \makecell[c]{0.004} & \makecell[c]{199.3} & \makecell[c]{No options} & \makecell[c]{0.010} & \makecell[c]{197.9}\\
\hline \hline
\cellcolor{gray} \makecell[c]{IQL} & \makecell[c]{Value} & \makecell[c]{Step} & \cellcolor{gray} \makecell[c]{IQL} & \makecell[c]{Value} & \makecell[c]{Step} \\
\hline
\makecell[c]{Multiple} & \makecell[c]{\textbf{0.828}} & \makecell[c]{\textbf{18.90}} &  \makecell[c]{Multiple} & \makecell[c]{\textbf{0.842}} & \makecell[c]{\textbf{17.66}}\\
\hline
\makecell[c]{Single} & \makecell[c]{0.783} & \makecell[c]{25.51} & \makecell[c]{Single} & \makecell[c]{0.685} & \makecell[c]{44.48}\\
\hline
\makecell[c]{No options} & \makecell[c]{0.693} & \makecell[c]{49.18} & \makecell[c]{No options} & \makecell[c]{0.827} & \makecell[c]{23.46}\\
\hline \hline
 \cellcolor{gray} \makecell[c]{DistQ} & \makecell[c]{Value} & \makecell[c]{Step} & \cellcolor{gray} \makecell[c]{DistQ} & \makecell[c]{Value} & \makecell[c]{Step}\\
\hline
\makecell[c]{Multiple} & \makecell[c]{\textbf{0.844}} & \makecell[c]{\textbf{16.89}} &  \makecell[c]{Multiple} & \makecell[c]{\textbf{0.883}} & \makecell[c]{\textbf{12.58}}\\
\hline
\makecell[c]{Single} & \makecell[c]{0.832} & \makecell[c]{19.57} & \makecell[c]{Single} & \makecell[c]{0.686} & \makecell[c]{52.92}\\
\hline
\makecell[c]{No options} & \makecell[c]{0.0} & \makecell[c]{200.0} & \makecell[c]{No options} & \makecell[c]{0.174} & \makecell[c]{162.7}\\
\hline \hline
 \cellcolor{gray} \makecell[c]{CentQ} & \makecell[c]{Value} & \makecell[c]{Step} & \cellcolor{gray} \makecell[c]{CentQ} & \makecell[c]{Value} & \makecell[c]{Step}\\
\hline
\makecell[c]{Multiple} & \makecell[c]{\textbf{0.800}} & \makecell[c]{\textbf{25.29}} &  \makecell[c]{Multiple} & \makecell[c]{\textbf{0.627}} & \makecell[c]{\textbf{66.05}}\\
\hline
\makecell[c]{Single} & \makecell[c]{0.589} & \makecell[c]{70.81} & \makecell[c]{Single} & \makecell[c]{0.402} & \makecell[c]{113.7}\\
\hline
\makecell[c]{No options} & \makecell[c]{0.0} & \makecell[c]{200.0} & \makecell[c]{No options} & \makecell[c]{0.0} & \makecell[c]{200.0}\\
\hline \hline
 \cellcolor{gray} \makecell[c]{CentQ+Force} & \makecell[c]{Value} & \makecell[c]{Step} & \cellcolor{gray} \makecell[c]{CentQ+Force} & \makecell[c]{Value} & \makecell[c]{Step}\\
\hline
\makecell[c]{Multiple} & \makecell[c]{\textbf{0.811}} & \makecell[c]{\textbf{21.06}} &  \makecell[c]{Multiple} & \makecell[c]{\textbf{0.867}} & \makecell[c]{\textbf{14.37}}\\
\hline
\makecell[c]{Single} & \makecell[c]{0.757} & \makecell[c]{32.80} & \makecell[c]{Single} & \makecell[c]{0.743} & \makecell[c]{36.81}\\
\hline
\makecell[c]{No options} & \makecell[c]{0.0} & \makecell[c]{200.0} & \makecell[c]{No options} & \makecell[c]{0.0} & \makecell[c]{200.0}\\
\hline
\end{tabular}
\caption{Numeric results on the two-room task and two-agent four-room task}
\end{table}

\begin{table}[htbp]
\centering
\begin{tabular}{ |p{2cm}|p{1.0cm}|p{1.0cm}||p{2cm}|p{1.0cm}|p{1.0cm}||p{2cm}|p{1.0cm}|p{1.0cm}|}
\hline
\multicolumn{3}{|c||}{Four-agent four-room task} & \multicolumn{3}{c||}{Six-agent four-room task} & \multicolumn{3}{c|}{Eight-agent four-room task}\\
\hline \hline
 \cellcolor{gray} \makecell[c]{IQL} & \makecell[c]{Value} & \makecell[c]{Step} & \cellcolor{gray} \makecell[c]{IQL} & \makecell[c]{Value} & \makecell[c]{Step} & \cellcolor{gray} \makecell[c]{IQL} & \makecell[c]{Value} & \makecell[c]{Step}\\
\hline
\makecell[c]{Multiple} & \makecell[c]{\textbf{0.774}} & \makecell[c]{\textbf{30.82}} &  \makecell[c]{Multiple} & \makecell[c]{\textbf{0.329}} & \makecell[c]{\textbf{126.0}} & \makecell[c]{Multiple} & \makecell[c]{0.0} & \makecell[c]{200.0}\\
\hline
\makecell[c]{Single} & \makecell[c]{0.427} & \makecell[c]{105.5} & \makecell[c]{Single} & \makecell[c]{0.0} & \makecell[c]{200.0} & \makecell[c]{Single} & \makecell[c]{0.0} & \makecell[c]{200.0}\\
\hline
\makecell[c]{No options} & \makecell[c]{0.095} & \makecell[c]{179.0} & \makecell[c]{No options} & \makecell[c]{0.0} & \makecell[c]{200.0} & \makecell[c]{No options} & \makecell[c]{0.0} & \makecell[c]{200.0}\\
\hline \hline
\cellcolor{gray} \makecell[c]{CentQ+Force} & \makecell[c]{Value} & \makecell[c]{Step} & \cellcolor{gray} \makecell[c]{CentQ+Force} & \makecell[c]{Value} & \makecell[c]{Step} & \cellcolor{gray} \makecell[c]{CentQ+Force} & \makecell[c]{Value} & \makecell[c]{Step}\\
\hline
\makecell[c]{Multiple} & \makecell[c]{\textbf{0.842}} & \makecell[c]{\textbf{17.96}} &  \makecell[c]{Multiple} & \makecell[c]{\textbf{0.753}} & \makecell[c]{\textbf{31.82}} & \makecell[c]{Multiple} & \makecell[c]{\textbf{0.605}} & \makecell[c]{\textbf{64.90}}\\
\hline
\makecell[c]{Single} & \makecell[c]{0.447} & \makecell[c]{102.3} & \makecell[c]{Single} & \makecell[c]{0.0} & \makecell[c]{200.0} & \makecell[c]{Single} & \makecell[c]{0.0} & \makecell[c]{200.0}\\
\hline
\makecell[c]{No options} & \makecell[c]{0.0} & \makecell[c]{200.0} & \makecell[c]{No options} & \makecell[c]{0.0} & \makecell[c]{200.0} & \makecell[c]{No options} & \makecell[c]{0.0} & \makecell[c]{200.0}\\
\hline
\end{tabular}
\caption{Numeric results on the $n$-agent four-room tasks with random grouping}
\end{table}

\begin{table}[htbp]
\centering
\begin{tabular}{ |p{2cm}|p{1.0cm}|p{1.0cm}||p{2cm}|p{1.0cm}|p{1.0cm}||p{2cm}|p{1.0cm}|p{1.0cm}|}
\hline
\multicolumn{3}{|c||}{Four-agent four-room task} & \multicolumn{3}{c||}{Six-agent four-room task} & \multicolumn{3}{c|}{Eight-agent four-room task}\\
\hline \hline
\cellcolor{gray} \makecell[c]{CentQ+Force} & \makecell[c]{Value} & \makecell[c]{Step} & \cellcolor{gray} \makecell[c]{CentQ+Force} & \makecell[c]{Value} & \makecell[c]{Step} & \cellcolor{gray} \makecell[c]{CentQ+Force} & \makecell[c]{Value} & \makecell[c]{Step}\\
\hline
\makecell[c]{Multiple} & \makecell[c]{\textbf{0.835}} & \makecell[c]{\textbf{18.90}} &  \makecell[c]{Multiple} & \makecell[c]{\textbf{0.784}} & \makecell[c]{\textbf{27.40}} & \makecell[c]{Multiple} & \makecell[c]{\textbf{0.717}} & \makecell[c]{\textbf{39.52}}\\
\hline
\makecell[c]{Single} & \makecell[c]{0.351} & \makecell[c]{122.3} & \makecell[c]{Single} & \makecell[c]{0.0} & \makecell[c]{200.0} & \makecell[c]{Single} & \makecell[c]{0.0} & \makecell[c]{200.0}\\
\hline
\makecell[c]{No options} & \makecell[c]{0.0} & \makecell[c]{200.0} & \makecell[c]{No options} & \makecell[c]{0.0} & \makecell[c]{200.0} & \makecell[c]{No options} & \makecell[c]{0.0} & \makecell[c]{200.0}\\
\hline
\end{tabular}
\caption{Numeric results on the $n$-agent four-room tasks with random grouping and dynamic influence among the agents}
\end{table}

\begin{table}[htbp]
\centering
\begin{tabular}{ |p{2cm}|p{1.0cm}|p{1.0cm}||p{2cm}|p{1.0cm}|p{1.0cm}|}
\hline
\multicolumn{3}{|c||}{Four-agent maze task} & \multicolumn{3}{c|}{Six-agent maze task} \\
\hline \hline
 \cellcolor{gray} \makecell[c]{CentQ+Force} & \makecell[c]{Value} & \makecell[c]{Step} & \cellcolor{gray} \makecell[c]{CentQ+Force} & \makecell[c]{Value} & \makecell[c]{Step}\\
\hline
\makecell[c]{Multiple} & \makecell[c]{\textbf{0.602}} & \makecell[c]{\textbf{53.15
}} &  \makecell[c]{Multiple} & \makecell[c]{\textbf{0.512}} & \makecell[c]{\textbf{72.06}}\\
\hline
\makecell[c]{Single} & \makecell[c]{0.0} & \makecell[c]{200.0} & \makecell[c]{Single} & \makecell[c]{0.0} & \makecell[c]{200.0}\\
\hline
\makecell[c]{No options} & \makecell[c]{0.0} & \makecell[c]{200.0} & \makecell[c]{No options} & \makecell[c]{0.0} & \makecell[c]{200.0}\\
\hline \hline
\multicolumn{3}{|c||}{$2\times2$ maze task} & \multicolumn{3}{c|}{$3\times2$ maze task} \\
\hline \hline
 \cellcolor{gray} \makecell[c]{CentQ+Force} & \makecell[c]{Value} & \makecell[c]{Step} & \cellcolor{gray} \makecell[c]{CentQ+Force} & \makecell[c]{Value} & \makecell[c]{Step}\\
\hline
\makecell[c]{Multiple} & \makecell[c]{\textbf{0.450}} & \makecell[c]{\textbf{82.89
}} &  \makecell[c]{Multiple} & \makecell[c]{\textbf{0.380}} & \makecell[c]{\textbf{100.8}}\\
\hline
\makecell[c]{Single} & \makecell[c]{0.0} & \makecell[c]{200.0} & \makecell[c]{Single} & \makecell[c]{0.0} & \makecell[c]{200.0}\\
\hline
\makecell[c]{No options} & \makecell[c]{0.0} & \makecell[c]{200.0} & \makecell[c]{No options} & \makecell[c]{0.0} & \makecell[c]{200.0}\\
\hline
\end{tabular}
\caption{Numeric results on the the $n$-agent maze tasks with random grouping or sub-task grouping}
\end{table}


\end{document}